\documentclass[11pt]{cui}

\usepackage{graphicx}
\usepackage{amsfonts}
\usepackage{amssymb}
\usepackage[mathscr]{eucal}
\usepackage{amsthm,amsmath,bm,color,ctable}
\usepackage{enumerate,natbib}
\usepackage{dsfont, titling}
\usepackage{multirow}
\usepackage{secdot}
\usepackage{mathrsfs}
\usepackage{epsfig}
\usepackage{ctable}

\newtheorem{exm}{\hspace{1mm}Example}

\newcommand{\etal}{{\em et al}}

\newcommand{\bX}{\mbox{\bf X}}

\newcommand{\rE}{\mbox{E}}

\newtheorem{theorem}{Theorem}

\newtheorem{proposition}{Proposition}

\sectiondot{subsection}
\sectiondot{subsubsection}
\makeatletter \@addtoreset{equation}{section}
\renewcommand\@biblabel[1]{}
\renewcommand{\cite}{\citeasnoun}
\def\baselinestretch{1.25}

\def\Y{\textbf{Y}}
\def\X{\textbf{X}}

\def\pen{\mathscr{P}}
\def \betabf{\mbox{\boldmath $\beta$}}
\def \spc{{\cal S}}

\def\B{\mbox{\boldmath $\beta$}}
\def\BB{\mbox{\boldmath $\scriptstyle \beta$}}
\def\F{\mathscr{F}}
\def \alphabf{\mbox{\boldmath $\alpha$}}

\def\t{\mbox{\boldmath $t$}}

\newcommand{\boxdiv}{\,\hbox{\raise-.4ex\hbox{\Large $\Box$}\kern-2.1ex$\div$}\,}

\begin{document}

\title{\Large\bfseries Component Selection in the Additive Regression Model}
\vspace{-0.8cm}
\author{ Xia Cui     (\tt{\it\small cuixia@mail.sysu.edu.cn })\\
                    {\it\small School of Mathematics and Computer Science, Sun Yat-Sen University, Guangzhou, China}\\
        Heng Peng   (\tt{\it\small hpeng@hkbu.edu.hk })\\
                    {\it\small Department of Mathematics, Hong Kong Baptist University, Hong Kong} \\
        Songqiao Wen (\tt{\it\small wensongqiao@gmail.com})\\
                     {\it\small College of Mathematics and Computational Science, Shenzhen University, China}\\
        Lixing Zhu (\tt{\it\small lzhu@hkbu.edu.hk})\\
                   {\it\small Department of Mathematics, Hong Kong Baptist University, Hong Kong}\\
                    {\it\small School of Finance and Statistics, East China Normal University, Shanghai, China} }

\date{March 22, 2010}

\maketitle

\begin{abstract} Similar to variable selection in  the linear regression model,
selecting significant components  in the popular additive regression
model is of great interest. However, such components are unknown
smooth functions of independent variables, which are unobservable.
As such, some approximation is needed. In this paper, we suggest a
combination of penalized regression spline approximation and group
variable selection, called the lasso-type spline method (LSM), to
handle this component selection problem with a diverging number of
strongly correlated variables in each group. It is shown that the
proposed method can select significant components  and estimate
nonparametric additive function components simultaneously with an
optimal convergence rate simultaneously. To make the LSM stable in
computation and able to adapt its estimators  to the level of
smoothness of the component functions, weighted power spline bases
and projected weighted power spline bases are proposed. Their
performance is examined by simulation studies across two set-ups
with independent predictors and correlated predictors, respectively,
and appears  superior to the performance of competing methods. The
proposed method is extended to a partial linear regression model
analysis with real data, and gives reliable results.
\end{abstract}


\def\baselinestretch{1.5}

\section{\label{sec-data} Introduction}

Consider the additive regression model
\begin{eqnarray}\label{addmodel}
Y_i=\alpha+\sum_{k=1}^K f_k(X_{ki})+\varepsilon_{i},
\end{eqnarray}
where $X_{ki}$ are the components of  $X_i=(X_{1i}, \ldots,
X_{Ki})$, $Ef_k(X_{ki})=0$, $\{ f_k(\cdot),~k=1,\ldots,K \}$ are
unknown smooth functions, and $\{\varepsilon_i\}$ is a sequence of
i.i.d random variables with a mean of 0 and a finite variance
$\sigma^2$. This model was first proposed by Friedman and Stuetzle
(1981), and  has become a popular multivariate nonparametric
regression model in practice. Hastie and Tibshirani (1990) gave a
comprehensive review of this model and showed that it could be
widely used in multivariate nonparametric modeling.

The additive model provides an approximation, with an additive
structure, for multivariate nonparametric regression. There are at
least two benefits of such an additive approximation.  First, as
every single individual additive component can be estimated using a
univariate smoother in an iterative manner, the so-called ``curse of
dimensionality'' that besets multivariate nonparametric regression
is largely avoided. Stone (1985, 1986) theoretically confirmed this
by showing that one can construct an estimator of $f$ that achieves
the same optimal convergence rate for a general value of $K$ as for
$K=1$. Second, the estimate of each individual component explains
how the dependent variable changes with the corresponding
independent variables; essentially, the simpler structure improves
the interpretability of the model.

There are several methods available in the literature for fitting
the additive model. These include the  backfitting algorithm
(Friedman and Stuetzle 1981; Buja, Hastie and Tibshirani 1989;
Opsomer and Ruppert 1998), the smooth backfitting algorithm (Mammen,
Linton and Nielsen 1999, Mammen and Park 2005, Nielsen and Sperlich
2005; Mammen and Park 2006; Yu, Park and Mammen 2008), marginal
integration estimation methods (Tj{\o}stheim and Auestad 1994;
Linton and Nielsen 1995; Fan \etal. 1998), the Fourier series or
wavelets approximation approach (Amato, Antoniadis and De Feis 2002;
Amato and Antoniadis 2001; Sardy and Tseng 2004), the penalized
B-splines method (Eilers and Marx 2002),  among others.

To make the additive model more efficient, the search for a
parsimonious version is clearly of importance. Although estimation
has been intensively investigated,
 insignificant independent variables and
function components increase the complexity of the model, which
leads to a great computational burden and numerical unstability.
Hence, deriving a method for obtaining  estimations in a
parsimonious additive model that still achieve an optimal
convergence rate, as is the case with only one nonparametric
component, is an interesting issue.

We use a real data example to demonstrate why selecting significant
components and searching for a parsimonious additive model is of
importance for statistical additive modeling. Fan and Peng (2004)
used an additive model and penalized SCAD least-squares to analyze
the employee dataset of the Fifth National Bank of Springfield based
on data from 1995 (see Example 11.3 in Albright {\sl et al.} 1999).
The bank, whose name has since changed, was charged in court with
paying its female employees substantially smaller salaries than its
male employees. For each of its 208 employees, the dataset includes
the following variables. \vspace{-0.5cm}
\begin{itemize}
\item  EduLev:  education level, a categorical variable with categories
        1 (finished high school), 2 (finished some college courses),
        3 (obtained a bachelor's degree), 4 (took some graduate courses),
        5 (obtained a graduate degree).
\vspace{-0.35cm}
\item  JobGrade:  a categorical variable indicating the current job
        level, the possible levels being 1-6 (6 is the highest).
\vspace{-0.3cm}
\item YrHired:  the year that an employee was hired.
\vspace{-0.3cm}
\item YrBorn: the year that an employee was born.
\vspace{-0.3cm}
\item Gender:  a categorical variable that takes the value ``Female'' or
``Male''.
\vspace{-0.3cm}
\item YrsPrior: the number of years of work experience  that employee had at another bank
        before working at the Fifth National Bank.
\vspace{-0.3cm}
\item PCJob:  a dummy variable that takes the value of 1 if the empolyee's current
        job is computer-related, and 0 otherwise.
\vspace{-0.3cm}
\item Salary:  the current (year 1995) annual salary in thousands of dollars.
\end{itemize}

\begin{figure}\label{BK1}
\vspace{0in} \centerline{
\psfig{file=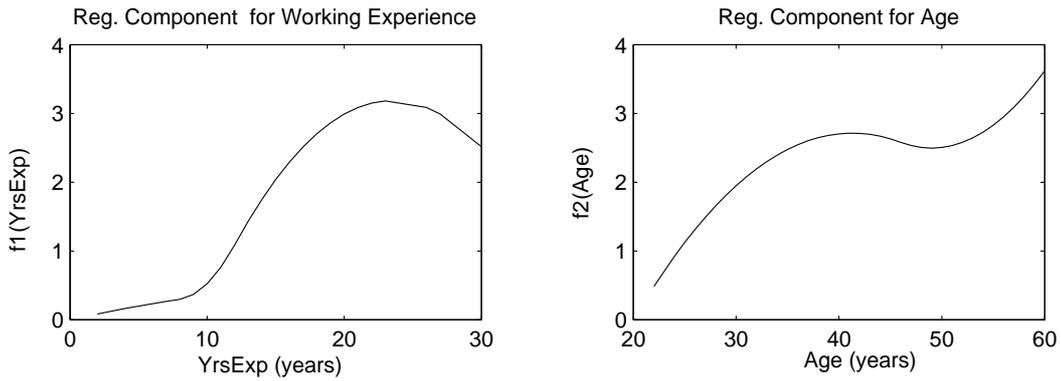,width=5.5in,height=2.0in} }
\vspace{0in} \caption{The estimated  regression functions of  ``Age'' or
``YrsExp'' respectively when both are included in the additive model.}
\end{figure}

\begin{figure}\label{BK2}
\vspace{0in} \centerline{
\psfig{file=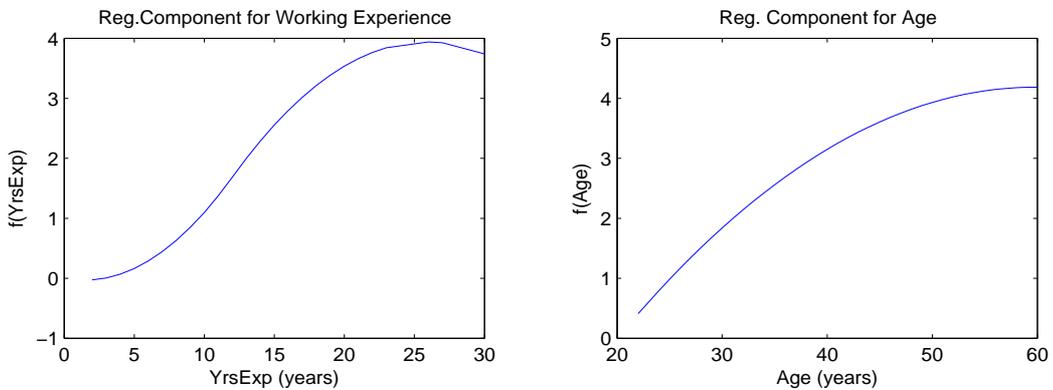,width=6.5in,height=2.0in} }
\vspace{0in} \caption{The estimated  regression functions of ``Age'' or
``YrsExp'' respectively when both are included in the respective
additive model.}
\end{figure}

Based on  the discussions of Lam and Fan (2008) and Zhang (2008),
both YrsExp and Age should have a nonlinear relationship with
``Salary'', and an additive model should be an appropriate model to
fit the data. The $R^2$ is $0.8123$. In the model, the nonparametric
components of Age and YrsExp are included. This is informally
confirmed by Figure \ref{BK1}, which presents the estimated curves
of ``YrsExp'' and ``Age'', respectively. However, the estimated
function $f_1(\mathrm{YrsExp})$ is  not an increasing function. This
is inconsistent with the general intuition that salary should
increase with ``YrsExp''. It is natural  to explore the reasons
behind this inconsistency. As we might suppose that ``Age'' and
``YrsExp'' will be strongly correlated, we may naturally ask whether
the phenomenon regarding ``YrsExp'' is caused by inappropriately
including insignificant variables or components
in the model.  
To demonstrate the necessity of component selection, we manually
remove one component to see what happens. That is, we consider two
additive models, each of which includes either ``Age'' or
``YrsExp''. We find that the model without the ``Age'' component has
a larger $R^2(0.8144)$ than the model with both ``Age''
and``YrsExp'' , and the model without the ``YrsExp'' component has a
smaller $R^2$ value of $0.8052$. This indicates  that we should keep
``YrsExp'' in the model. More importantly, from Figure \ref{BK2}, we
can see that when the ``Age'' component is selected out, the
estimated function of ``YrsExp'' is an increasing function of
``YrsExp'', which fits the intuition. This also suggests that when
insignificant components are selected out, the remaining components
have a better explanatory power. As such,  the means of
automatically selecting the ``Age'' component out from the model is
of importance, because  we
 need to select out a nonparametric component rather than a
 variable that is observable. We thus need a new
 method to handle this modeling issue.

\subsection{Goals of the paper}

There have been some studies on variable selection in additive
modeling. Smith and Kohn (1996) proposed a Bayesian approach to
select significant variables. Chen and H$\ddot a$rdle (1995) used a
simple threshold method to select significant independent variables
for the additive model, in which the function components are
estimated by the marginal integration method. Shively, Kohn and Wood
(1999) proposed a hierarchical Bayesian approach to variable
selection and function estimation that uses a data-driven prior, and
estimated their functions by model averaging. Lin and Zhang (2006)
penalized the norm of the two-order derivative of component
functions to obtain sparse additive model in which the functions are
estimated by using a smoothing spline technique. Ravikumar \etal.
(2009) proposed a new method to produce sparse additive model based
on the idea of group variable selection and nonnegative garrote
variable selection. All of these methods involve decoupled smoothing
and sparsity, and penalize the norm of the estimated additive
component functions to produce a sparse additive model. Most are
based on classical variable selection methods for linear models, and
hence cannot select significant independent variables and estimate
the components simultaneously. The statistical properties of the
estimates are also difficult to analyze. Furthermore, these methods
impose a large computational burden, especially when there are  many
nonparametric function components to be estimated.

Recently, some attempts have been made to resolve these problems in
the additive models (see, for example, Meier, van de Geer and
B\"{u}hlmann 2009 and Huang, Horowitz and Wei 2009). These  methods
are  based on the B-spline and group variable selection techniques
and are capable of estimating and selecting component functions
simultaneously, even in high-dimension situations. However, the
approach developed by Meier, van de Geer and B\"{u}hlmann (2009)
seems to be unstable in the selection process, because it uses every
observation as a knot, which results in much fluctuation. The method
proposed by Huang, Horowitz and Wei (2009) does not provide optimal
estimates for the component functions. This is well known problem
with spline regression because the efficiency of the estimates
depends on the number and the position of the knots.

In this paper, we propose a lasso-type spline method (LSM) for
component selection and estimation. First, we use a penalized
regression spline approximation to parametrize the nonparametric
components in the additive model, and then consider the spline
approximation as a group of variables for selection. It is worth
mentioning that in our setting, the design matrix in each group is
formed from the truncated power spline basis functions. Hence, there
is a diverging number of strongly correlated variables in each
group, which makes the study more complicated and difficult.
Nevertheless,  the estimate of every single function component
achieves the same optimal convergence rate as that in univariate
local adaptive nonparametric regression splines, and our final
selected model is rather parsimonious.  To make the LSM in stable in
computation and able to adapt its estimates to the level of
smoothness of the component function, weighted penalized regression
splines method and projected weighted penalized regression splines
method are proposed. The two-stage estimation is obtained by using
one-dimensional non-parametric techniques to refine the estimates in
the first stage, which serve as initial approximations for the
additive components. Our proposed procedure depends on only one
parameter, which controls both prediction error and
misclassification error. Hence, to a certain degree, it reduces the
computational burden and attains computational stability. Simulation
results illustrate that the method is superior in a set-up with
independent predictors, and is comparable when the predictors are
correlated.

The outline of the remainder of this paper is as follows. In Section
2, we describe our new method, study its asymptotic properties, and
propose an approximation algorithm. In section 3, simulations and a
real data application are presented to illustrate the performance of
the proposed method. A brief conclusion and discussion are given in
Section 4. The technical details of the proof are relegated to
Section \ref{addproof}.

\section{Methodology}

\subsection{Penalized regression splines}
As the components in the additive model are unobservable
nonparametric functions, it is impossible to perform selection
directly, and an approximation is needed. To this end, we first
examine the univariate nonparametric regression model with only one
independent variable as a basis for our method.
$$
Y_i=m_0(X_i)+\epsilon_i,~i=1,\ldots,n,
$$
where $X_i$ is in $[0,~1]$. Mammen and Van de Geer (1997) proposed
the use of  the total variation $TV(m_0^{(p-1)})$ of the function
$m_0(\cdot)$ as a penalty and to minimize the following penalized
sum of the squared residuals to obtain the estimation of
$m_0(\cdot)$,
$$
F_{p,\lambda}=\sum_{i=1}^n(Y_i-m(X_i))^2+\lambda TV(m^{(p-1)}).
$$
As with the smoothing spline, Mammen and Van de Geer (1997) proved
that the minimizer of this equation falls into the spline space such
that the estimate of $m_0(\cdot)$ itself is also a spline function.
They also showed that the estimate of $m_0(\cdot)$ has some good
asymptotic properties, such as local adaption and an optimal
convergence rate.

To implement their idea, consider the following spline space
$\mathcal{S}(p,\t)$ with knots
$$
\t=\{0=t_0<t_1<\ldots<t_{k+1}=1\}.
$$
For  $p\geq2$, $\spc(p,\t)$ is defined as
$$
\spc(p,\t)=\{s(x)\in C^{p-2}[0,1] {~\rm where~} s(x)~~ {\rm is ~ a
~polynimial~ of~ the~order~} p {\rm~ on~ each~ subinterval~}
[t_i,t_{i+1}]\}.
$$
When $p=1$, $\spc(p,\t)$ is the set of step functions with jumps at
the knots.

It is known that the space $\spc(p,\t)$ is a $k+p$ dimensional
linear function space, and that the truncated power function series
$$
\X_x=\{1, x, x^2, \ldots, x^{p-1}, (x-t_1)_+^{p-1}, \ldots,
(x-t_k)_+^{p-1}\}
$$
forms its basis (see de Boor, 1978). Thus, if the number of knots
$k$ is sufficiently large, then we can approximate $m_0(x)$ by a
spline function with the form
\begin{eqnarray}
m(x,\betabf)=\X_x\betabf=\beta_0+\beta_1x+\ldots+\beta_{p-1}x^{p-1}+\sum_{i=1}^k\beta_{p+i-1}(x-t_i)_+^{p-1}.
\end{eqnarray}
Note that
$$
TV(m^{(p-1)}(x,\betabf))=\sum_{i=1}^k|m^{(p-1)}(t_i,\betabf)-m^{(p-1)}(t_{i-1},\betabf)|=
(p-1)!\sum_{i=1}^k|\beta_{p-1+i}|.
$$
By minimizing
\begin{eqnarray}
\min_{\betabf}\sum_{i=1}^n(Y_i-m(x_i,\betabf))^2+\lambda\sum_{i=1}^p|\beta_{p-1+i}|,
\end{eqnarray}
we can obtain an estimate $m(x,\hat\betabf)$ for the function
$m_0(\cdot)$.

\subsection{Component  selection for the additive model}
We now return to the additive regression model
$$
Y_i=\alpha+\sum_{k=1}^Kf_k(X_{ki})+\varepsilon_i,~~ i=1,\ldots,n.
$$
For every function component, we assume that
$Ef_k(\cdot)=0,~k=1,\ldots,K$, is approximated by the spline
function
\begin{eqnarray}
f_k^*(x)=\beta_{k0}+\beta_{k1}x+\ldots+\beta_{k(p-1)}x^{p-1}
+\sum_{j=1}^{p_k}\beta_{k(p_k-1+p)}(x-t_{kj})_+^{p-1};
\end{eqnarray}
where $\{t_{kj},~j=1,\ldots,p_k\}$ is the series of knots  for the
$k$th function component.

For any $k$, let $\{B_{kj}(\cdot), ~j=1,\ldots,p_k\}$ be the spline
bases (note that although the number of bases should be $p_k+p-1$,
for convenience, we still denote it as $p_k$). The additive model
can then be approximated by the following linear model.
\begin{eqnarray}
Y_i=\alpha^*+\sum_{k=1}^K\Big
(\sum_{j=1}^{p_k}\beta_{kj}B_{kj}(X_{ki})\Big
)+\varepsilon_i,~i=1,\ldots,n.
\end{eqnarray}
For any $k$ with $1\le k\le K$, the basis series $B_{kj}(\cdot)$,
$j=1,\ldots,p_k$ can be regarded as a natural group of variables in
the foregoing linear model, and the group variable selection can be
used to estimate $\beta_{kj}$ and to select the grouped variables.
We combine the hierarchical LASSO method (Zhou and Zhu's 2007), the
group bridge approach (Huang \etal. 2009), and the ideas of Mammen
and Van de Geer (1997) and propose the criterion
\begin{eqnarray}\label{addspline}
\min_{\alpha^*, ~ \beta_{kj}}\sum_{i=1}^n\left\{Y_i-\alpha^*-
\sum_{k=1}^K\sum_{j=1}^{p_k}\beta_{kj}B_{kj}(X_{ki})\right\}^2
+\lambda\sum_{k=1}^K\sqrt{|\beta_{k1}|+|\beta_{k2}|+\ldots+|\beta_{kp_k}|}.\nonumber\\
\end{eqnarray}
to select  the  groups. That is, we simultaneously to select the
significant components, and estimate the parameters $\beta_{kj}$.

However, this linear approximation does not mean that the problem is
exactly identical to the case in the classical linear model. First,
to make a good approximation of the function $f_k(\cdot)$, $p_k$,
the number of basis functions in the spline approximation, must be
sufficiently large, and theoretically, increases with the sample
size $n$. Thus, even when $K$, the number of function components, is
of a moderate size, the linear structure derived has a diverging
number of predictors if we do want to regard the model as linear.
Second, the grouped variables $B_{kj}(\cdot),j=1,\ldots,p_k$, are
all related to the variable $X_k$, and  are thus strongly
correlated, especially when the power basis functions are used.
Third, distinct from Zhou and Zhu (2007), in our setup, the
estimation accuracy of the whole function, rather than the
estimation accuracy of a particular coefficient, is of interest and
importance. Thus,  as the objective here is to find a good
approximation of each function component, the asymptotical results
obtained by Zhou and Zhu (2007), and Huang \etal. (2009) can not be
directly applied to the additive model.

\subsection{Asymptotic theory}
To study the asymptotic behavior of the model, we first consider  a
more general situation. Let $\F$ be a class of functions on $[0,1]$.
For a linear subspace  $\F_n$ of $\F$, we consider a penalty
$\mathcal{P}: \F_n \to [0,\infty)$ that satisfies
$$
\mathcal{P} (f_1+f_2) \le \mathcal{P}(f_1)+\mathcal{P}(f_2), \quad
f_1,f_2 \in \F_n,
$$
and
$$
\mathcal{P}(\alpha f) \le |\alpha| \mathcal{P} (f), \quad f \in
\F_n, \alpha \in \mathbf{R}.
$$

Consider the additive model (1.1) with $f_k \in \F, k=1,\ldots,K.$
For a tuning variable $\lambda_n$, $\hat{f}_k, k=1,\ldots,K$ are
estimated by minimizing the penalized sum of squares over $\F^K_n$:
$$
\{\hat{f}_1,\ldots,\hat{f}_k\}=\mbox{arg}\min\limits_{f_1,\ldots,f_K
\in \F_n} \left\{ \frac{1}{n}\sum\limits_{i=1}^n
\left(Y_i-\alpha-\sum\limits_{k=1}^K f_k(x_i)\right)^2+\lambda_n
\sum\limits_{k=1}^K \mathcal{P}^{\frac12}(f_k)\right\}.
$$
Write $\F_n(1)=\{f \in \F_n; \mathcal{P} \le 1\}$. For a subset
$\mathscr{A}$ of $\F$, we denote the $\delta$ entropy of
$\mathscr{A}$ by $\log N_2(\delta, \|\cdot\|_n, \mathscr{A})$. This
is the logarithm of the minimal number of  balls of a radius
$\delta$
 covering $\mathscr{A}$,  where $\|\cdot\|_n$ is
the $L_2$-norm with respect to the empirical probability measure of
$x_i$ $i=1, \ldots, n$ with the form  $
\parallel g \parallel_n^2=\frac{1}{n}\sum_{i=1}^n g^2(x_i).
$ To obtain the required result, we must first assume the following
condition first.

\noindent {\sc Condition 1} ~~~The errors $\varepsilon_1,
\varepsilon_2, \ldots, \varepsilon_n$ are independent, with
$E\varepsilon_i=0,$  and have subgaussian tails. That is, there
exist some positive $\beta$ and $ \Gamma$ such that
$$\rE[\exp(\beta\varepsilon_i^2)]\leq\Gamma\leq\infty,~~~~\mathrm{for}~i=1,~\ldots,~n.$$

\begin{theorem}
Assume that Condition 1 holds. Let $c_n$ be a positive number
sequence such that for the functions $f_{k,n},k=1,\ldots K$ in
$\F_n$ we have
\begin{equation}
\|f_{k,n}-f_k\|_n=O(n^{-{1/(2+w)}}c_n^{w/{(2+w)}}) \quad \mbox{and}
\quad  \mathcal{P}(f_{k,n}) \le c_n, \ k=1,\ldots,K. \end{equation}
Let $\lambda_n=Cn^{-2/(2+w)}c_n^{1/2-(2-w)/(2+w)}.$ Furthermore,
assume that for some $C>0$ and $0<w<2$, the following entropy bound
condition is satisfied.
\begin{equation}
\log N_2(\delta, \|\cdot\|_n, \F_n(1)) \le C\delta^{-w} \quad
\mbox{for all} \ \delta>0.
\end{equation}
We then have
 $$  \|\hat{f}_{k}-f_k\|_n=O_p(n^{-1/(2+w)}c_n^{w/(2+w)}), \quad \mbox{and} \quad \mathcal{P}(\hat{f}_k)=O_p(c_n). $$
\end{theorem}

From (2.3)---(2.5), we can define the penalty functional as the
$L_1$ norm of the coefficients for the spline approximation
$f^\ast_k(\cdot)$, that is,
$$\mathcal{P}(f_k^*)=\sum_{k=1}^{p+p_k-1}|\beta_k|.$$

In fact, this gives
$\mathcal{P}(f_k^*)=\sum_{k=0}^{p-1}|\beta_k|+(p-1)!\cdot
\mathrm{TV}(f_k^{*(p-1)})$, where $\mathrm{TV}$ denotes the total
variation. We obtain the following result for the entropy of the
total variation space.
\begin{proposition}\label{lemm1}
Define $\mathscr F_n(1)=\{f\in\mathscr
F_n:\mathcal{P}(f)=\sum\limits_{k=1}^K \mathcal{P}(f_k) \leq 1\}$.
There then exists a constant $M>0$ such that
\begin{eqnarray}
\log N_2(\delta,\|\cdot\|_n,\mathscr{F}_n(1))\leq M \delta^{-1/p}.
\end{eqnarray}

\end{proposition}

To state our results for the asymptotic behavior of the
penalized least-squares estimate (2.5), we need  some further conditions. \\
\noindent {\sc Condition 2} For any $j$
$$\max_{1\leq i\leq k_j}|h_{ji+1}-h_{ji}|=o(k_j^{-1}),~~\frac{\max_{1\leq i\leq k_j}h_{ji}}{\min_{1\leq i\leq k_j}h_{ji}}\leq M,$$
where $h_i^j=t_{ji}-t_{ji-1}$, $k_j$ is the number of knots, and
$M>0$ is a predetermined constant.\\

\begin{theorem}
Assume that  $f_k\in\F, k=1,2,\ldots,K$, and that the total
variation of its $(p-1)$-th derivative is bounded. Let
$\lambda_n=Cn^{-\frac{2p}{2p+1}}$, where $C$ is a large constant.
Then, under  Conditions 1 and 2, we have
\begin{equation}
 \parallel \hat
{f}_{k}-f_{k}\parallel_{n}=O_p(n^{-\frac{p}{1+2p}})
\end{equation}
for $\min\limits_{1 \le k \le K} p_k>n^{\frac{1}{2p+1}}$.
\end{theorem}

\subsection{Computation}
\subsubsection{\label{OPRS} Algorithms} The
penalty function in (\ref{addspline}) can be regarded as nonconcave.
Hence, the quadratic approximation method and the iterative
algorithm proposed by Fan and Li (2001) can be used to define
estimates of the coefficients. First,  consider the derivative of
penalty function for $\beta_{kj}$. Let
$\pen(\betabf_k)=\sqrt{|\beta_{k1}|+\ldots+|\beta_{kp_k}|}$.  This
gives
\begin{eqnarray}
\pen'(\beta_{kj})=\frac{\lambda\cdot {\rm
sgn}(\beta_{kj})}{\sqrt{|\beta_{k1}|+\ldots+|\beta_{kp_k}|}}
=\frac{\lambda\cdot\beta_{kj}}{|\beta_{kj}|\sqrt{|\beta_{k1}|+\ldots+|\beta_{kp_k}|}}.
\end{eqnarray}
To simplify the notation, we rewrite (\ref{addspline}) in matrix
form as
\begin{eqnarray}\label{addspline2}
(\Y-\alphabf-\X\betabf)^2+\lambda\pen(\betabf),
\end{eqnarray}
where $\X=(X_1,X_2,\ldots,X_n)^T$,
\begin{eqnarray}
X_i&=&\{B_{11}(X_{1i}),\ldots,B_{1p_1}(X_{1i}),B_{21}(X_{2i}),\ldots,B_{2p_2}(X_{2i}),\ldots,B_{Kp_K}(X_{Ki})\}^T\nonumber\\
&\triangleq&\{1,X_{\beta_{11}i},\ldots,X_{\beta_{1p_1}i},X_{\beta_{21}i},\ldots,X_{\beta_{2p_2}i},\ldots,X_{\beta_{Kp_K}i}\}^T\nonumber
\end{eqnarray}
and $\pen(\betabf)=\sum_{k=1}^K\pen(\betabf_k)$.

If $\betabf^*$ with nonzero coefficients
$(\alpha^*,\betabf_1^*,\ldots,\betabf_K^*)$ minimizes the equation
(\ref{addspline2}), then the following equation is satisfied.
\begin{eqnarray}\label{addspline3}
\betabf^*=(\X_{\BB^*}^T\X_{\BB^*}+\Sigma_{\lambda}(\betabf^*))^{-1}\X_{\BB^*}^T\Y,
\end{eqnarray}
where $\X_{\BB^*}=(X_1^*,X_2^*,\ldots,X_n^*)^T$, and
$$
X_i^*=\{1,X_{\beta_{11}^*i},\ldots,X_{\beta_{Kp_K}^*i}\}^T,
$$
and
$$
\Sigma_{\lambda}(\betabf^*)=\mbox{diag}\{0,\pen'(\betabf_1^*)/{\betabf_1^*},\ldots,
\pen'(\betabf_K^*)/{\betabf_K^*}\}.
$$
Hence, as in Fan and Li (2001), given an initial value $\betabf_0$,
(\ref{addspline3}) requires an iterative algorithm to update the
estimate to $\betabf_1$ according to the following equation
$$
\betabf_1=(\X_{\BB_0}^T\X_{\BB_0}+\Sigma_{\lambda}(\B_0))^{-1}\X_{\BB_0}^T\Y.
$$
Fan and Li (2001) suggested that this iterative step is similar to
the one-step MLE if the initial value is sufficiently good. If a
reasonable initial value of $\betabf$ is selected, then our
algorithm should converge within a few steps.

\subsubsection{Tuning parameter selection}

The tuning parameter $\lambda$ is very important for estimating
$\betabf$. Fan and Li (2001) proposed using generalized
cross-validation to select $\lambda$. Let $\hat{\B}(\lambda)$ be the
estimate of $\B$ with the tuning parameter $\lambda$. The
generalized cross-validation statistic is defined as
\begin{eqnarray}
\mbox{GCV}(\lambda)=\frac{1}{n}\frac{\parallel\Y-\X_{\hat{\BB}(\lambda)}\hat{\B}(\lambda)\parallel^2}{\{1-e(\lambda)/n\}^2}
\end{eqnarray}
and $$\hat\lambda={\rm argmin}_{\lambda}\{\mbox{GCV}(\lambda)\},$$
where $e(\lambda)={\rm trace}[P_{X}(\hat\betabf(\lambda))]$,
$P_X(\hat\betabf)=\X_{\hat{\BB}}(\X^T_{\hat{\BB}}\X_{\hat{\BB}}+\Sigma_{\lambda}(\hat\betabf))^{-1}\X_{\hat{\BB}}^T$.

According to  Wang, Li, and Tsai (2007), the log(GCV) is very
similar to the traditional model selection criterion AIC. Although
AIC is an efficient selection criterion that selects the best
finite-dimensional candidate model in terms of prediction accuracy,
it is not a consistent selection criterion because it does not
select a correct model with a probability approaching 1 as the
sample size goes to infinity. However, for our proposed method the
number of knots, or the dimension of $\betabf$ is very large and
increases with the sample size $n$,  and thus an adjustment for such
a criterion is necessary. Accounting for the effect of
dimensionality to correctly select the significant variables, we
suggest using the inflated factor for GCV. A modified generalized
cross-validation(MGCV) is defined as
\begin{eqnarray}\label{MGCV}
\mbox{MGCV}(\lambda)=\frac{1}{n}\frac{\parallel\Y-\X\hat{\B}(\lambda)\parallel^2}{\{1-\gamma
e(\lambda)/n\}^2}
\end{eqnarray}
where $\gamma$ is the inflated factor. When $\gamma=1$, the MGCV is
no different from the GCV proposed by Fan and Li (2001). Based on
our experience and the discussions of Luo and Wahba (1997) and
Friedman and Silverman (1989), we suggest selecting  $\gamma$ within
the interval $(1.2, 3)$ as an extra penalty.

\subsection{Further Considerations}

\subsubsection{\label{WPRS}Weighted penalized regression splines method}

Our method is based on the power spline regression. It is well known
that the power spline regression is not stable in computation
because of a strong correlation between power bases, and many base
functions are  related to only a few observations. To make our
numerical results more stable, we weight the power spline base for
every component function as
$$
\mathbf{X}^{*w}= \mathbf{X}^*\times \mathbf{W},
$$
where $\mathbf{W}^2 =
\mbox{diag}\{(\mathbf{X}^{*T}\mathbf{X}^*/n)^{-1}\}$ and
$\mathbf{X}^*$ is given in (\ref{addspline3}). (\ref{addspline2})
can then be rewritten as
\begin{eqnarray*}
(\Y-\mathbf{X}^{*wT}\betabf^{*w})^2+\lambda\pen(\betabf^{*w}).
\end{eqnarray*}


By some elementary calculations, it is easy to determine that when
all of the components of $\mathbf{X}^{*w}$ are independent, the
variance of the least-squares estimate of $\betabf^{*w}$ should be
of the order $1/n$. Also, when the sample points
$X_{ki},i=1,\ldots,n$ are equally spaced, our method is equivalent
to transferring the power base spline approximation to a B-spline
approximation with an $L_1$-norm penalty of a linear combination of
the coefficients in the B-spline approximation. Furthermore, as the
variance of the least-squares estimate of $\betabf^{\ast}$ is of the
order $1/n$, as to the wavelet approximation (Donoho and Johnstone
1994), the universal threshold $\sqrt{\log n /n}$ can be used to
penalize each coefficient. In other words, the tuning parameter can
be searched within a small interval with the length $O(\sqrt{\log n
/{n} })$. These modifications result in  a stable final penalized
component function estimate.

\subsubsection{\label{PWPRS}Projected weighted penalized regression splines method}

To make our final estimated model  parsimonious and easy to
interpret, in addition to selecting  significant component
functions, we also suggest the following procedure for the component
estimation. Note that the power spline approximation (see the
definition of $\mathbf{X}_x$  above (2.1)) expands  the component
function as the sum of a polynomial and a linear combination of
truncated power base functions. Divide $\mathbf{X}^{*w}$ by
$(\mathds{1}, B^{*w}_1, \ldots, B^{*w}_k, \ldots, B^{*w}_K)$, with
$B^{*w}_k$ being the block from the
$(2+\sum\limits_{l=1}^{k-1}p_l)$-th column to the
$(1+\sum\limits_{l=1}^{k}p_l)$-th column in matrix
$\mathbf{X}^{*w}$. Here, $\mathds{1}$ is an $n\times 1$ column
vector in which all the  elements are equal to 1. We then write
$B^{*w}_{k}=(B^{*w}_{k1} \quad B^{*w}_{k2})$, where $B^{*w}_{k1}$
are the coefficients of the polynomial  part and $B^{*w}_{k2}$ are
the coefficients of the truncated power base functions. We regard
these as two groups and then penalize each group separately, which
make it possible to adaptively estimate the component functions when
they are actually polynomial functions without any great effect from
the
truncated power base functions. 
This provides a way of adaptively estimating the component functions
if they are actually polynomial, and means that the estimation is
adaptive to the level of smoothness of the component functions. This
approach may result in a more parsimonious estimation than that
obtained by the previous estimation algorithm. However, we note that
the two groups are strongly correlated. To realize the approach and
to make the algorithm more efficient, we consider the following
empirical power base functions.
$$
B^{*pw}_k=\{ B^{*w}_{k1} \quad (I-P_{B^{*w}_{k1}})B^{*w}_{k2}\},
$$
where $P_{B^{*w}_{k1}}$ is the project matrix from $B^{*w}_{k2}$ to
$ B^{*w}_{k1}$. This projection method is able to reduce the
correlation between the two groups in the spline approximation. Let
$$
\mathbf{X}^{pw} = (\mathds{1}, B^{*pw}_1, \ldots, B^{*pw}_k, \ldots, B^{*pw}_K).
$$
(\ref{addspline2}) can also be written as
\begin{eqnarray*}
(\Y-\mathbf{X}^{*pw}\betabf^{*pw})^2+\lambda\pen(\betabf^{*pw}_1)+\lambda\pen(\betabf^{*pw}_2),
\end{eqnarray*}
where $\pen(\betabf^{*pw}_1)$ are the group penalty functions for
the polynomial coefficients for all of the component additive
functions and $\pen(\betabf^{*pw}_2)$ are the group penalty
functions for the coefficients of the truncated power bases.

\subsubsection{Two-stage estimation}

When the dimension of the additive regression model is very high,
selecting significant component functions becomes very  difficult.
The model selection and estimation may not  be consistent, and the
estimation procedure may also become unstable. To improve the
 estimation accuracy, we suggest a two-stage estimation approach. In
the first stage, we use our proposed methods to select and estimate
significant component functions  as initial approximations of all of
the  selected components. Let
$$\mathcal{M}=\{k: f_k(X_k)\mbox{\,\,
selected to be significant based on our methods}\}$$ and denote the
corresponding estimates by $\hat{f}^0_k,\,k\in\mathcal{M}$.  In the
second stage, we obtain refined estimates as follows. For the
$f_s(\cdot)$ selected in the first stage, define
$$
Y^\ast_i=Y_i-\sum\limits_{k\in\mathcal{M}, k\neq s}
\hat{f}^0_k(X_{ki}),\,\, i=1,2,\ldots,n
$$
and then estimate $f_s(X_{si})$ non-parametrically using the
following model
$$
Y^\ast_i=f_s(X_{si})+\varepsilon^\ast_i, \,\,i=1,\ldots,n.
$$
For this new model, we can again use the method applied in the
first-stage estimation to obtain the final estimator of
$f_s(X_{si})$.

\section{Numerical studies}
\subsection{Simulations}\label{addsimu}

We conduct simulations to examine the effectiveness of the proposed
lasso-type spline method for  component function  selection and
estimation in the additive regression model. The algorithm proposed
in Section \ref{OPRS} is called  the original lasso-type spline
method (OLSM), that in Section \ref{WPRS} the weighted lasso-type
spline method (WLSM),  and that in Section \ref{PWPRS} the projected
weighted lasso-type spline method (PWLSM). We also compare the
results with those obtained using the sparsity-smoothness penalty
(SSP) approach recently proposed by Meier, van de Geer, and
B\"{u}hlmann (2009) by using the R packages provided by the authors.
For selection performance, we compute the true positive ratio (TPR)
and false positive ratio (FPR); and for  estimation accuracy, we
compute the empirical prediction mean square error (MSE). Letting
$\hat{f}_k$ be the estimator of $f_k$, MSE is defined as
$$
\frac{1}{n}\sum\limits_{i=1}^n |\hat{f}_k(X_{ki})-f_k(X_{ki}))|^2,
$$
where $\{Y_i, X_{1i}, \ldots, X_{Ki}\}$ are the data points.

In the simulations, the sample size $n=400$ and a total of  $100$
simulation runs are used. To reduce the computational burden, the
knots are designed as follows.  Let the number of knots be $k$. For
each predictor $X_i$,  the knots are selected to be the $[nj/k]$-th
order statistics  $\{X_{i([nj/k])}, j=1,\ldots,k\}$ of
$\{X_{j1},\ldots, X_{jn}\}$. Quadratic splines are used, which gives
a total number of base functions with $K$ function components of
$(2+k)K+1$. To check the sensitivity of the methods to the knot
number selection, we tried the values $10, 15, 40$, and $60$ with a
fixed $\lambda$, and found that the numerical results did not differ
much. We thus posit that our proposed three procedures are
insensitive to the initial knot number as long as it is sufficiently
large. However, with a larger number knots, the computation time is
grated and the performance is a little worse, as the computation may
be less stable due to strongly correlated variables in the splines.
We thus set $k$ at $15$ in the simulations. The penalty parameter
$\lambda$ is found to be critical. We choose $\lambda$ by computing
the MGCV criterion defined in (\ref{MGCV}) for a grid of $\alpha$
values and choosing the minimizer over the grid. The inflated factor
$\gamma$ is taken to be $1.5$. The grid of $\lambda$ for all three
proposed procedures has 100 values and satisfies the condition that
the values of $\log_{10}(\lambda)$ are equally spaced between $-5$
and $2$. The sparsity-smoothness penalty approach (SSP) require teh
selection of two parameters $\lambda_1$ and $\lambda_2$,  where the
former serves to control the sparsity and the latter the smoothness.
Both parameters are chosen by using $100$ grid points for
$\lambda_1$ and $15$ grid points for $\lambda_2$ in the spirit of
Meier, van de Geer, and B\"{u}hlmann (2009). The simulation
experiments  are similar to those in Example 1 and Example 3 of
Meier, van de Geer, and B\"{u}hlmann (2009).  As our focus is on
simultaneous selection and estimation, $K$ is chosen to be $50$
rather than an ultra-high dimension.

{\exm \label{example-ind}(Covariates are independent). The data are
generated from
$$
Y = f_1(X_1) + f_2(X_2) + f_3(X_3) + f_4(X_4) +
\sum\limits_{k=5}^{50}f_k(X_k)+\varepsilon,
$$
where
$$
\begin{array}{rll}
&f_1(x) = -\sin(2x), \mbox{\quad} f_2(x) = x^2-25/12, & f_3(x) = x,\\
& f_4(x) = \exp(-x)-2\sinh(5/2)/5, & f_k(x)=0, \mbox{\quad if
\quad}k\geq 5,
\end{array}
$$
and $\varepsilon\sim N(0,1)$. The predictors are sampled from the
uniform distribution of $(-2.5, 2.5)$. }

{\exm \label{example-cor}(Covariates are correlated). The model is
$$
Y = f_1(X_1) + f_2(X_2) + f_3(X_3) + f_4(X_4) +
\sum\limits_{k=5}^{50}f_k(X_k)+\varepsilon,
$$
with
$$
\begin{array}{rll}
& f_1(x) = 5x, \quad f_2(x)=3(2x-1)^2,\quad f_3(x) =
\frac{4\sin(2\pi x)}{2-\sin(2\pi x)}, \\
&f_4(x) = 0.6\sin(2\pi x) + 1.2\cos(2\pi x) +1.8\sin^2(2\pi x) +
2.4\cos^3(2\pi x) + 3\sin^3(2\pi x),\\
&  f_k(x)=0, \mbox{\quad if \quad}k\geq 5,
\end{array}
$$
and $\varepsilon\sim N(0,1.74)$. The covariates $X=(X_1, \ldots,
X_K)$ are generated from
$$
X_k = \frac{W_k+0.5U}{1+0.5}, \quad k=1,\ldots, K,
$$
where $W_1,\ldots,W_K$ and $U$ are i.i.d uniform(0, 1). This
provides a design with a correlation coefficient of $0.5$ between
all of the covariates. }

\begin{figure}
\vspace{-0.5in} \centerline{
\psfig{file=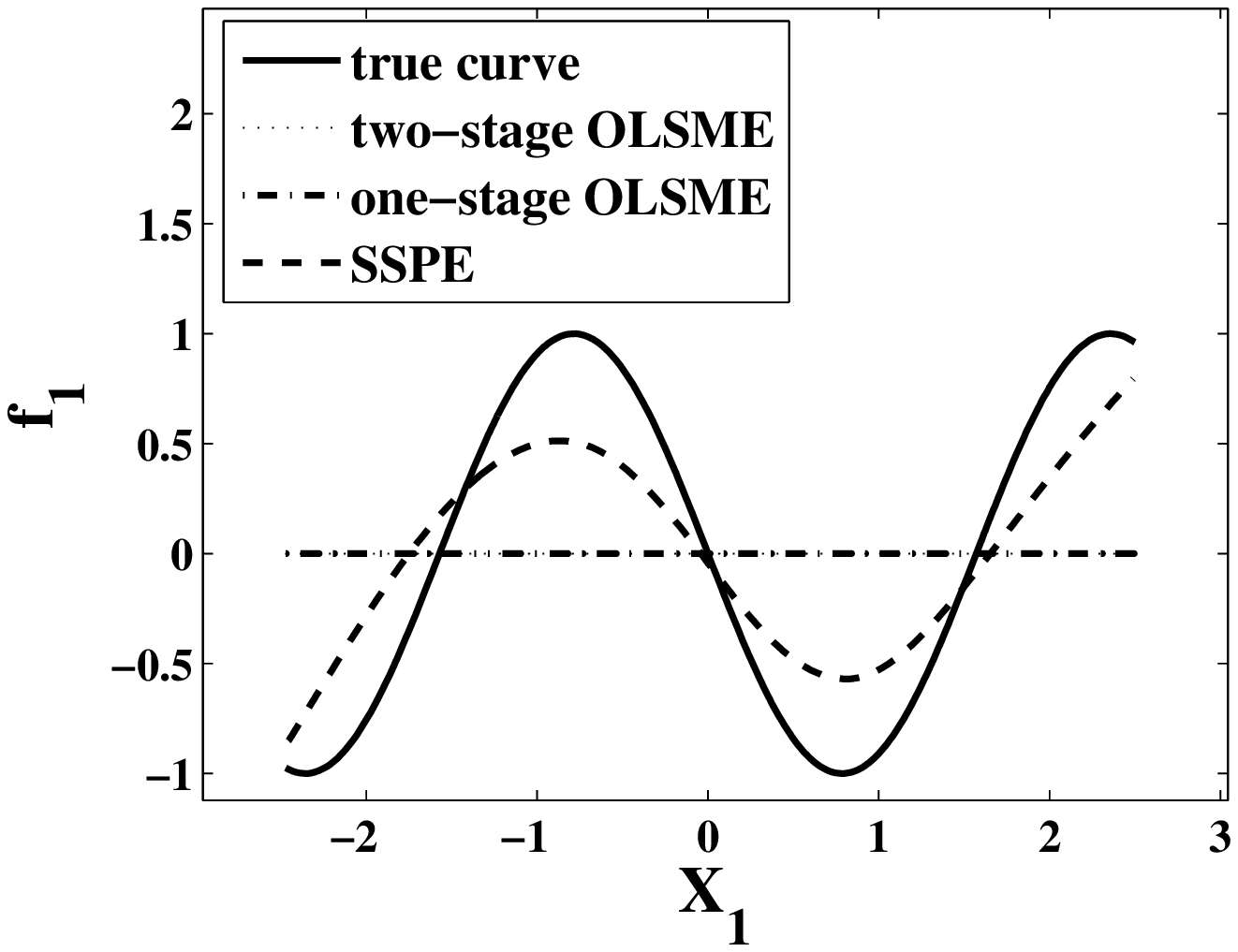,width=1.65in,height=1.85in}\hspace{-0.5cm}
\psfig{file=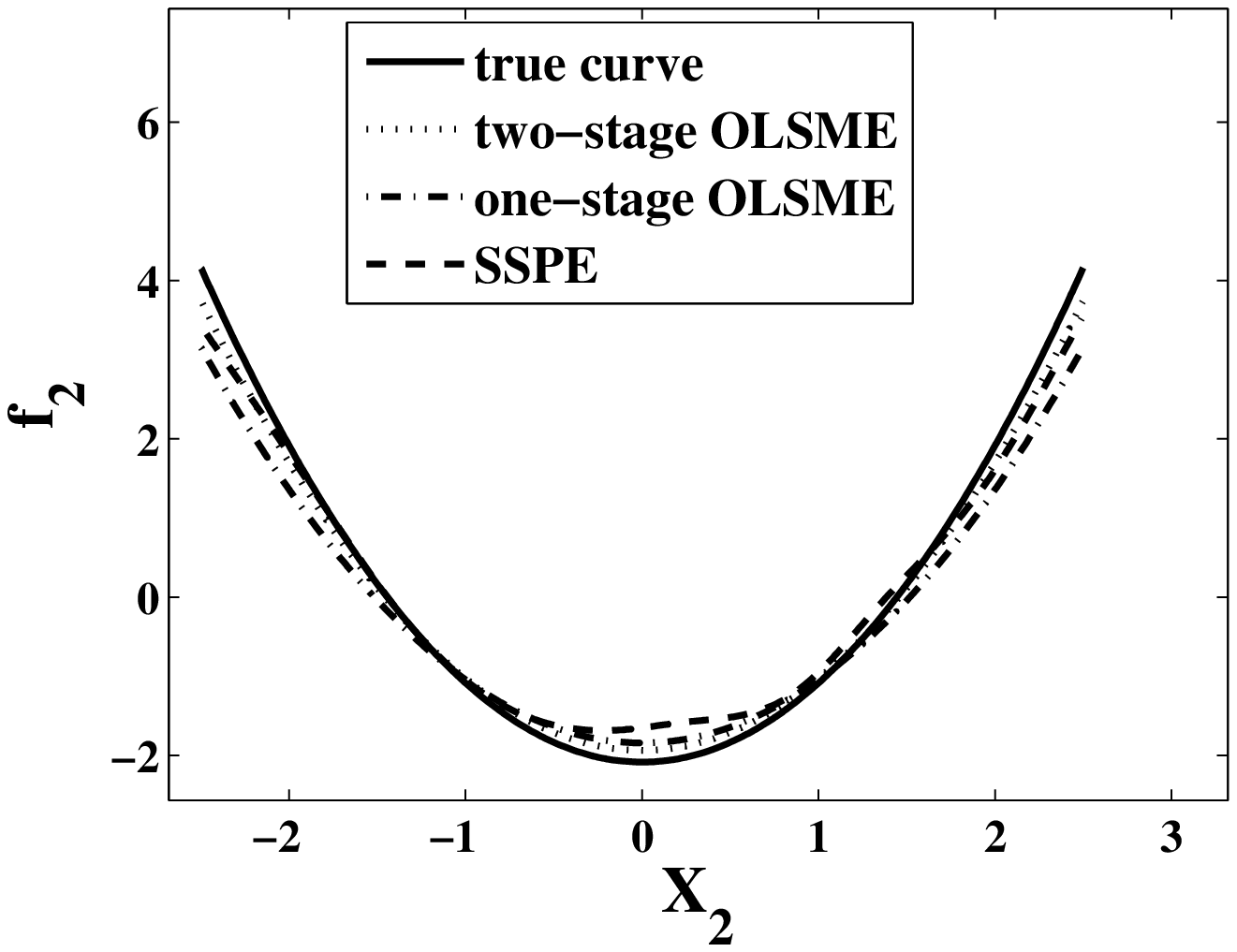,width=1.65in,height=1.85in}\hspace{-0.5cm}
\psfig{file=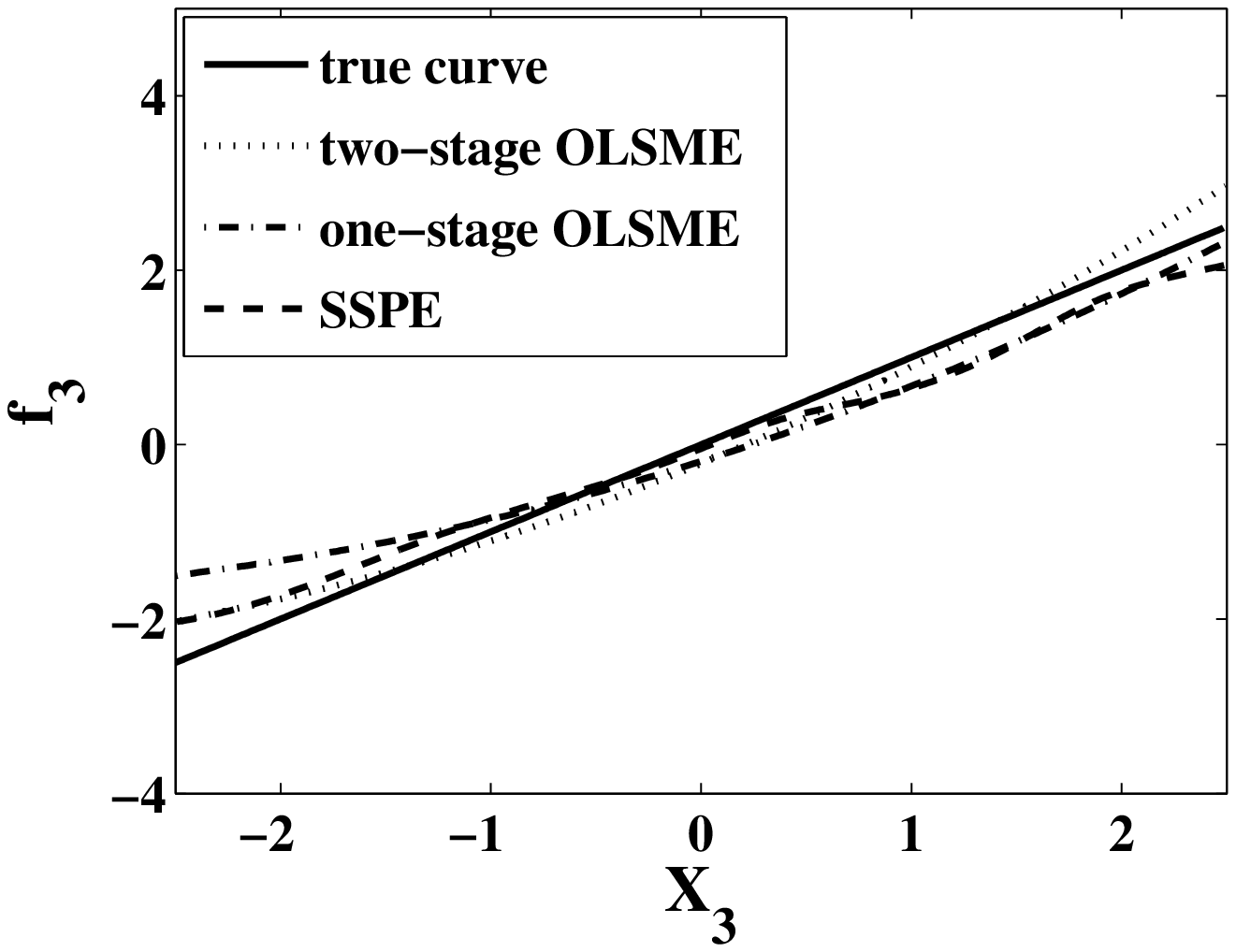,width=1.65in,height=1.85in}\hspace{-0.5cm}
\psfig{file=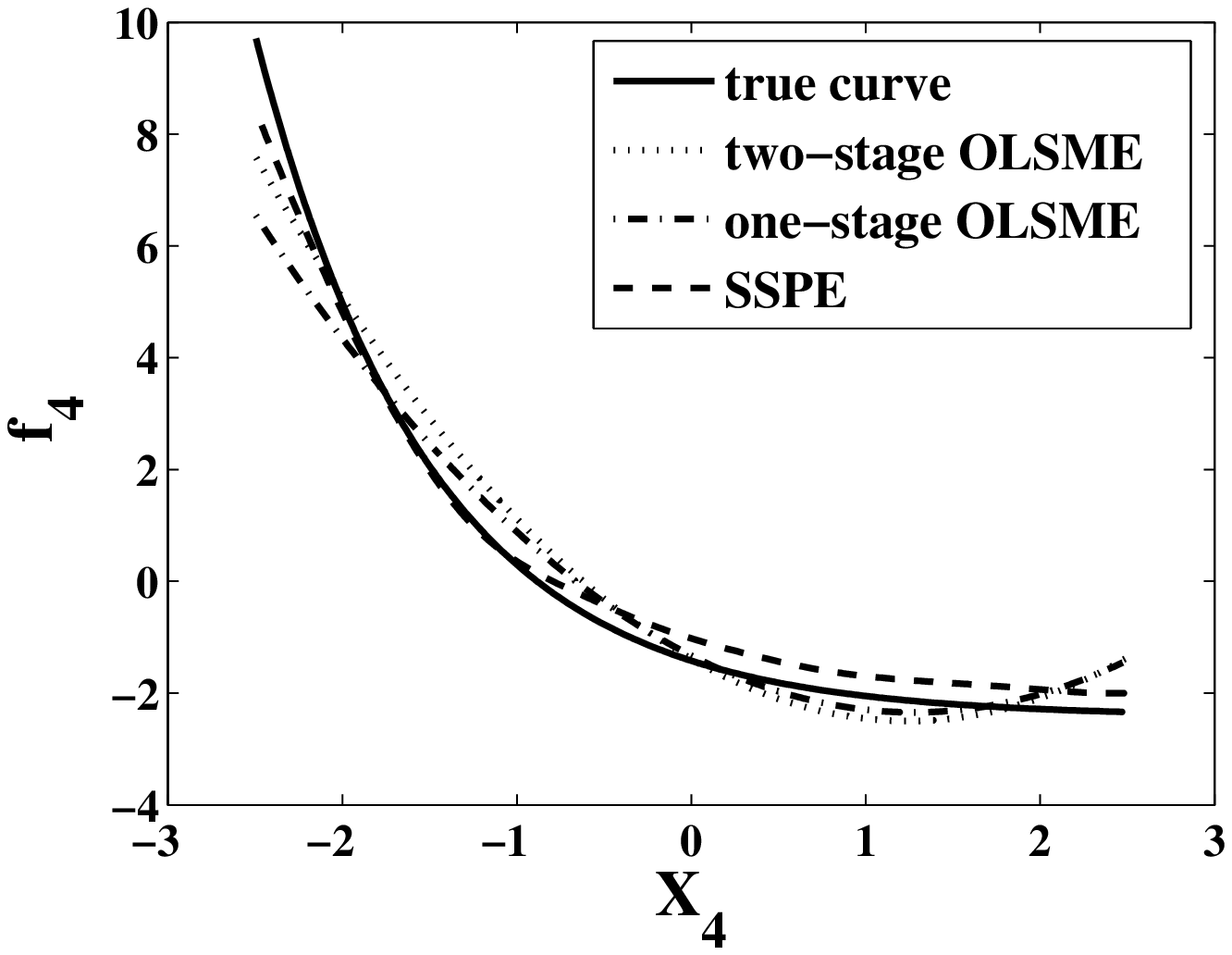,width=1.65in,height=1.85in}\hspace{-0.5cm}
} \centerline{
\psfig{file=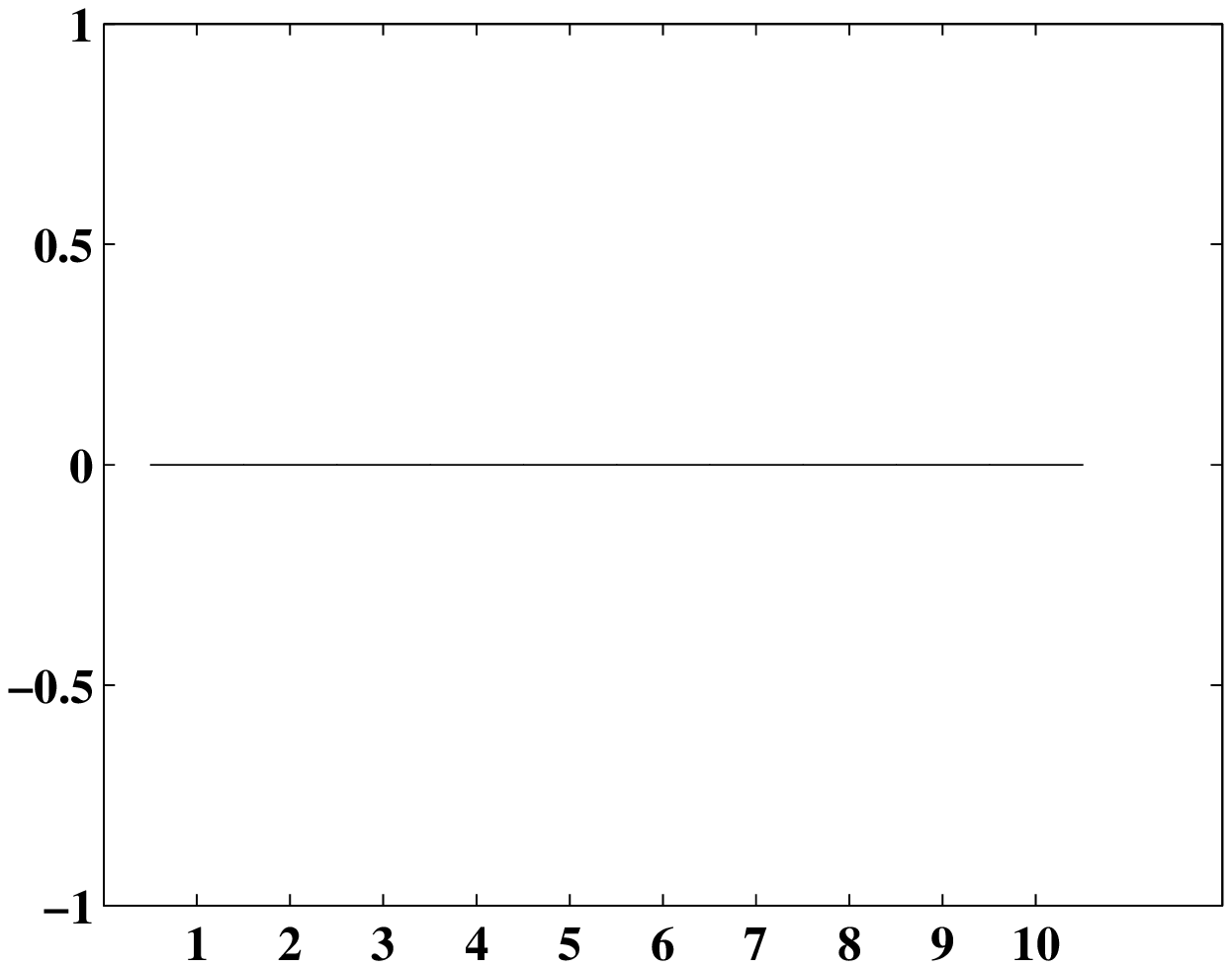,width=1.65in,height=1.35in}\hspace{-0.5cm}
\psfig{file=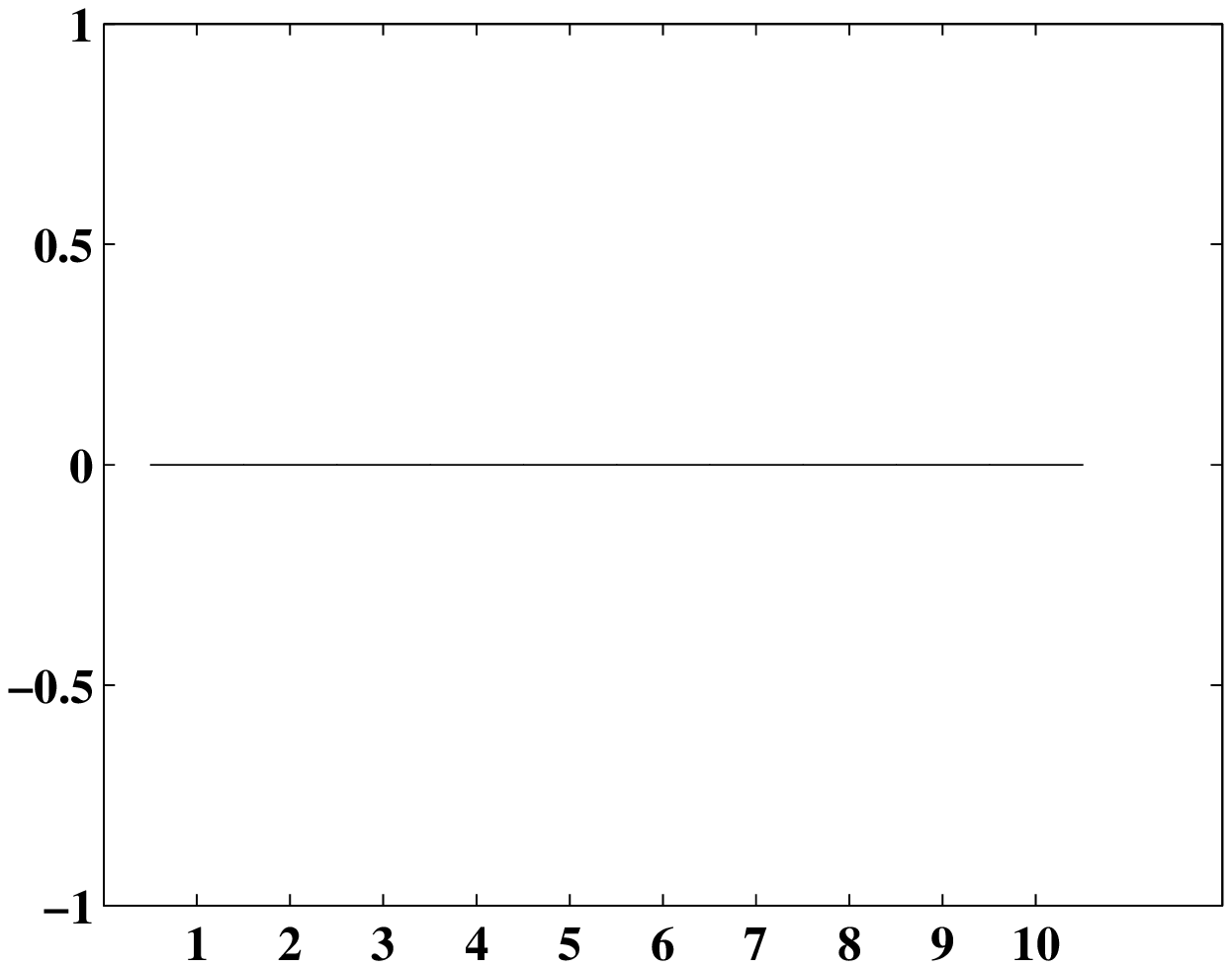,width=1.65in,height=1.35in}\hspace{-0.5cm}
\psfig{file=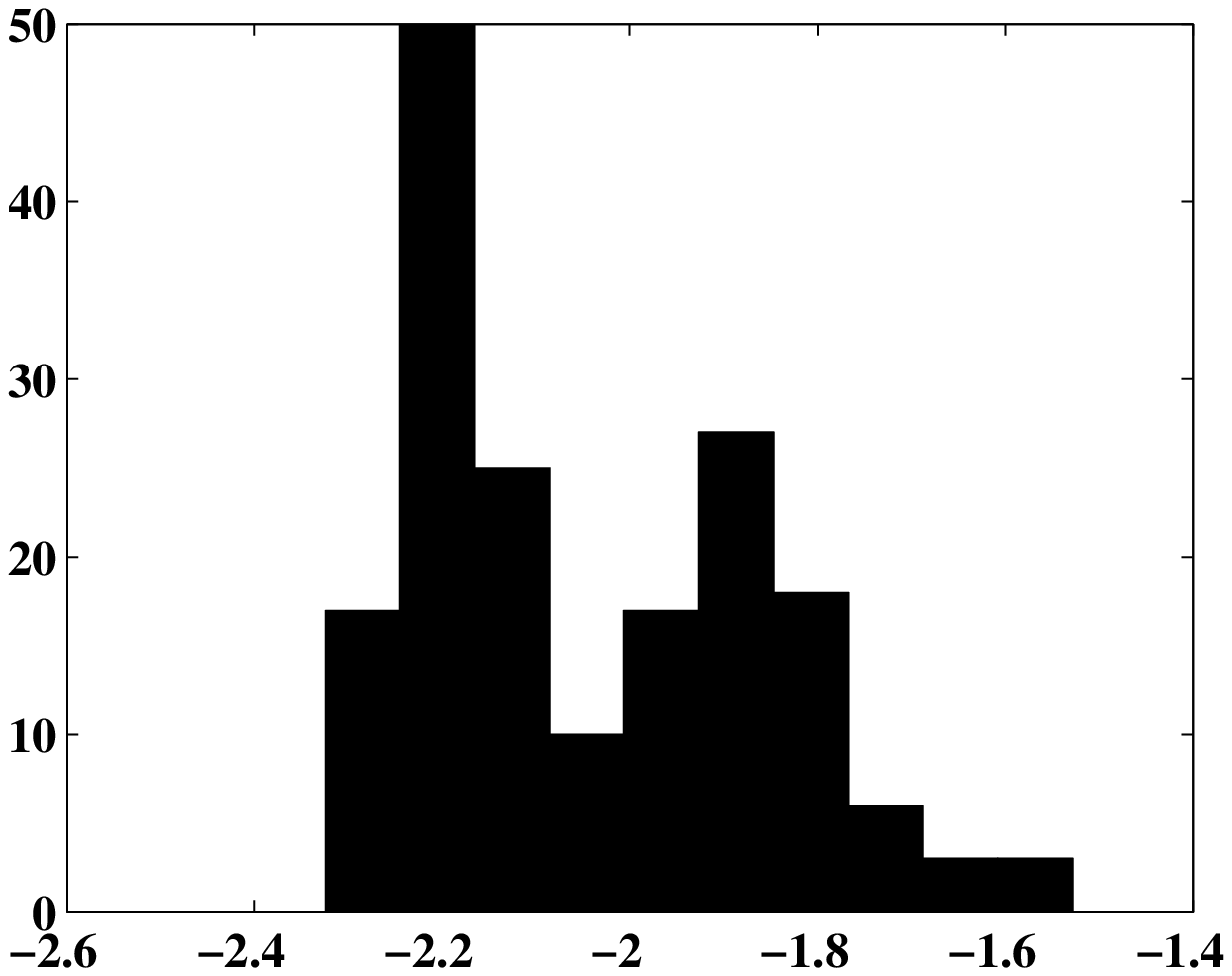,width=1.65in,height=1.35in}\hspace{-0.5cm}
\psfig{file=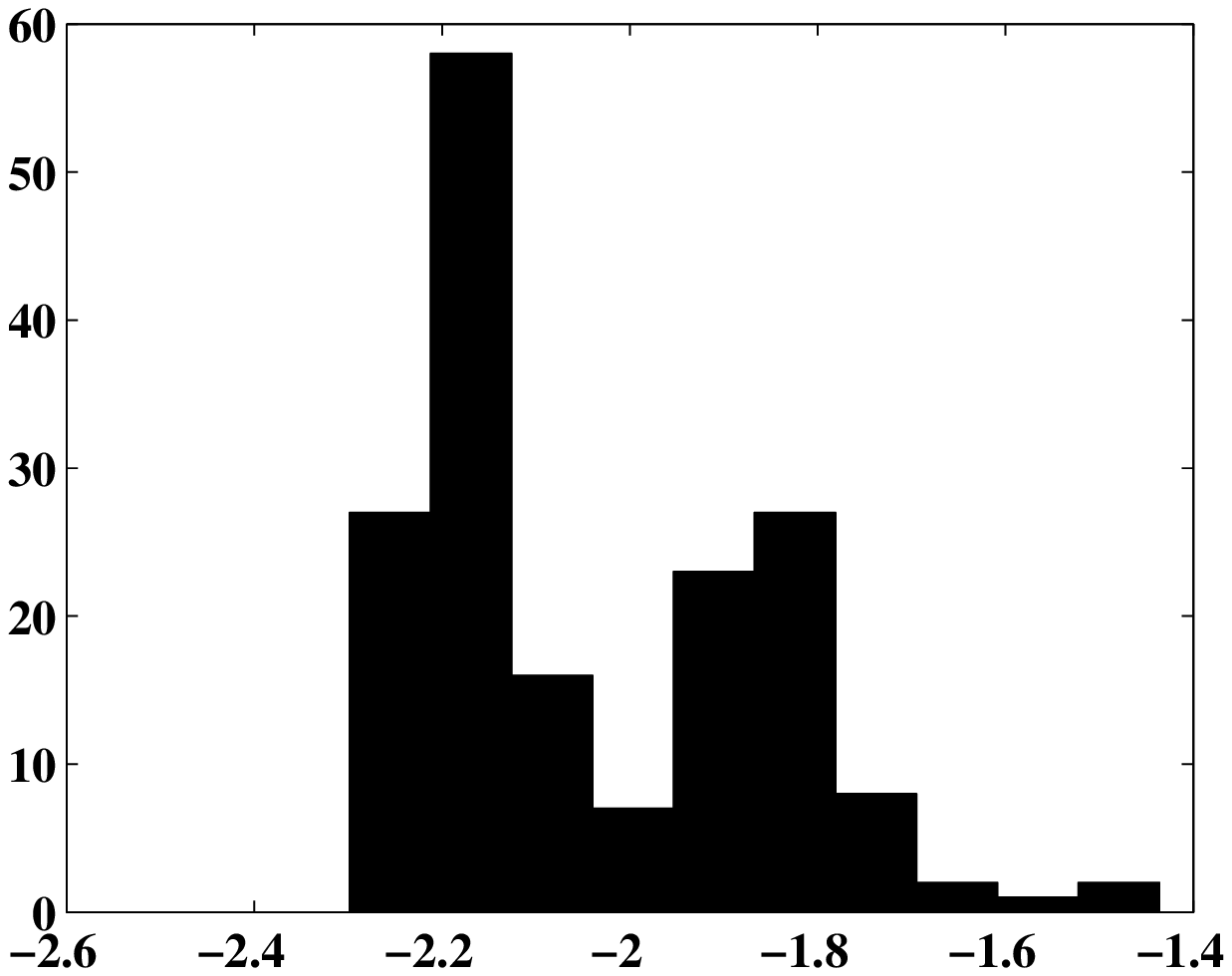,width=1.65in,height=1.35in}\hspace{-0.5cm}
} \vspace{-0.2in} \caption{\footnotesize Estimation curves from the
original Lasso-type spline method (OLSM) for Example
\ref{example-ind}. True functions $f_j$ and estimation curves
$\hat{f}_j$ for the first four components of the simulation run that
achieved the median of the MSE are presented.  The pictures in the
second row are histograms of the knots used in the one-stage OLSM
estimation for the first four components, respectively.}
\label{OLSM1}
\end{figure}

\begin{figure}
\vspace{0in} \centerline{
\psfig{file=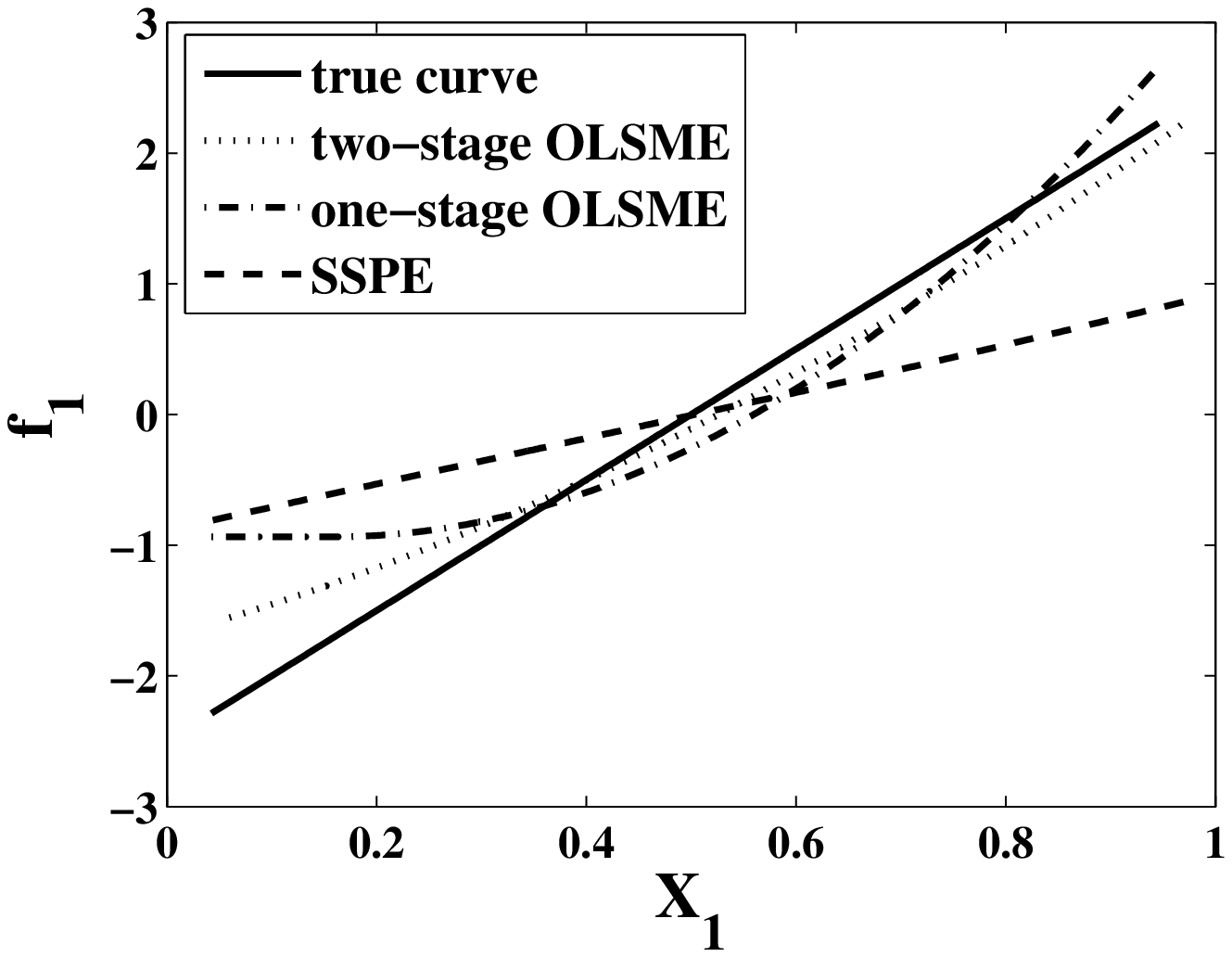,width=1.65in,height=1.85in}\hspace{-0.5cm}
\psfig{file=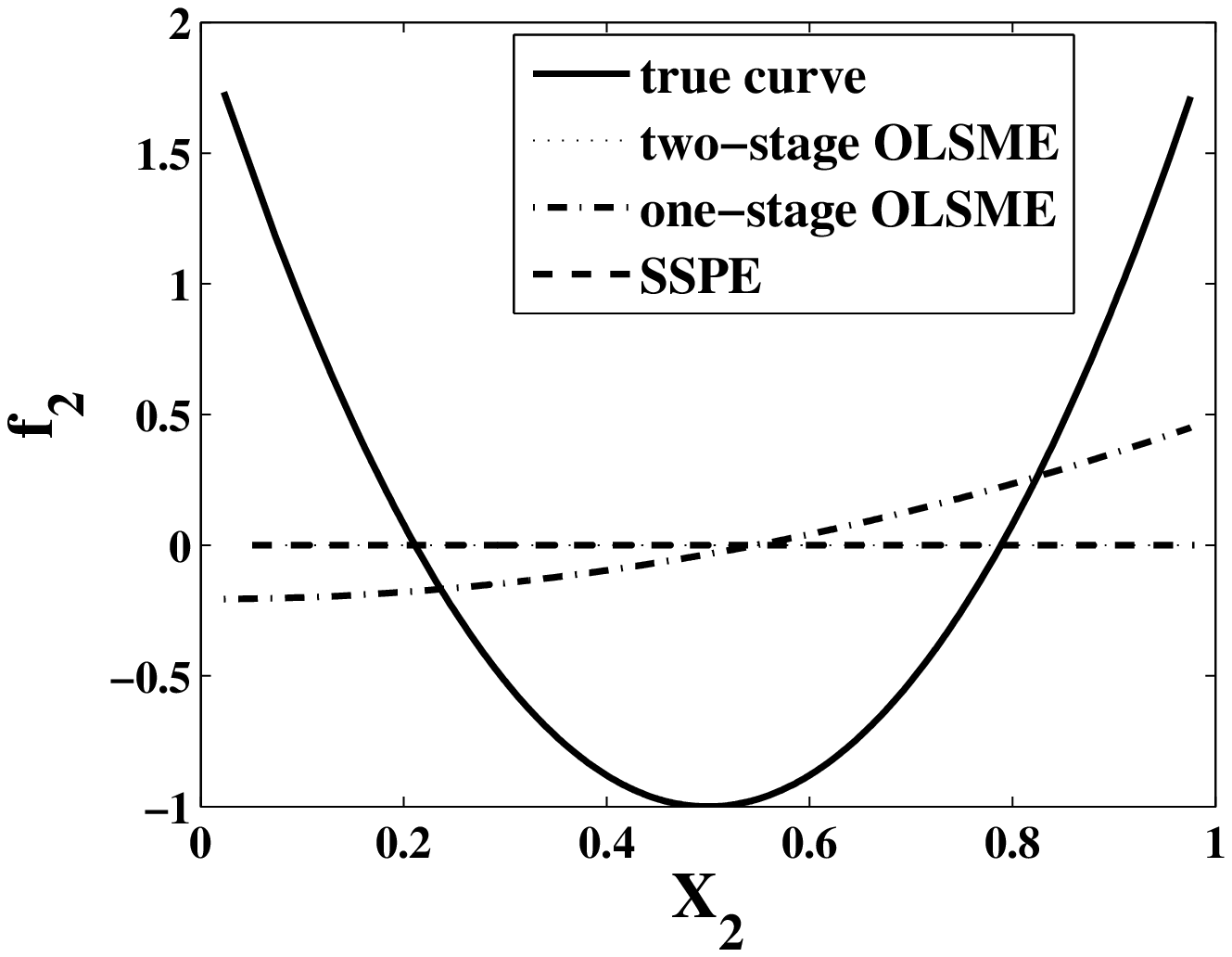,width=1.65in,height=1.85in}\hspace{-0.5cm}
\psfig{file=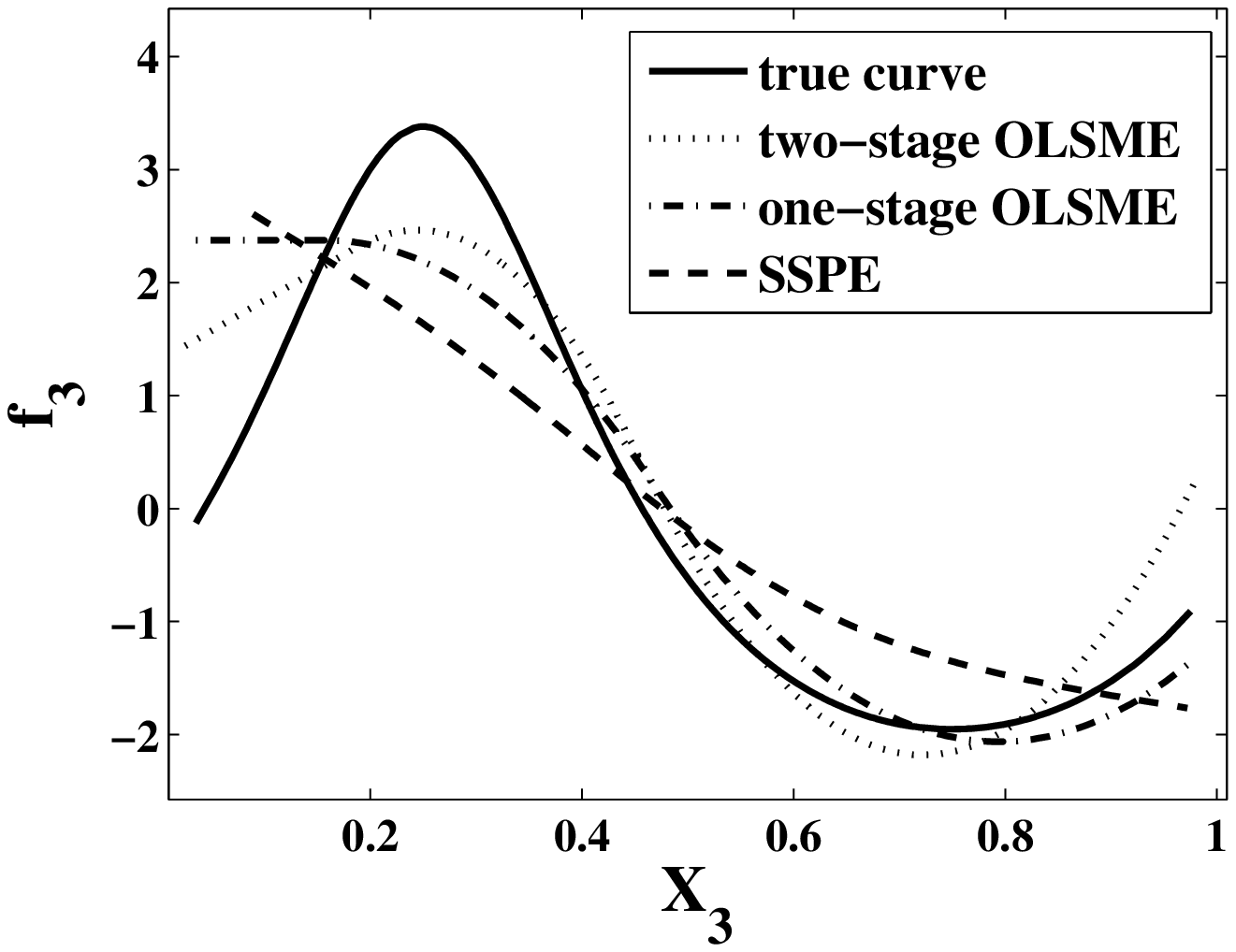,width=1.65in,height=1.85in}\hspace{-0.5cm}
\psfig{file=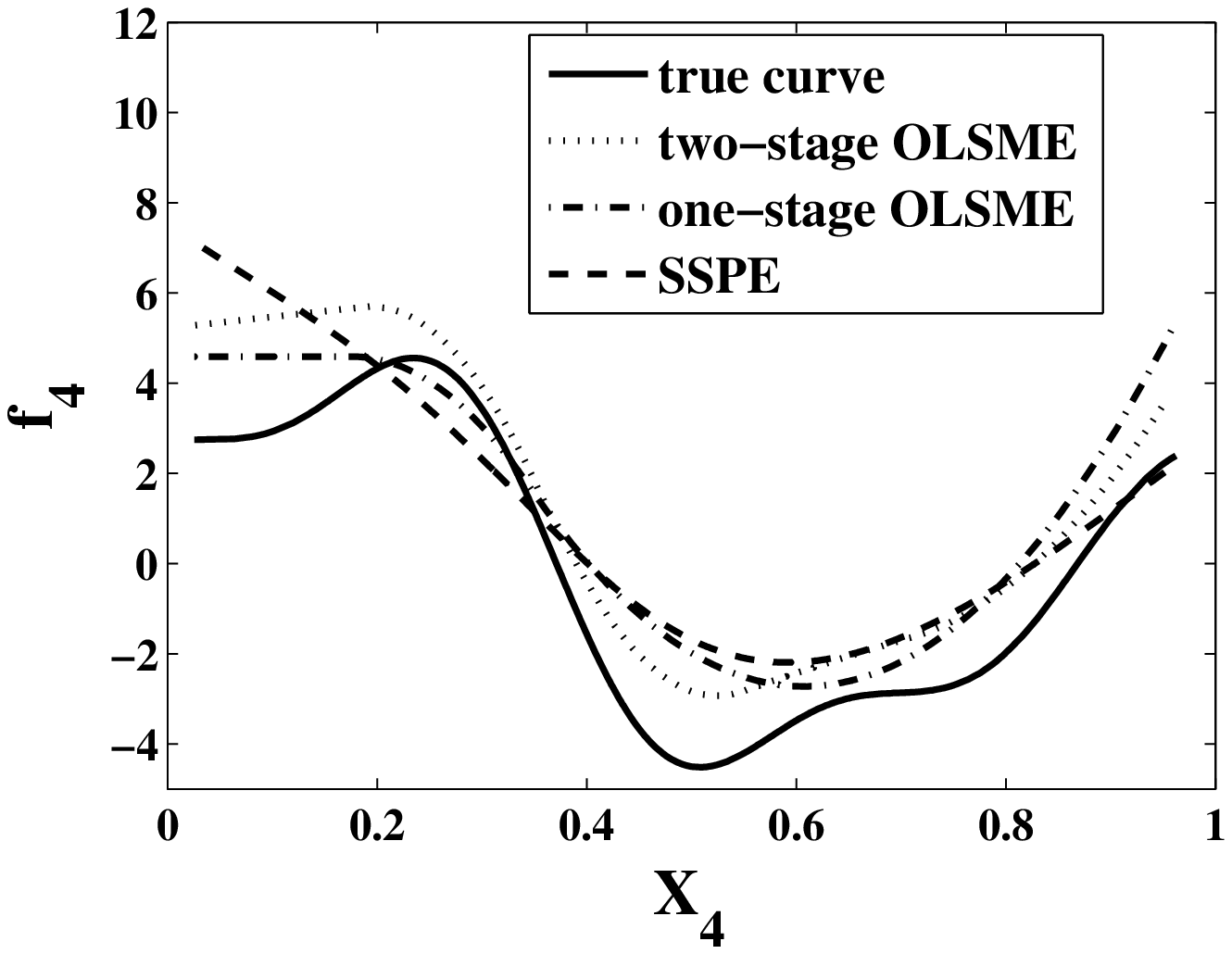,width=1.65in,height=1.85in}\hspace{-0.5cm}
} \centerline{
\psfig{file=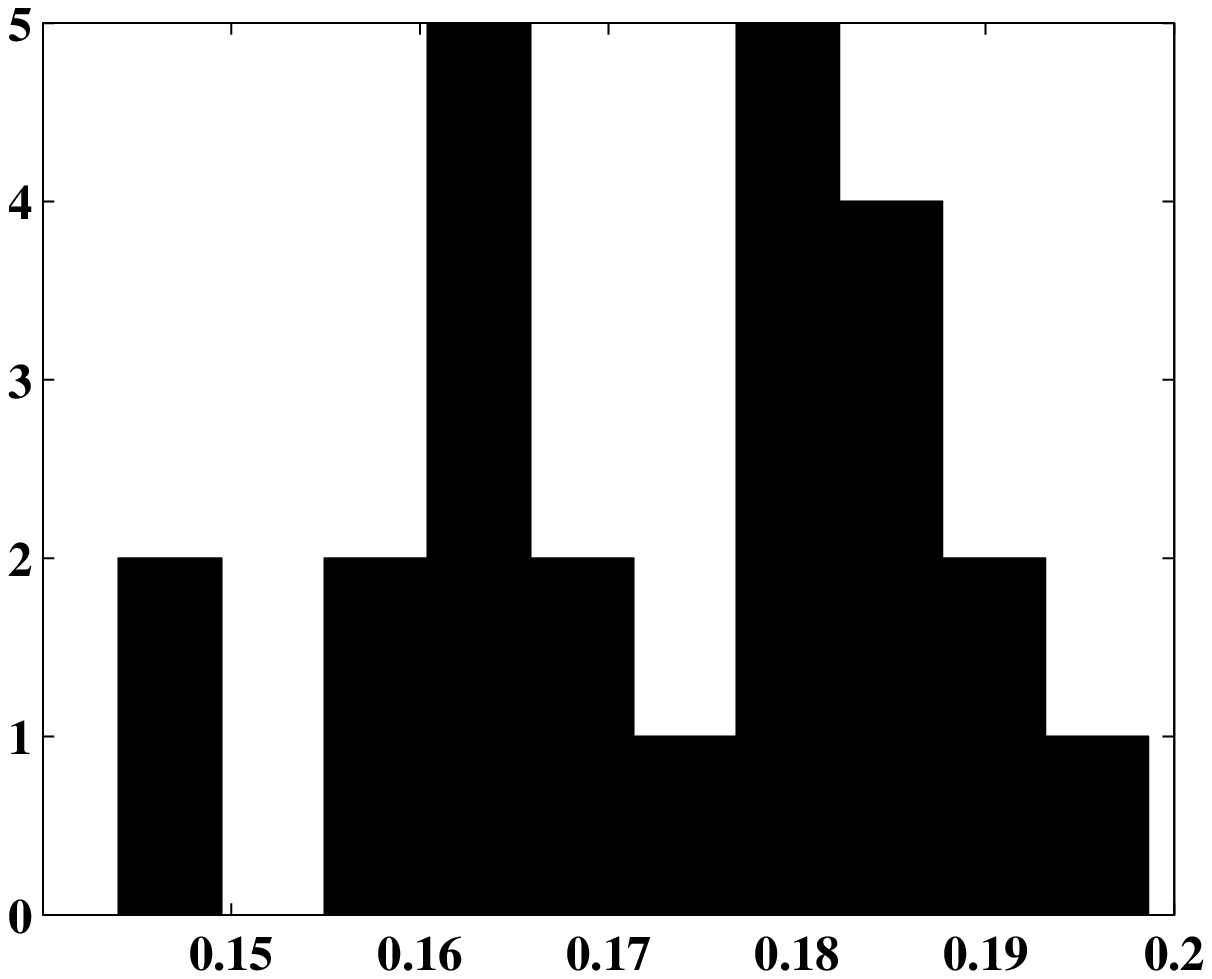,width=1.65in,height=1.35in}\hspace{-0.5cm}
\psfig{file=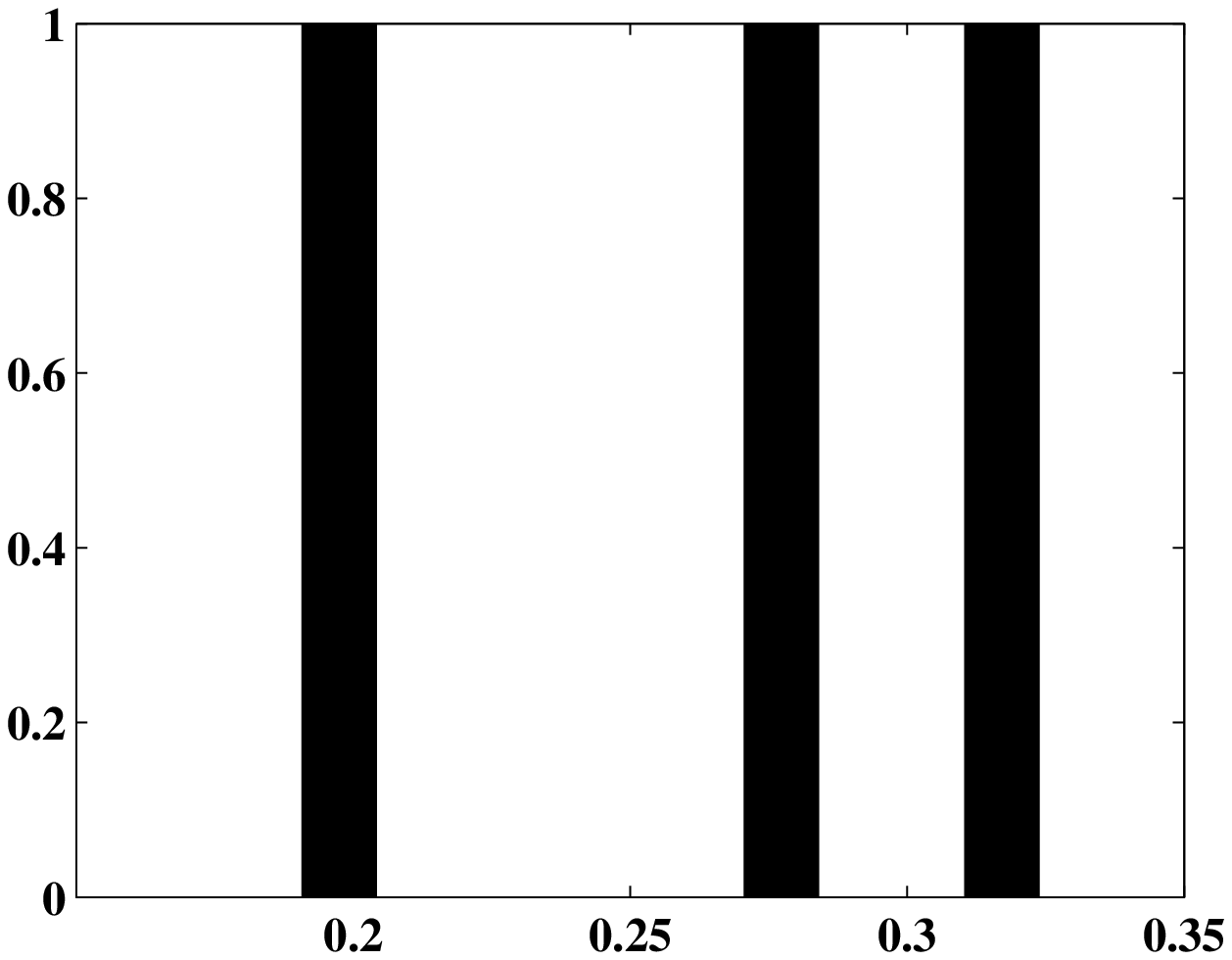,width=1.65in,height=1.35in}\hspace{-0.5cm}
\psfig{file=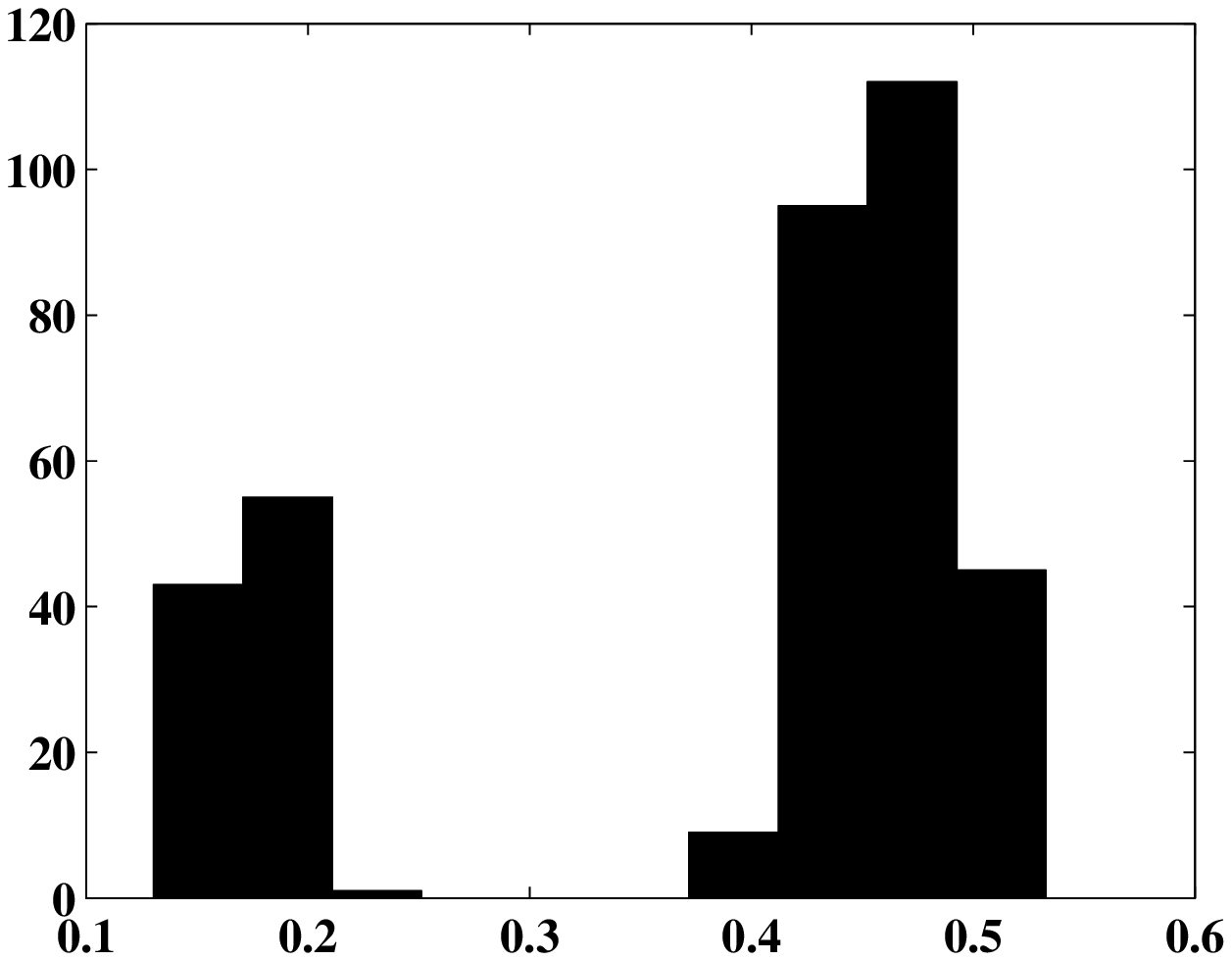,width=1.65in,height=1.35in}\hspace{-0.5cm}
\psfig{file=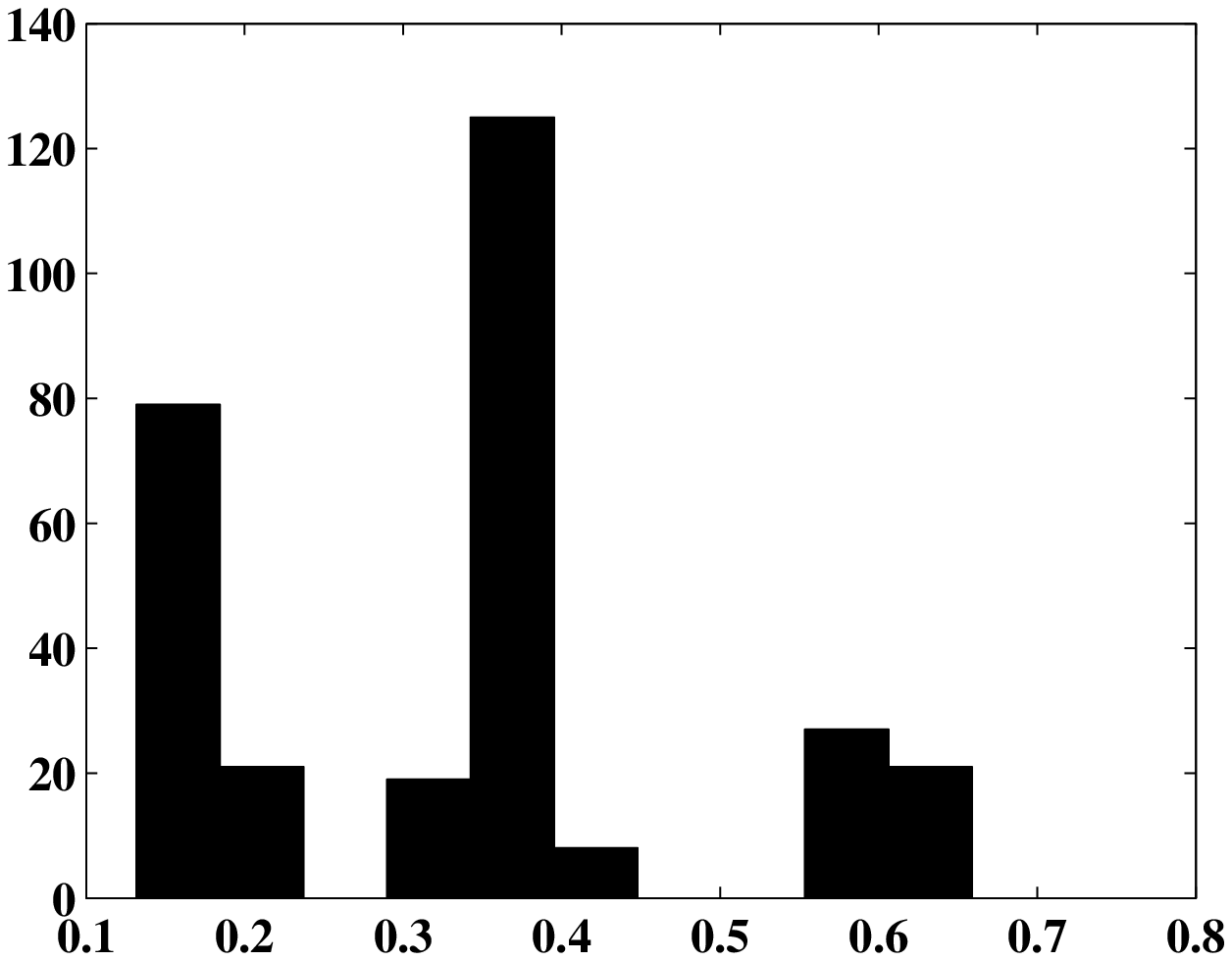,width=1.65in,height=1.35in}\hspace{-0.5cm}
} \vspace{-0.2in} \caption{\footnotesize Estimation curves from the
original Lasso-type spline method (OLSM) for Example
\ref{example-cor}. True functions $f_j$ and estimation curves
$\hat{f}_j$ for the first four components of the simulation run that
achieved the median of the MSE are presented.  The pictures in the
second row are histograms of the knots used in the one-stage OLSM
estimation for the first four components, respectively.}
\label{OLSM2}
\end{figure}

\begin{figure}
\vspace{-0.5in} \centerline{
\psfig{file=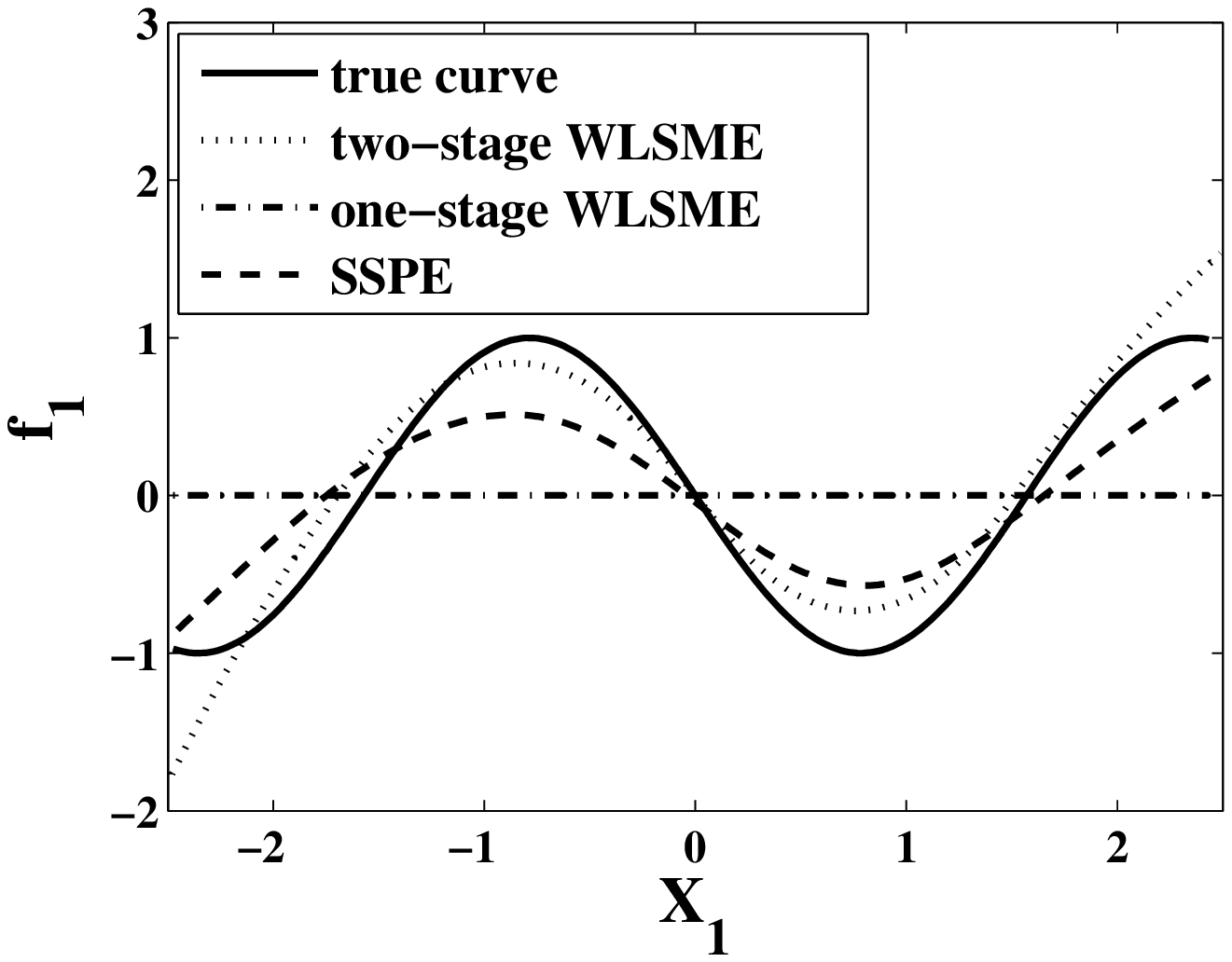,width=1.65in,height=1.85in}\hspace{-0.5cm}
\psfig{file=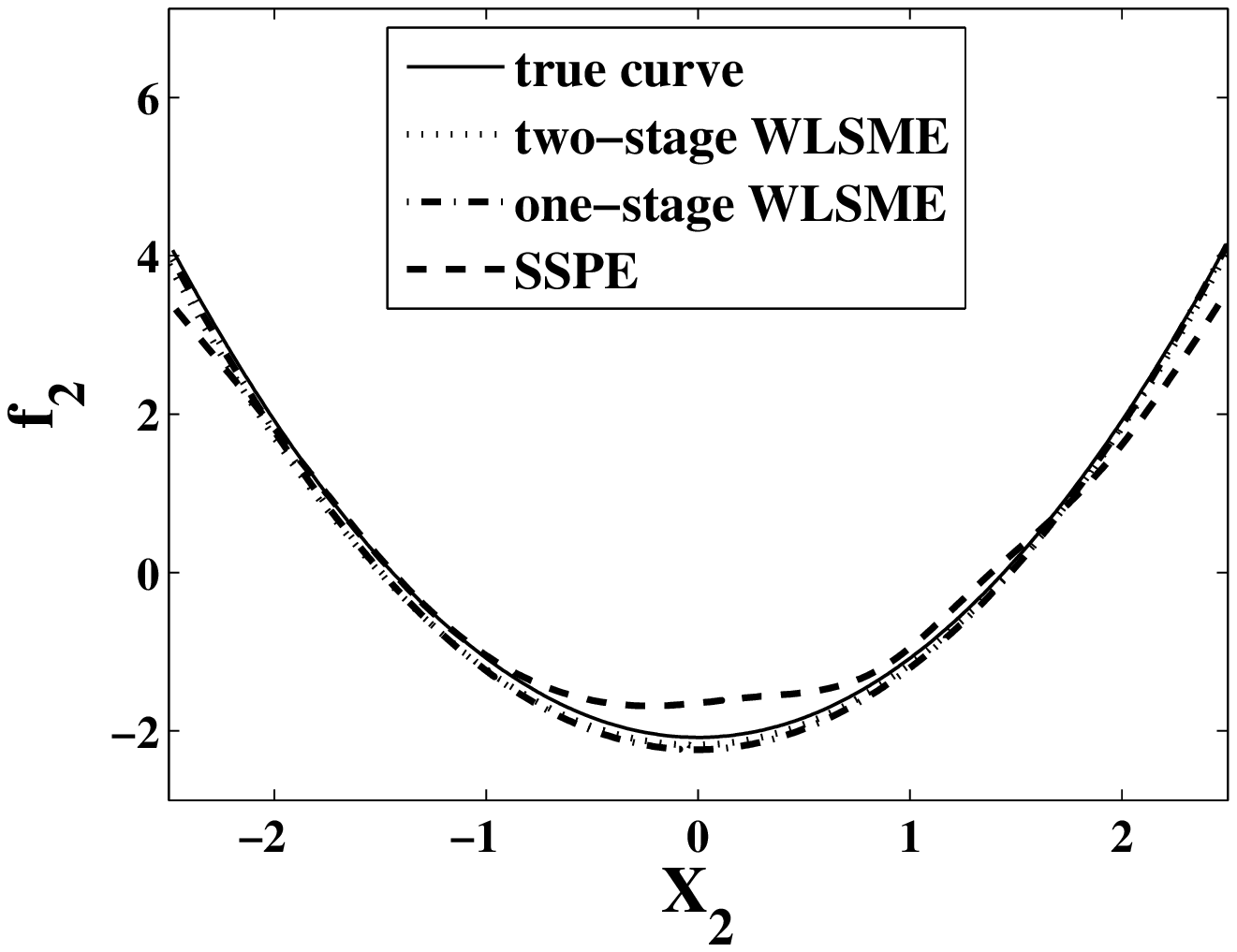,width=1.65in,height=1.85in}\hspace{-0.5cm}
\psfig{file=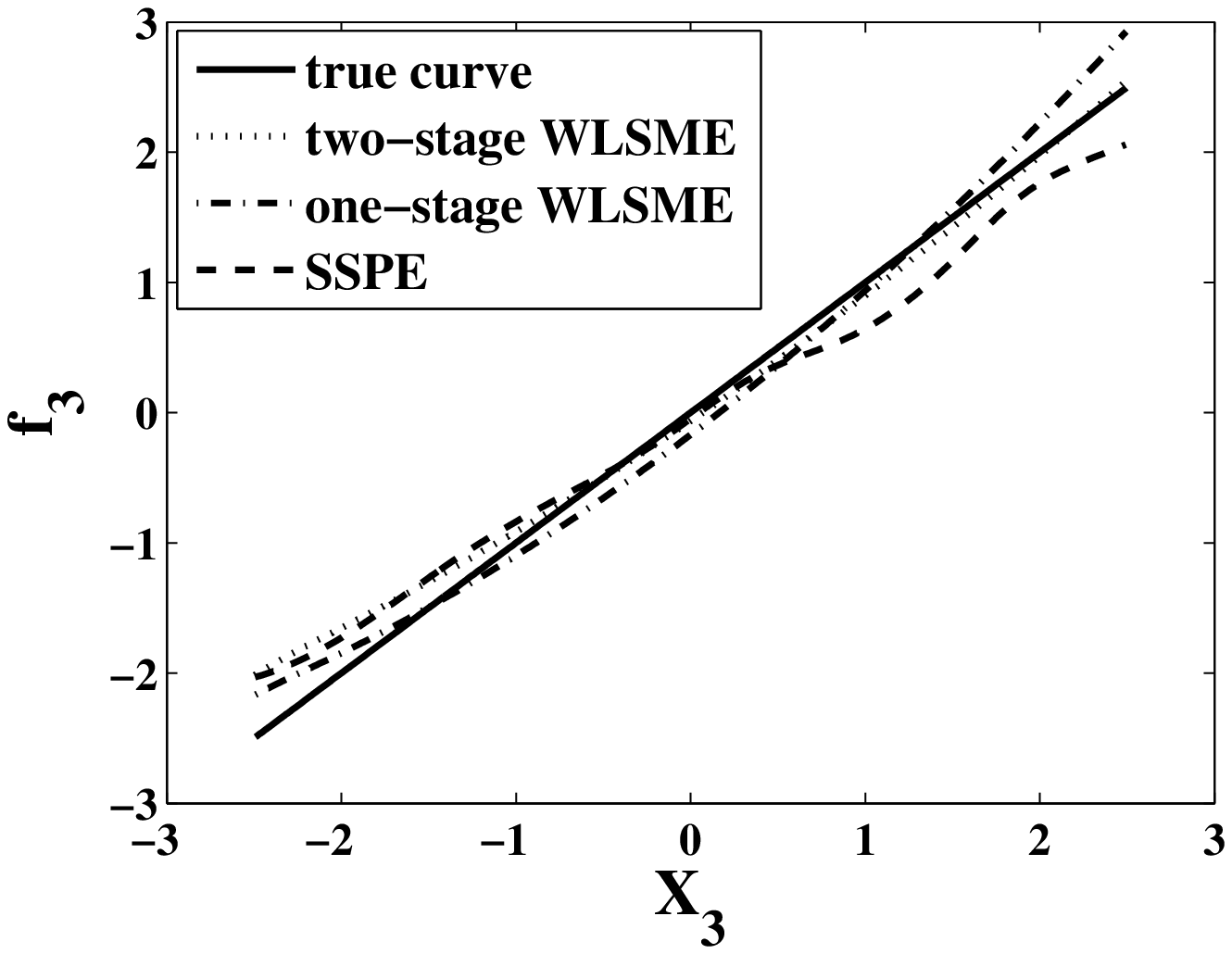,width=1.65in,height=1.85in}\hspace{-0.5cm}
\psfig{file=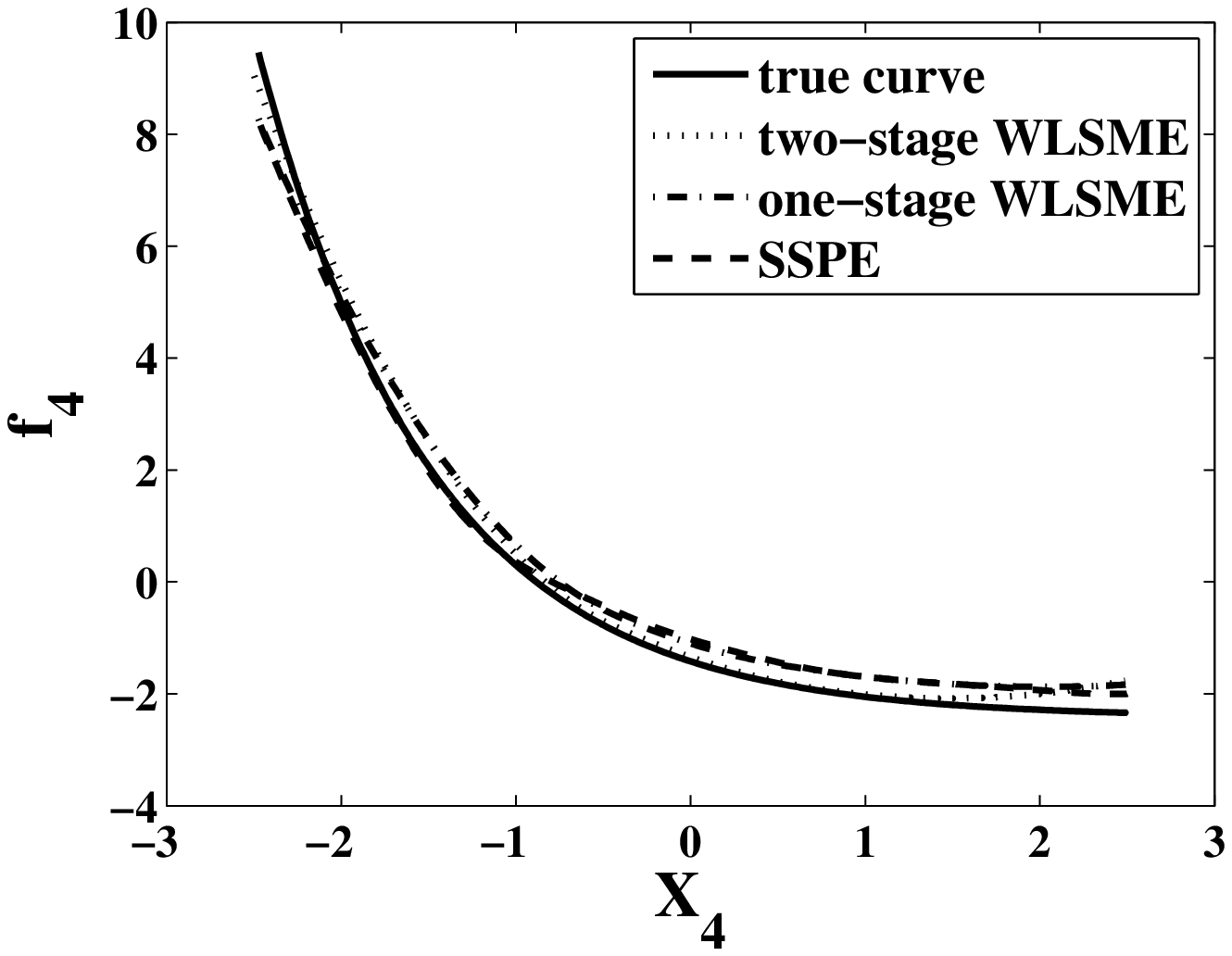,width=1.65in,height=1.85in}\hspace{-0.5cm}
} \centerline{
\psfig{file=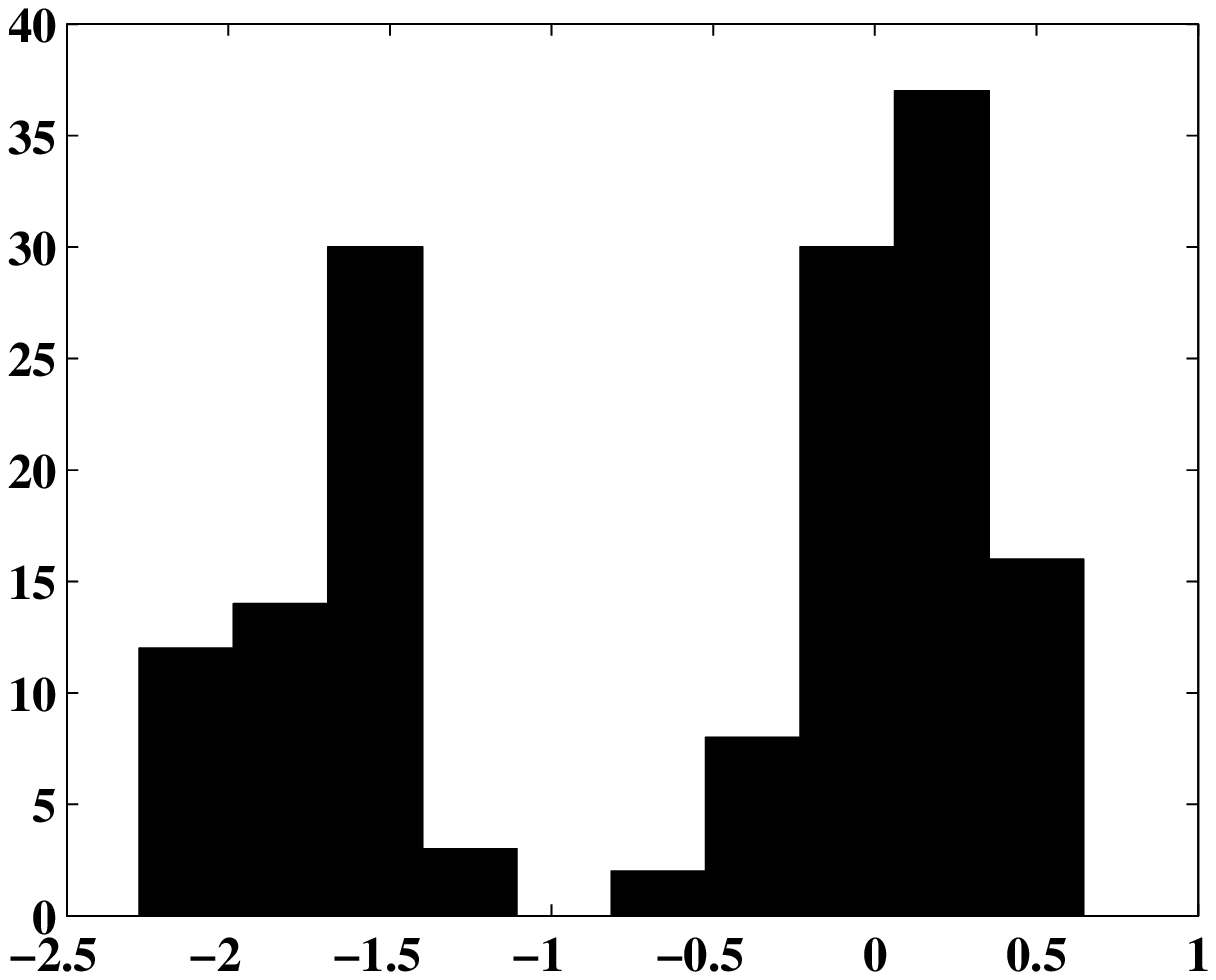,width=1.65in,height=1.35in}\hspace{-0.5cm}
\psfig{file=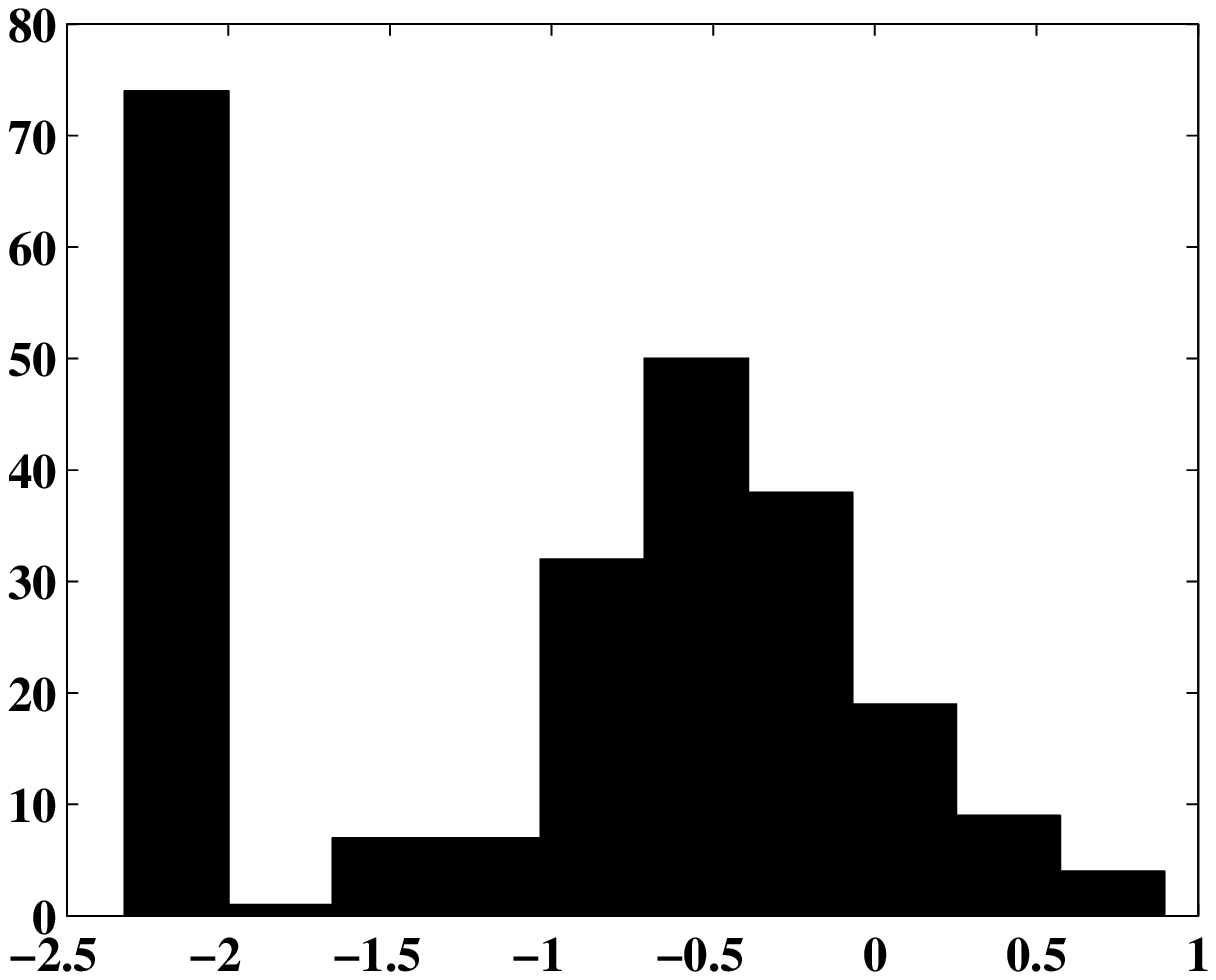,width=1.65in,height=1.35in}\hspace{-0.5cm}
\psfig{file=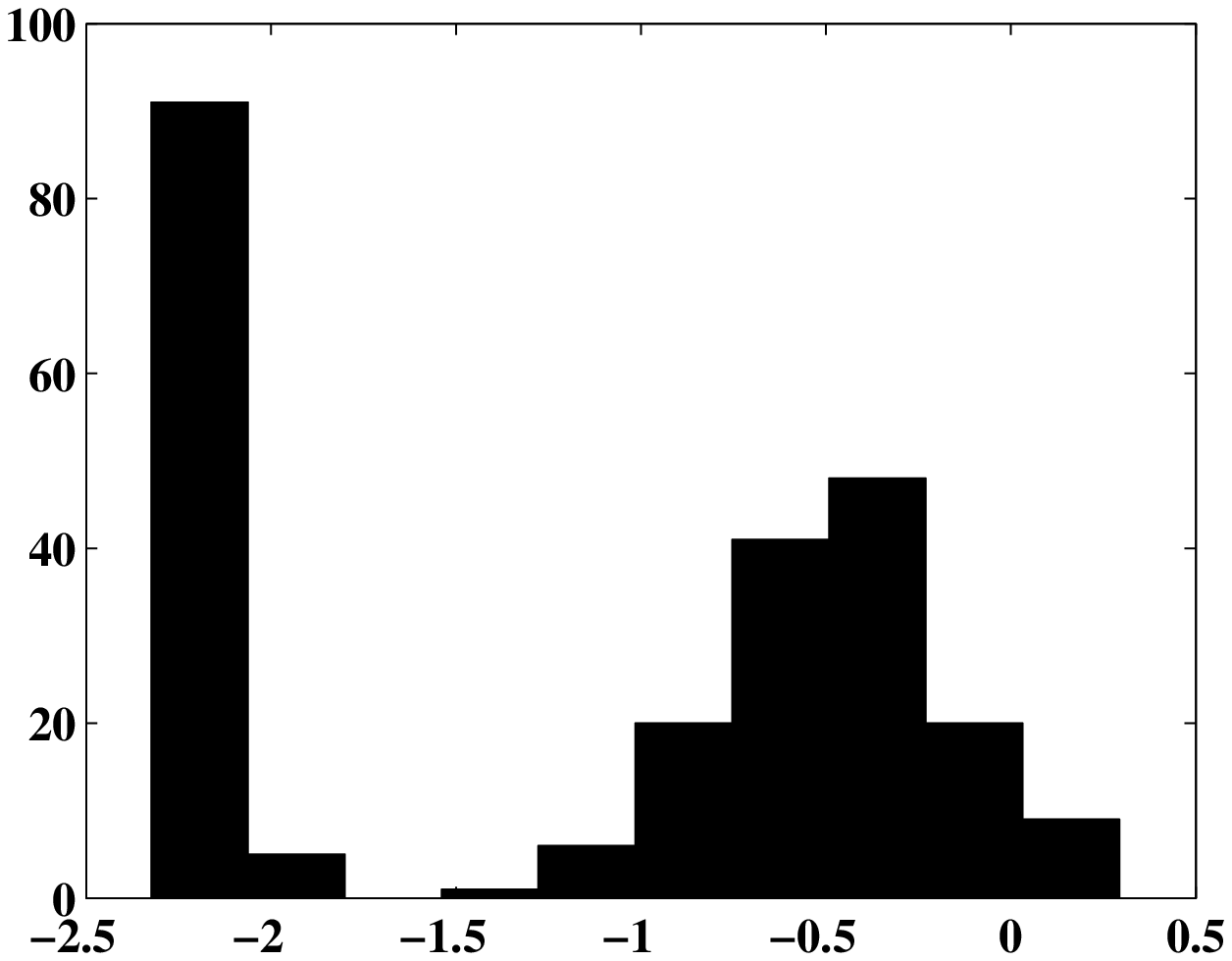,width=1.65in,height=1.35in}\hspace{-0.5cm}
\psfig{file=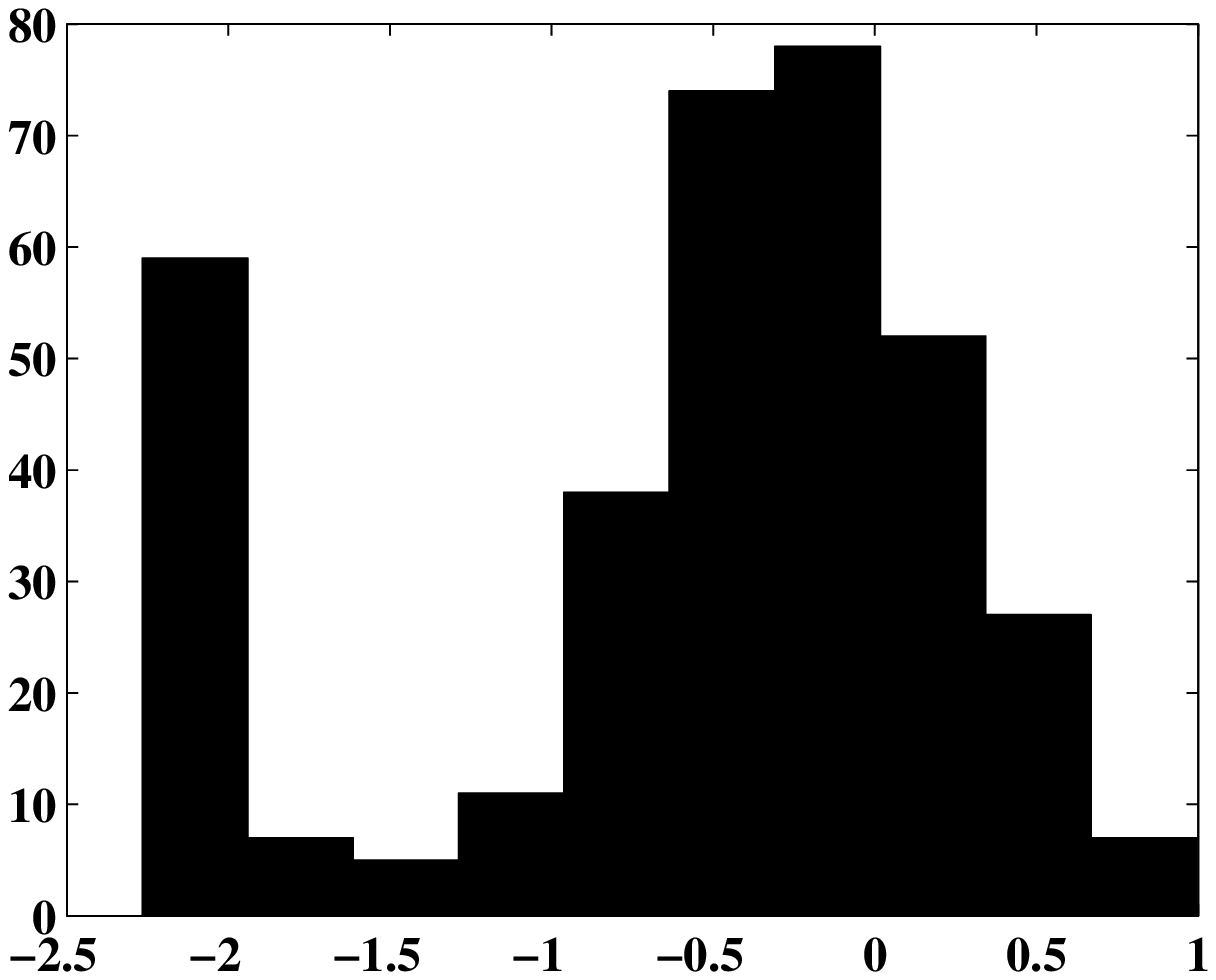,width=1.65in,height=1.35in}\hspace{-0.5cm}
} \vspace{-0.2in} \caption{\footnotesize Estimation curves from the
weighted Lasso-type spline method (WLSM) for Example
\ref{example-ind}. True functions $f_j$ and estimation curves
$\hat{f}_j$ for the first four components of the simulation run that
achieved the median of the MSE are presented.  The pictures in the
second row are histograms of the knots used in the one-stage WLSM
estimation for the first four components, respectively.}
\label{WLSM1}
\end{figure}

\begin{figure}
\vspace{0in} \centerline{
\psfig{file=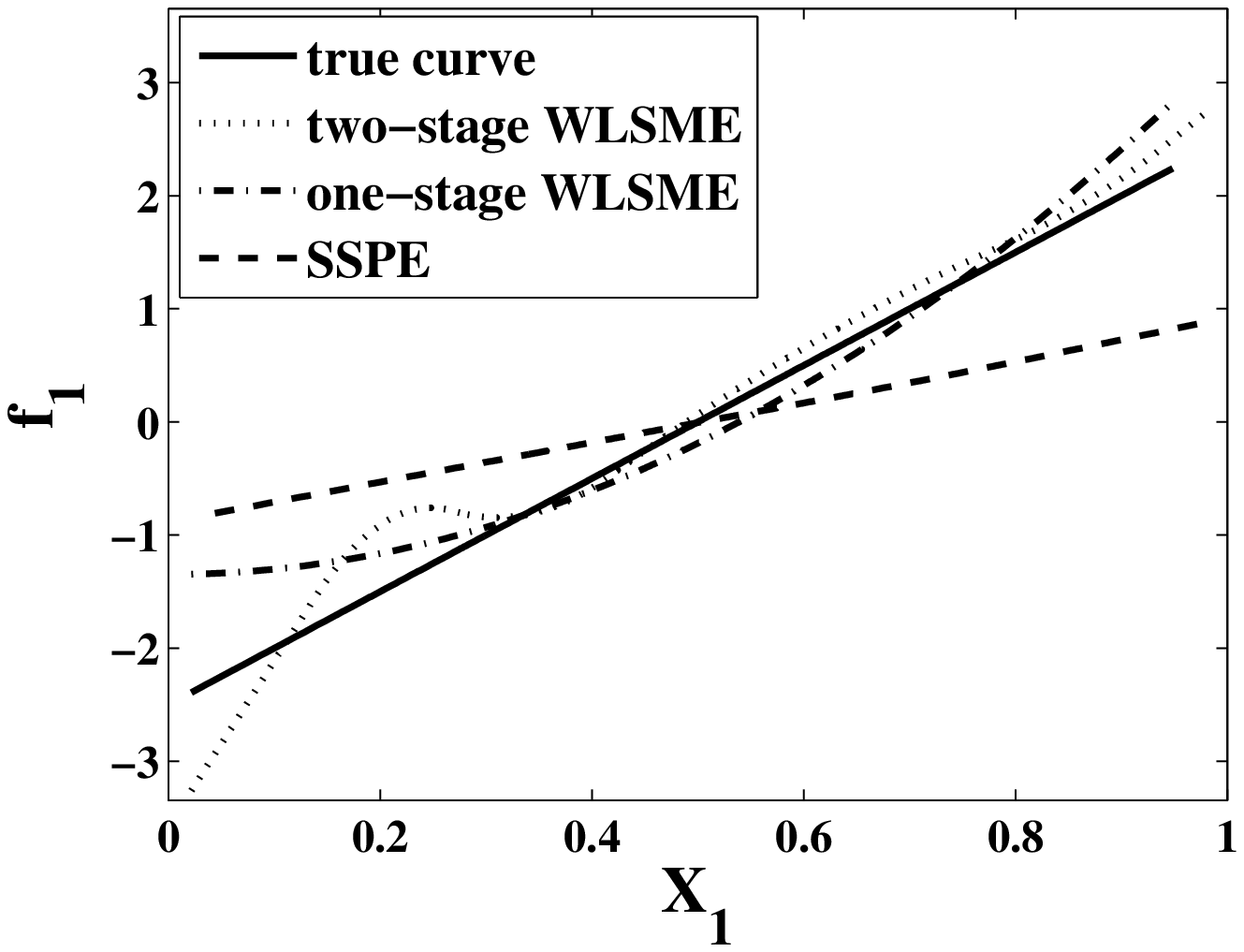,width=1.65in,height=1.85in}\hspace{-0.5cm}
\psfig{file=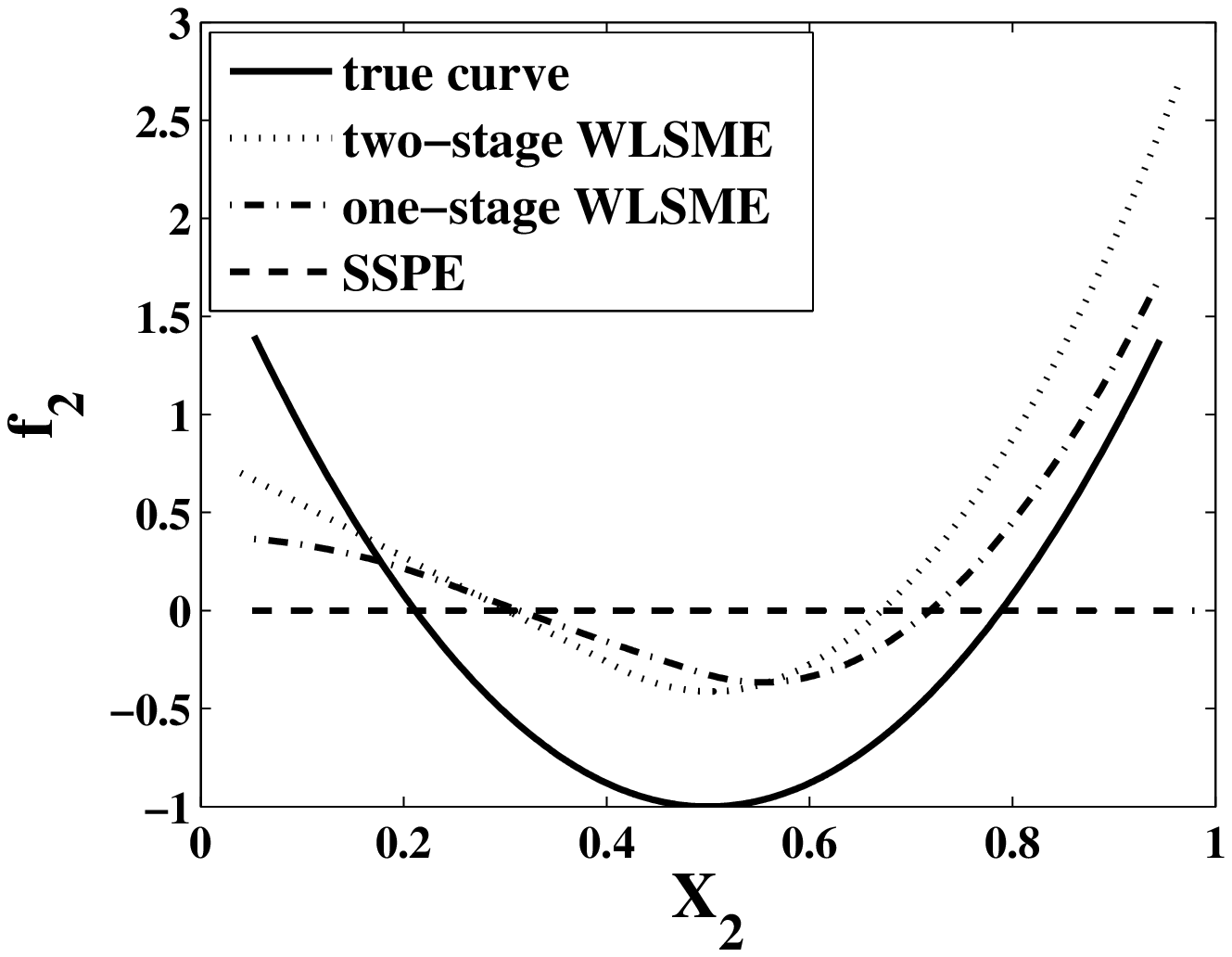,width=1.65in,height=1.85in}\hspace{-0.5cm}
\psfig{file=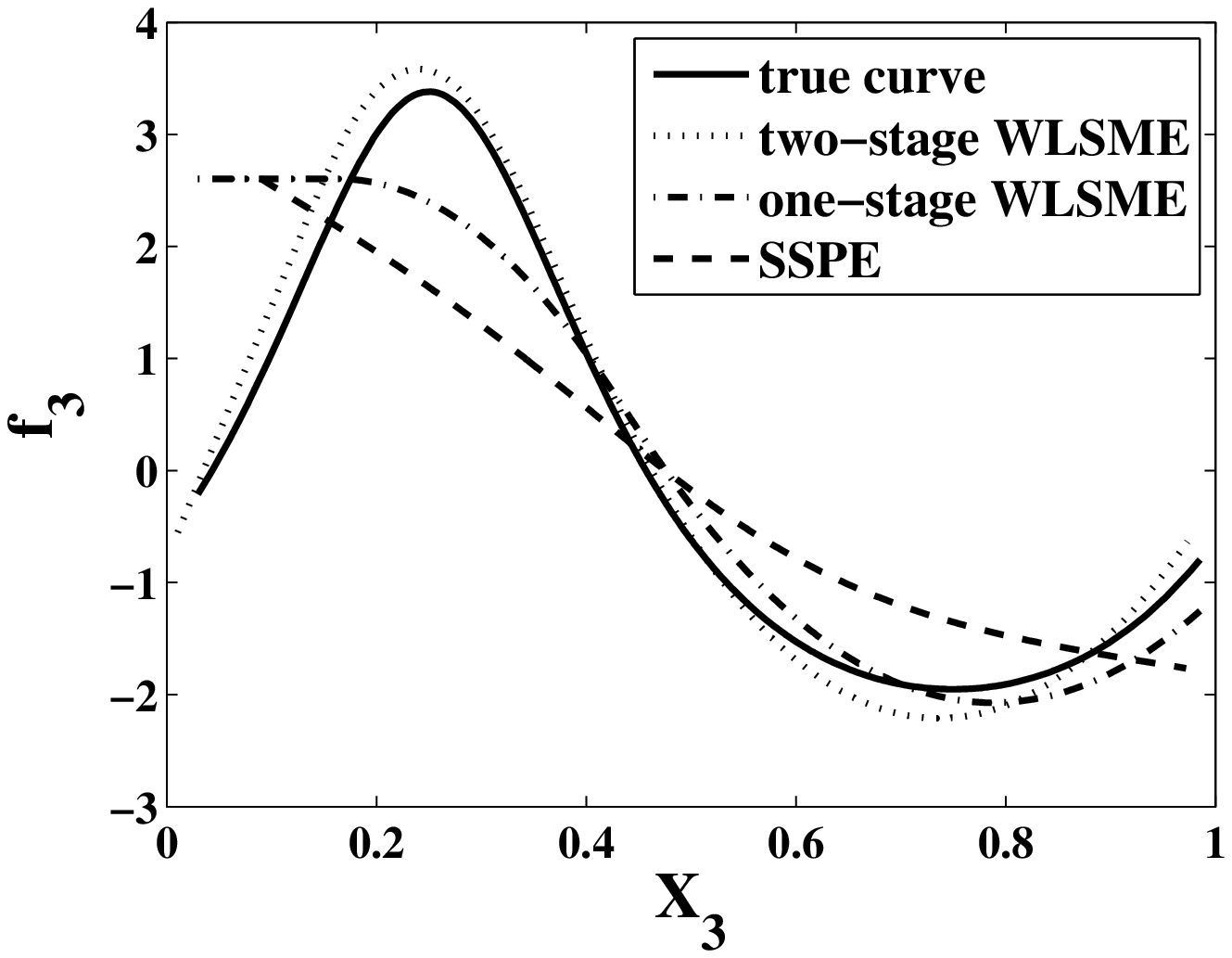,width=1.65in,height=1.85in}\hspace{-0.5cm}
\psfig{file=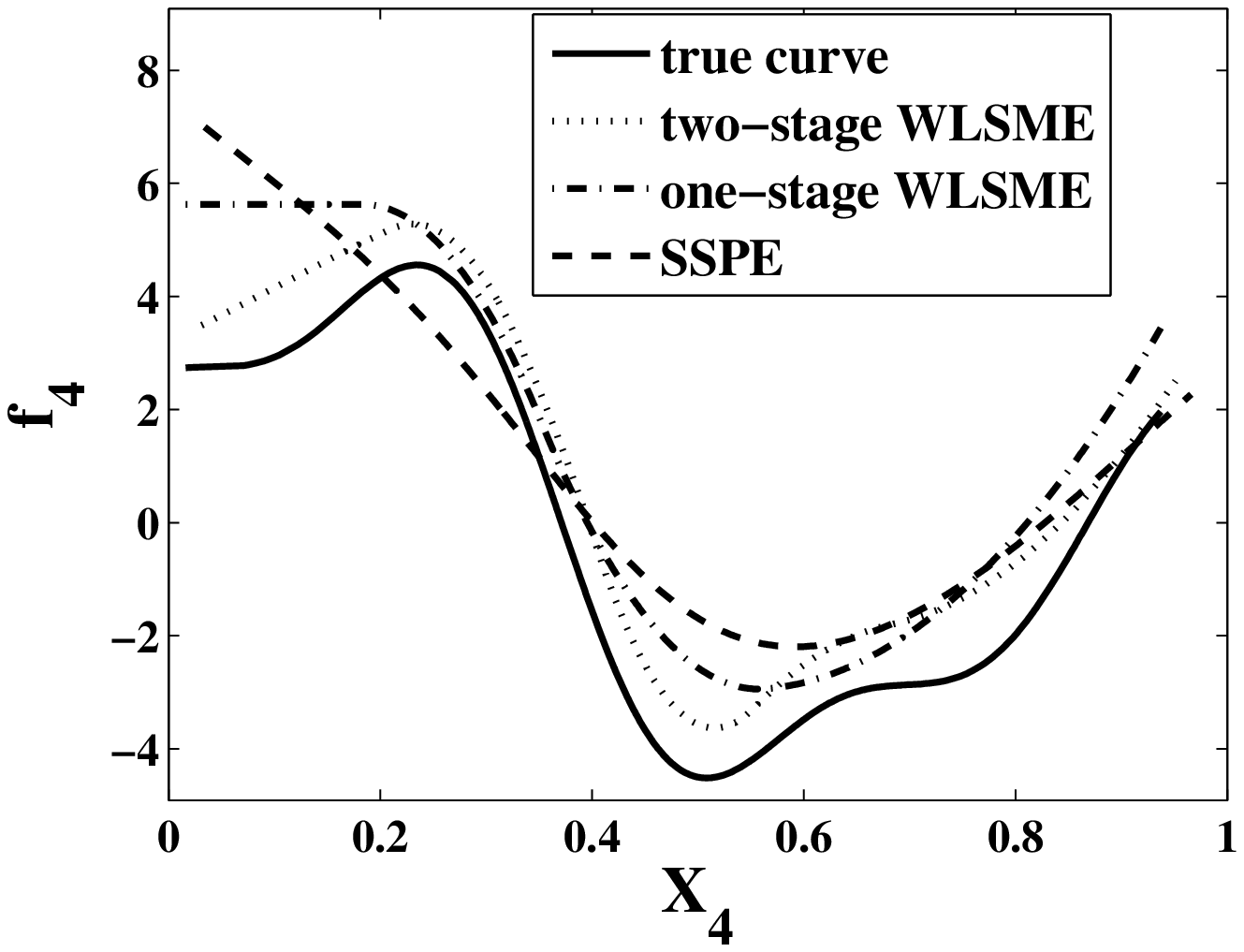,width=1.65in,height=1.85in}\hspace{-0.5cm}
} \centerline{
\psfig{file=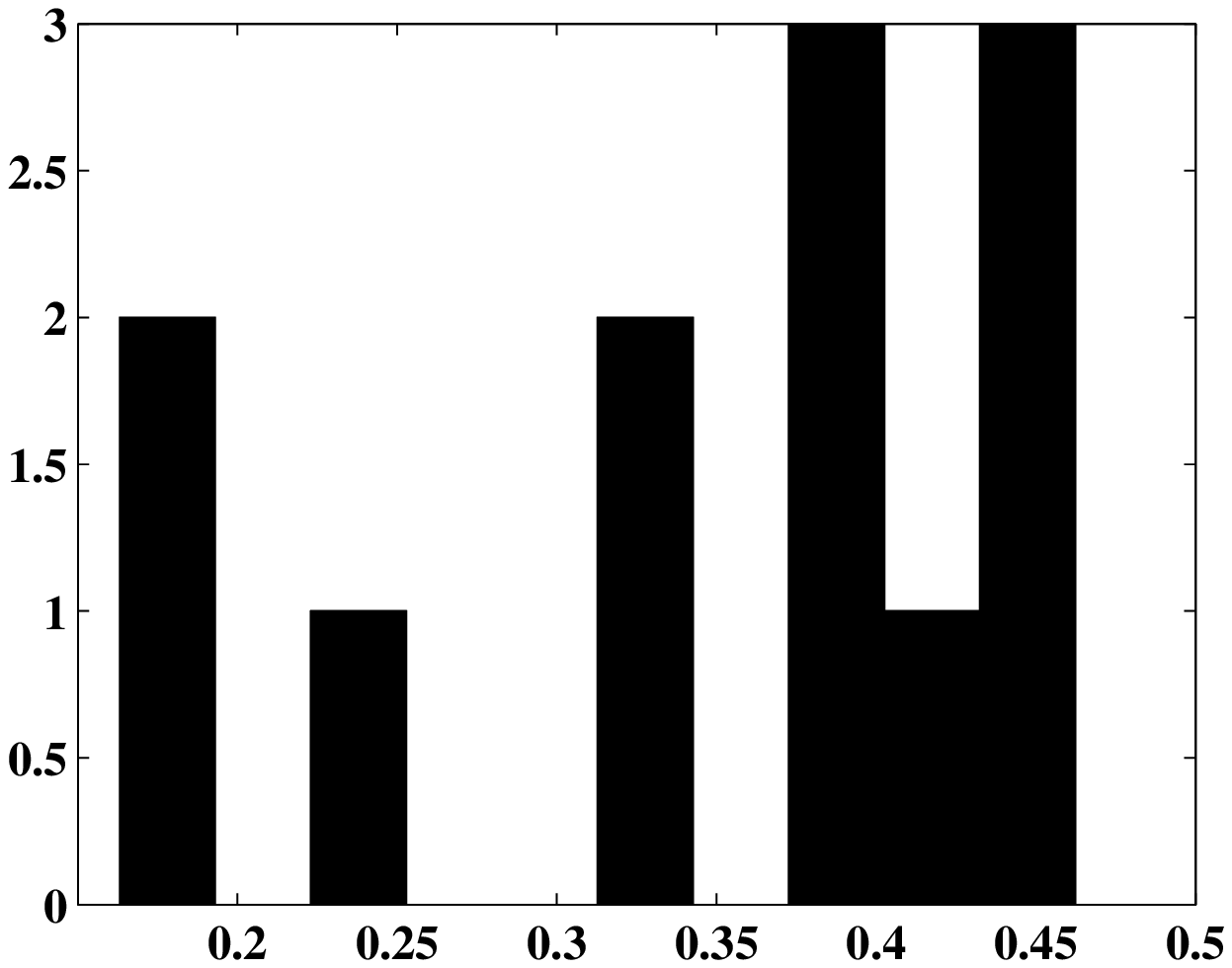,width=1.65in,height=1.35in}\hspace{-0.5cm}
\psfig{file=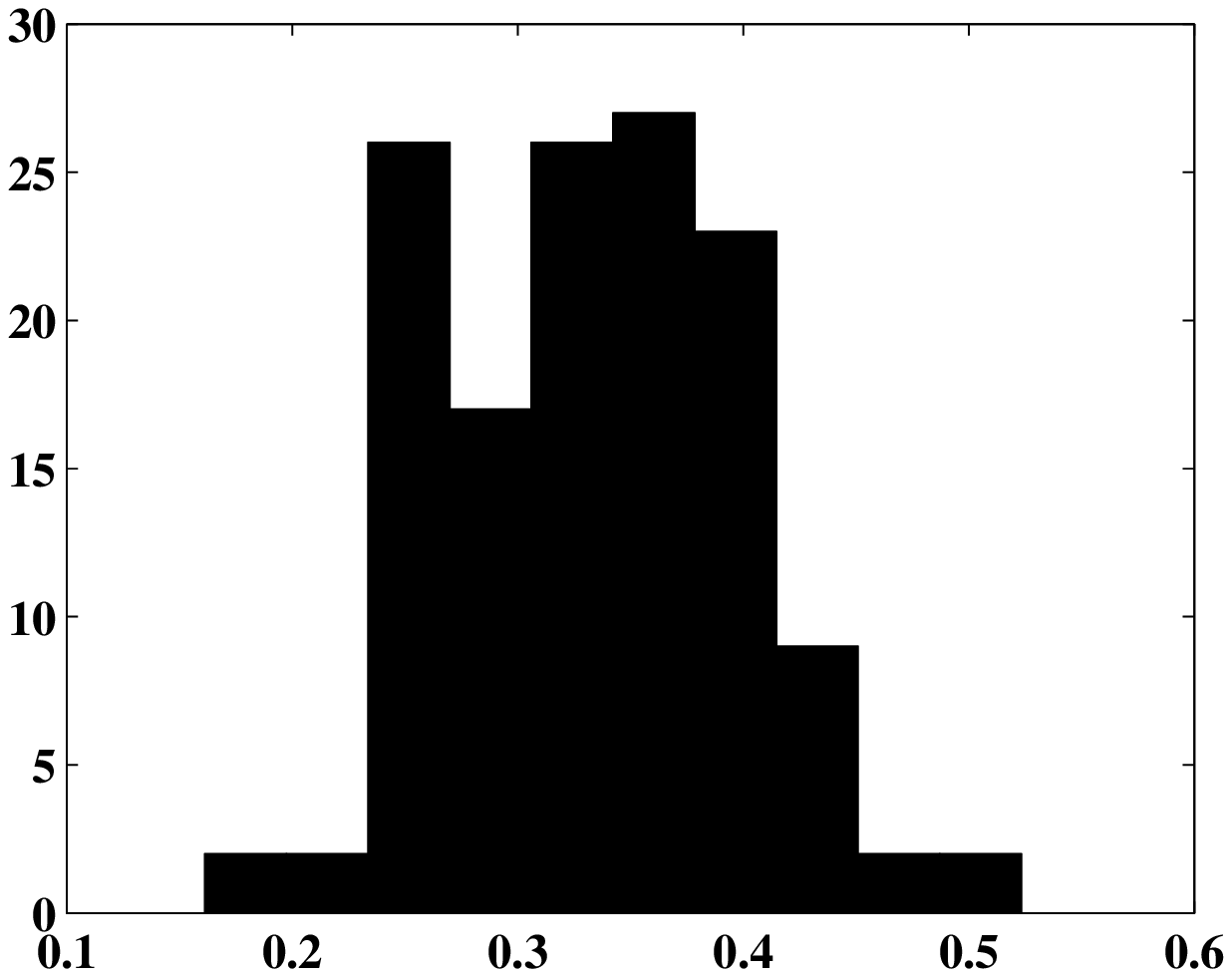,width=1.65in,height=1.35in}\hspace{-0.5cm}
\psfig{file=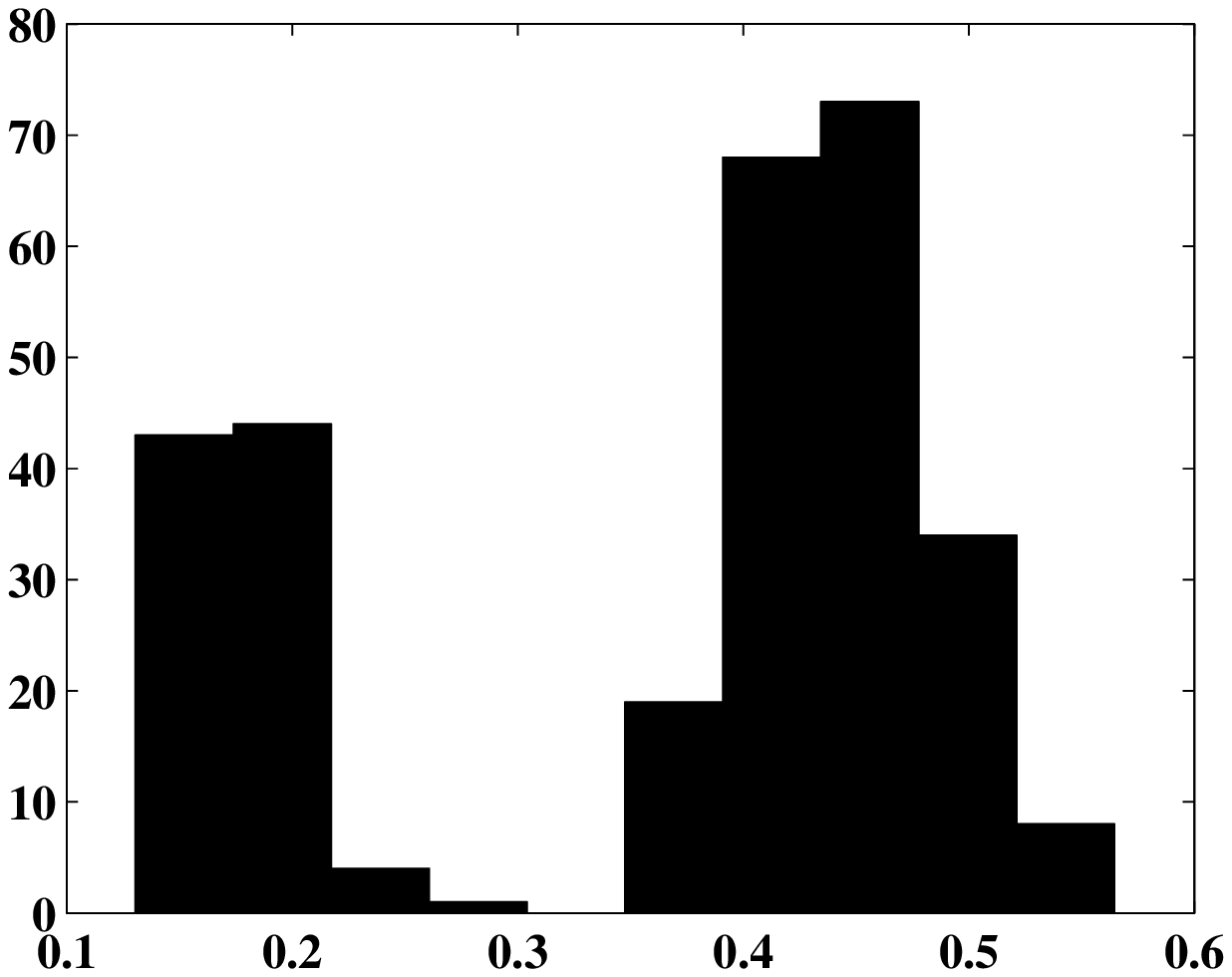,width=1.65in,height=1.35in}\hspace{-0.5cm}
\psfig{file=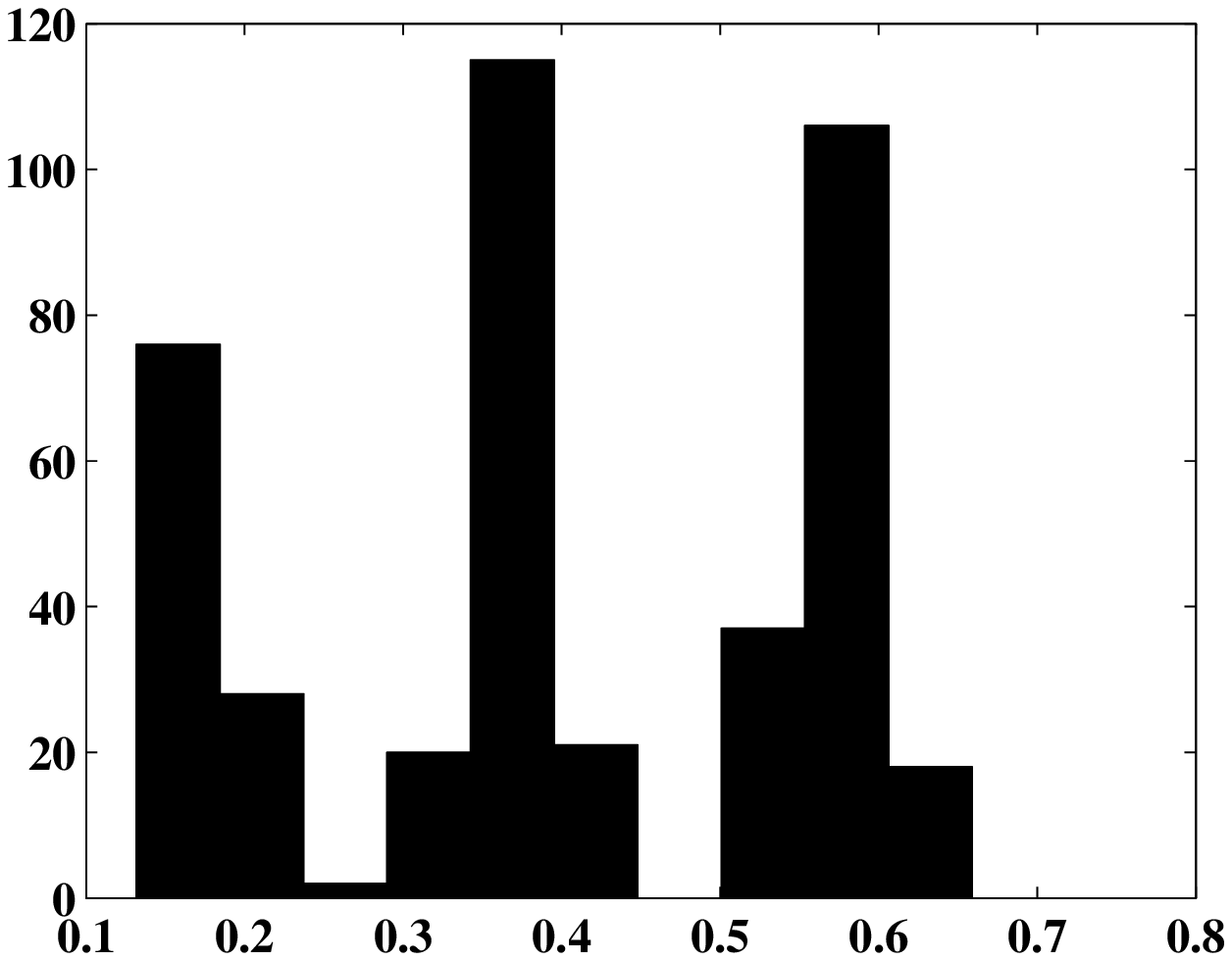,width=1.65in,height=1.35in}\hspace{-0.5cm}
} \vspace{-0.2in} \caption{\footnotesize Estimation curves from the
weighted Lasso-type spline method (WLSM) for Example
\ref{example-cor}. True functions $f_j$ and estimation curves
$\hat{f}_j$ for the first four components of the simulation run that
achieved the median of MSE are presented.  The pictures in the
second row are histograms of the knots used in the one-stage WLSM
estimation for the first four components, respectively.}
\label{WLSM2}
\end{figure}

\begin{figure}
\vspace{-0.5in} \centerline{
\psfig{file=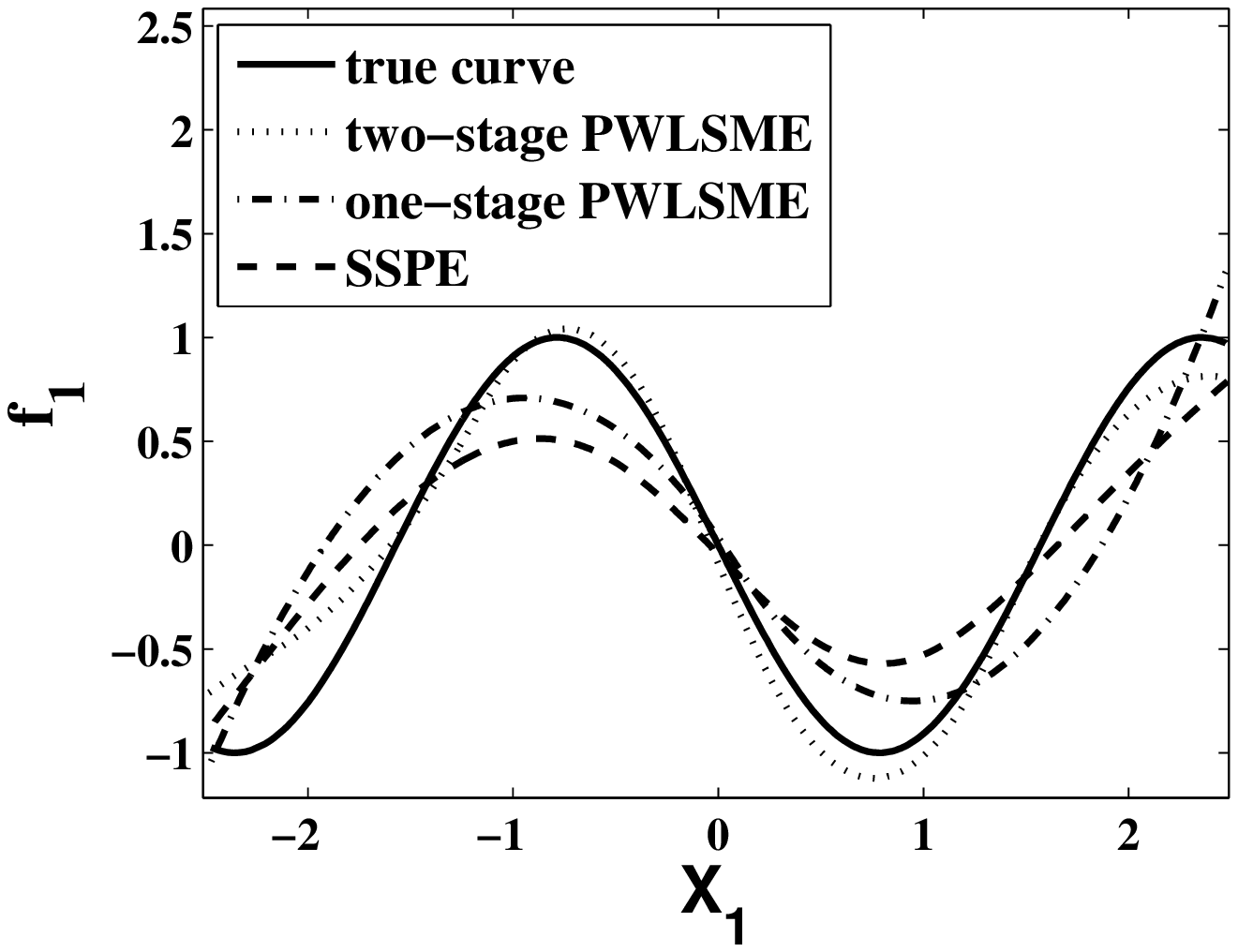,width=1.65in,height=1.85in}\hspace{-0.5cm}
\psfig{file=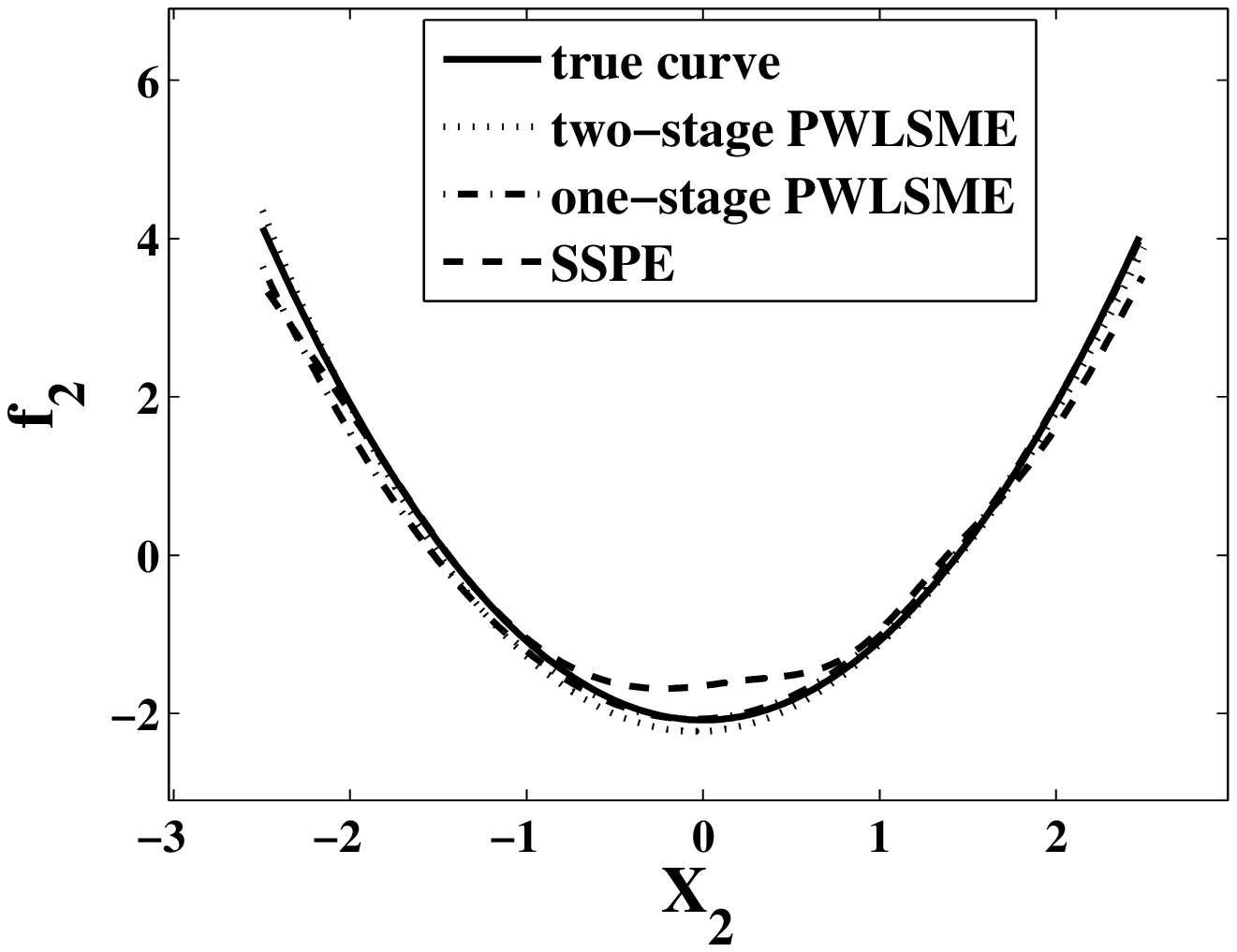,width=1.65in,height=1.85in}\hspace{-0.5cm}
\psfig{file=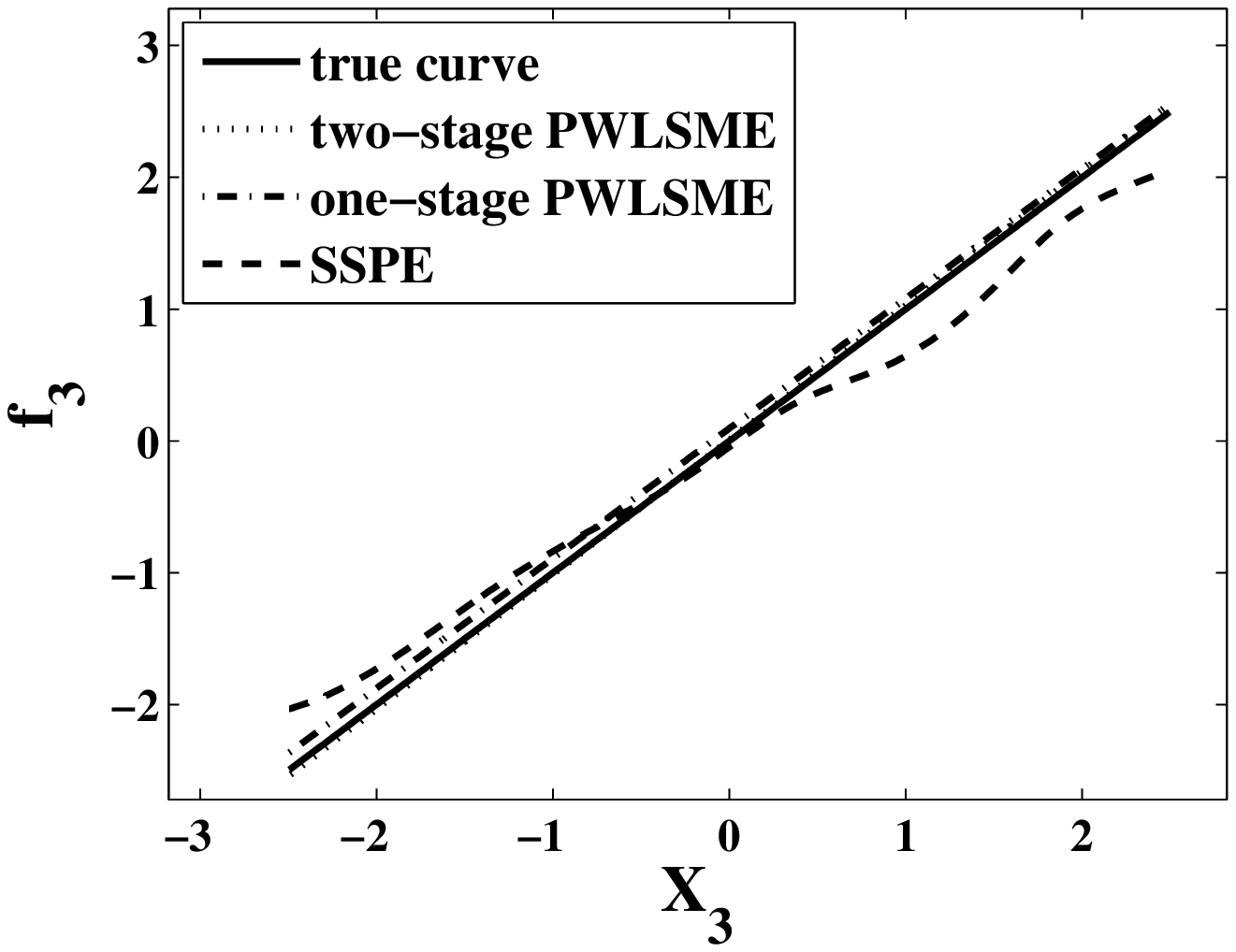,width=1.65in,height=1.85in}\hspace{-0.5cm}
\psfig{file=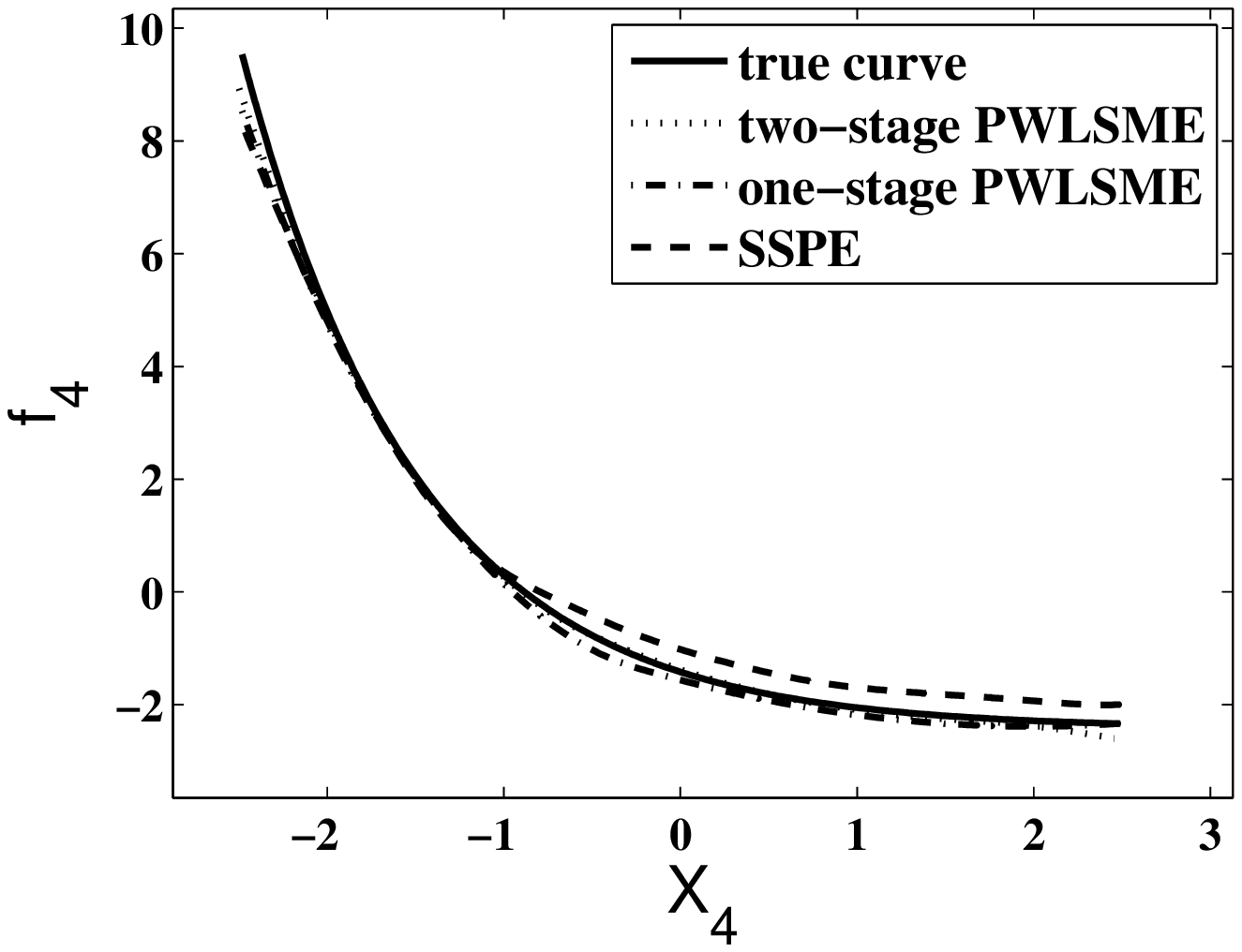,width=1.65in,height=1.85in}\hspace{-0.5cm}
} \centerline{
\psfig{file=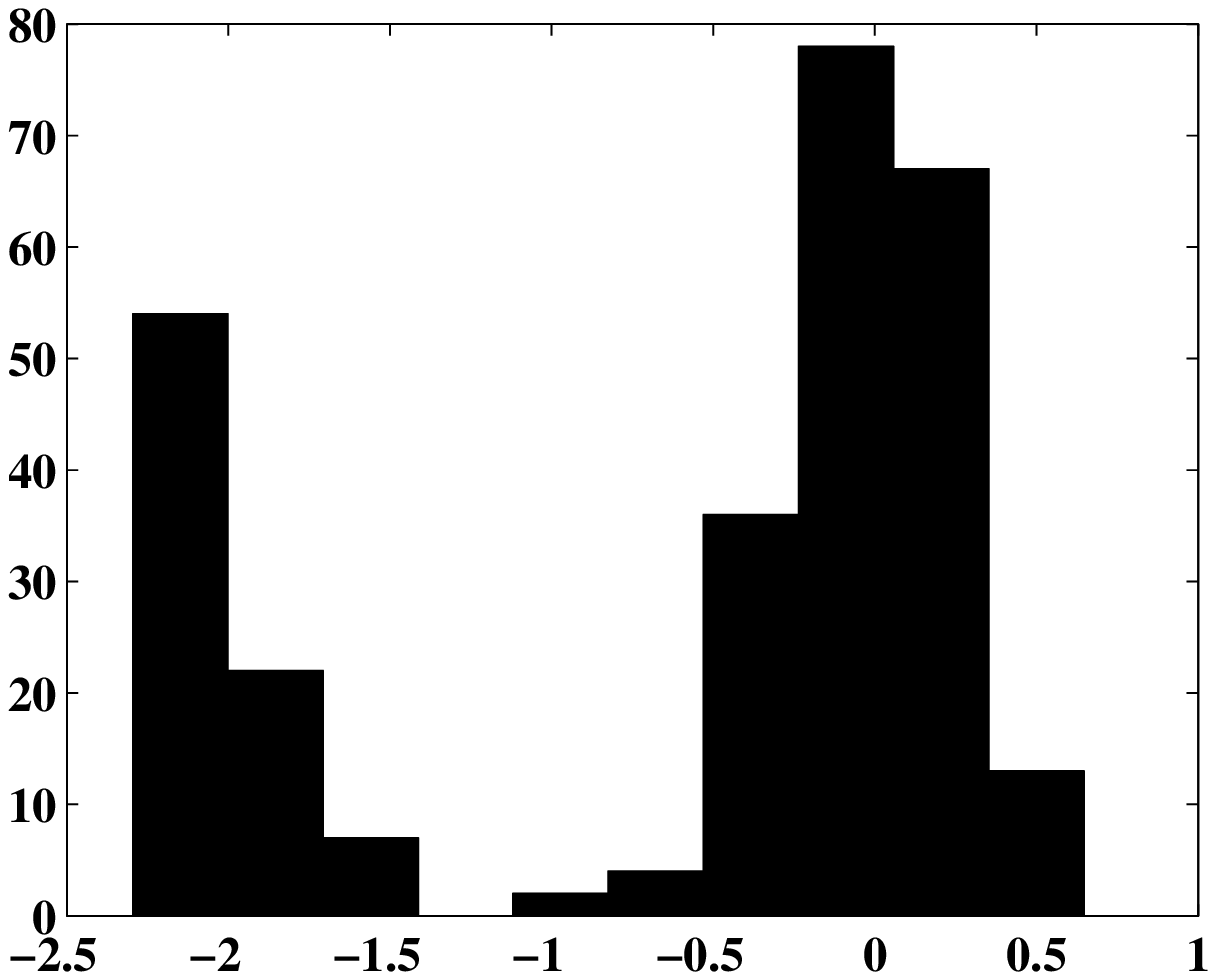,width=1.65in,height=1.35in}\hspace{-0.5cm}
\psfig{file=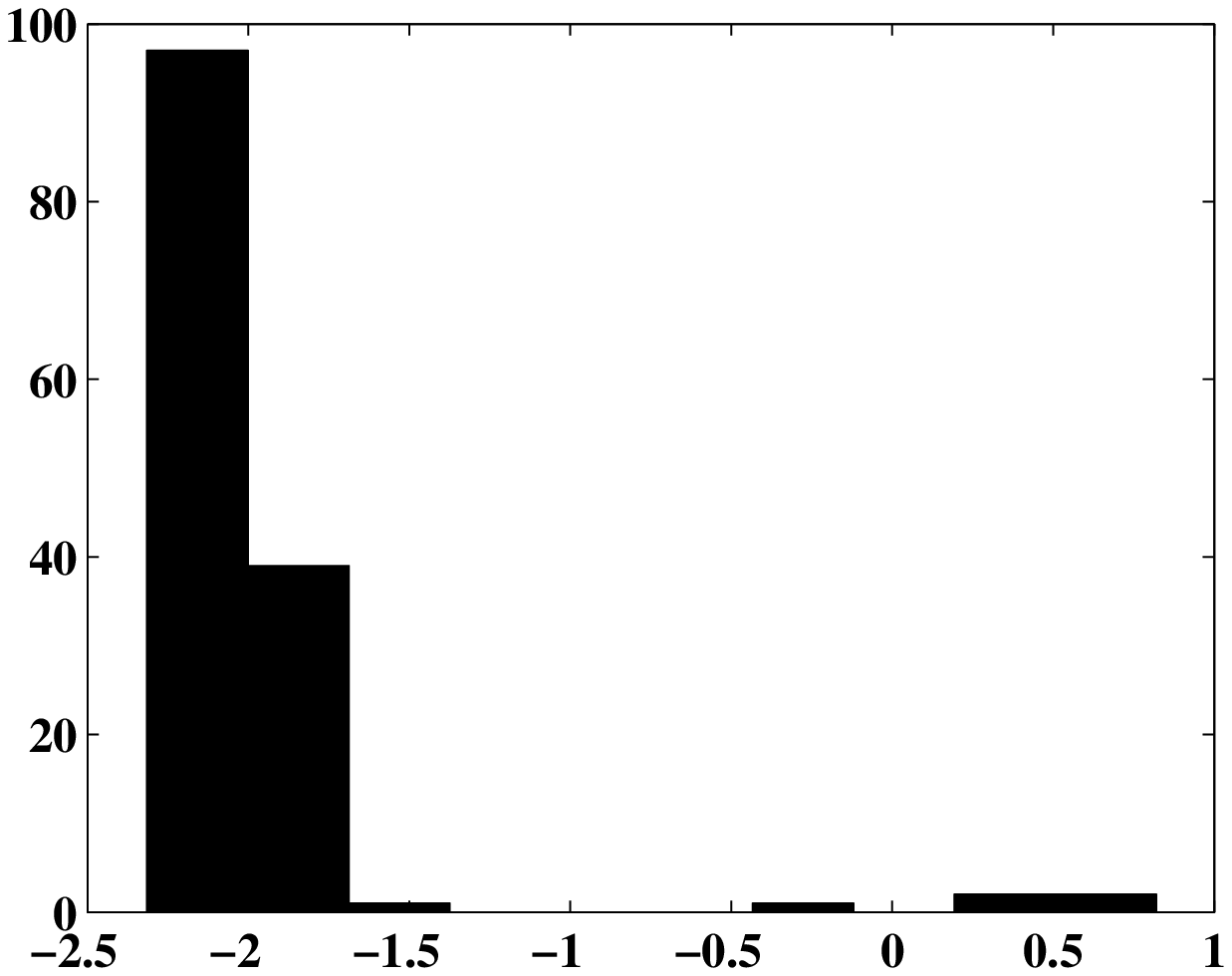,width=1.65in,height=1.35in}\hspace{-0.5cm}
\psfig{file=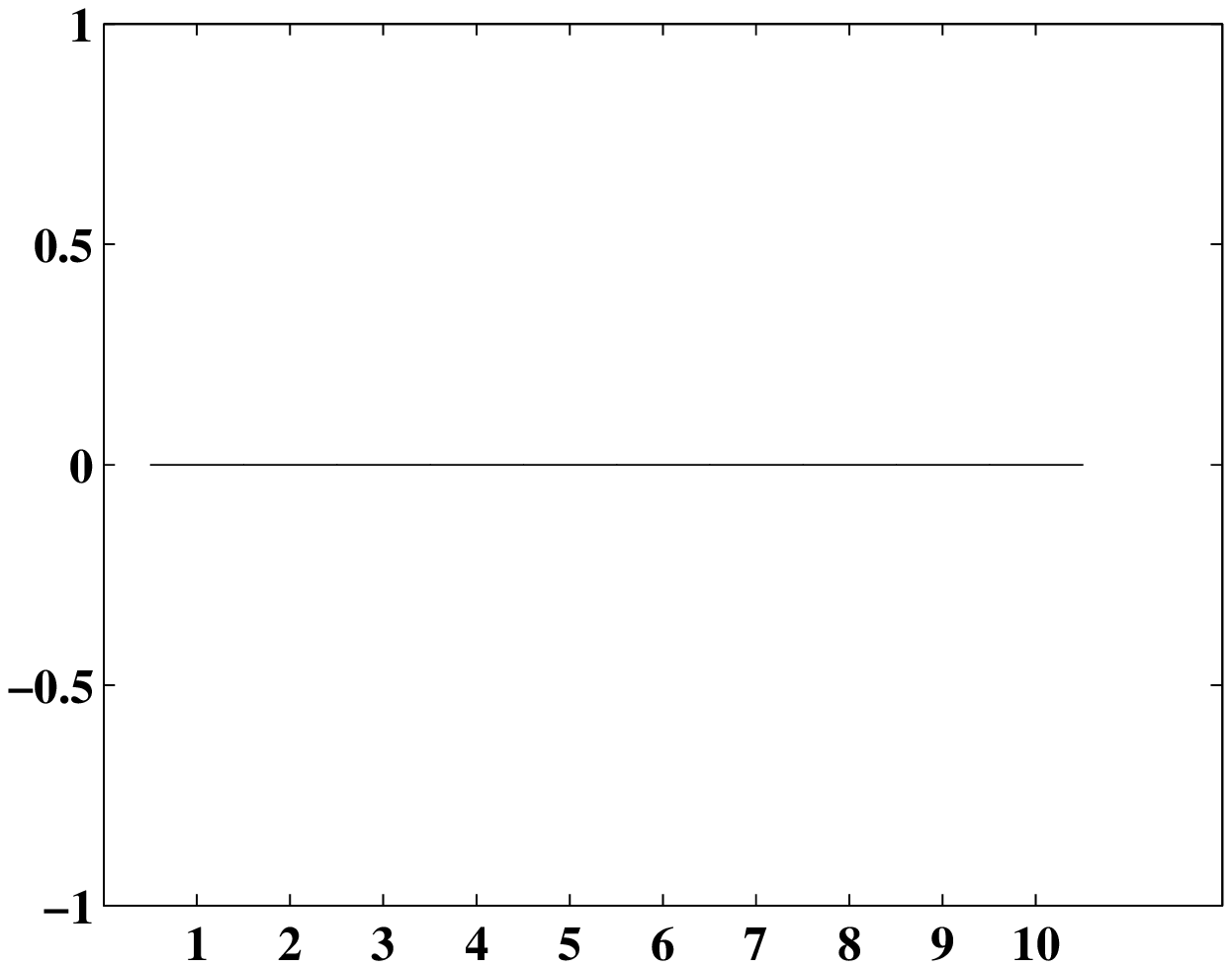,width=1.65in,height=1.35in}\hspace{-0.5cm}
\psfig{file=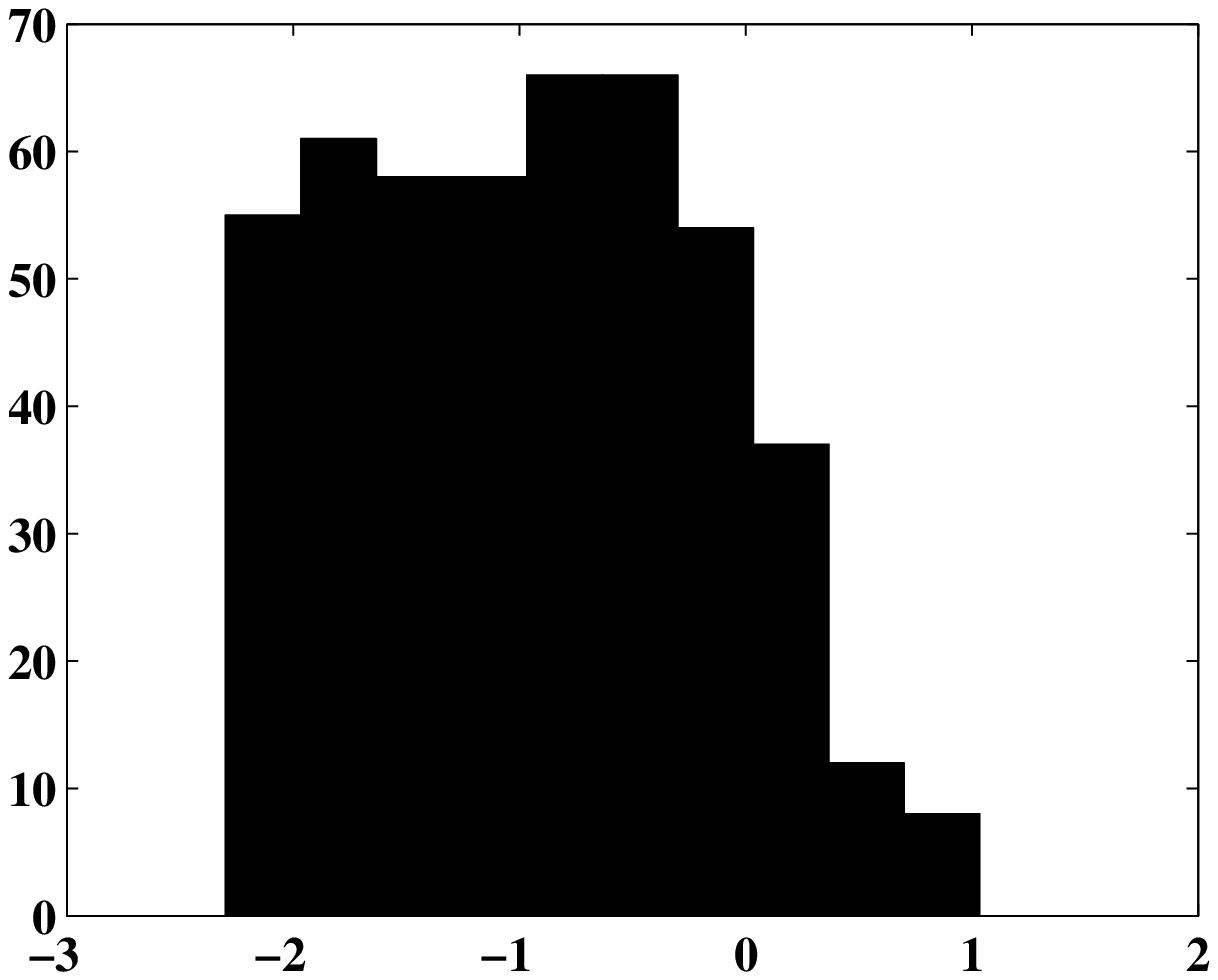,width=1.65in,height=1.35in}\hspace{-0.5cm}
} \vspace{-0.2in} \caption{\footnotesize Estimation curves from the
projected weighted Lasso-type spline method (PWLSM) for Example
\ref{example-ind}. True functions $f_j$ and estimation curves
$\hat{f}_j$ for the first four components of the simulation run that
achieved the median of the MSE are presented.  The pictures in the
second row are histograms of the knots used in the one-stage PWLSM
estimation for the first four components, respectively.}
\label{PWLSM1}
\end{figure}

\begin{figure}
\vspace{0in} \centerline{
\psfig{file=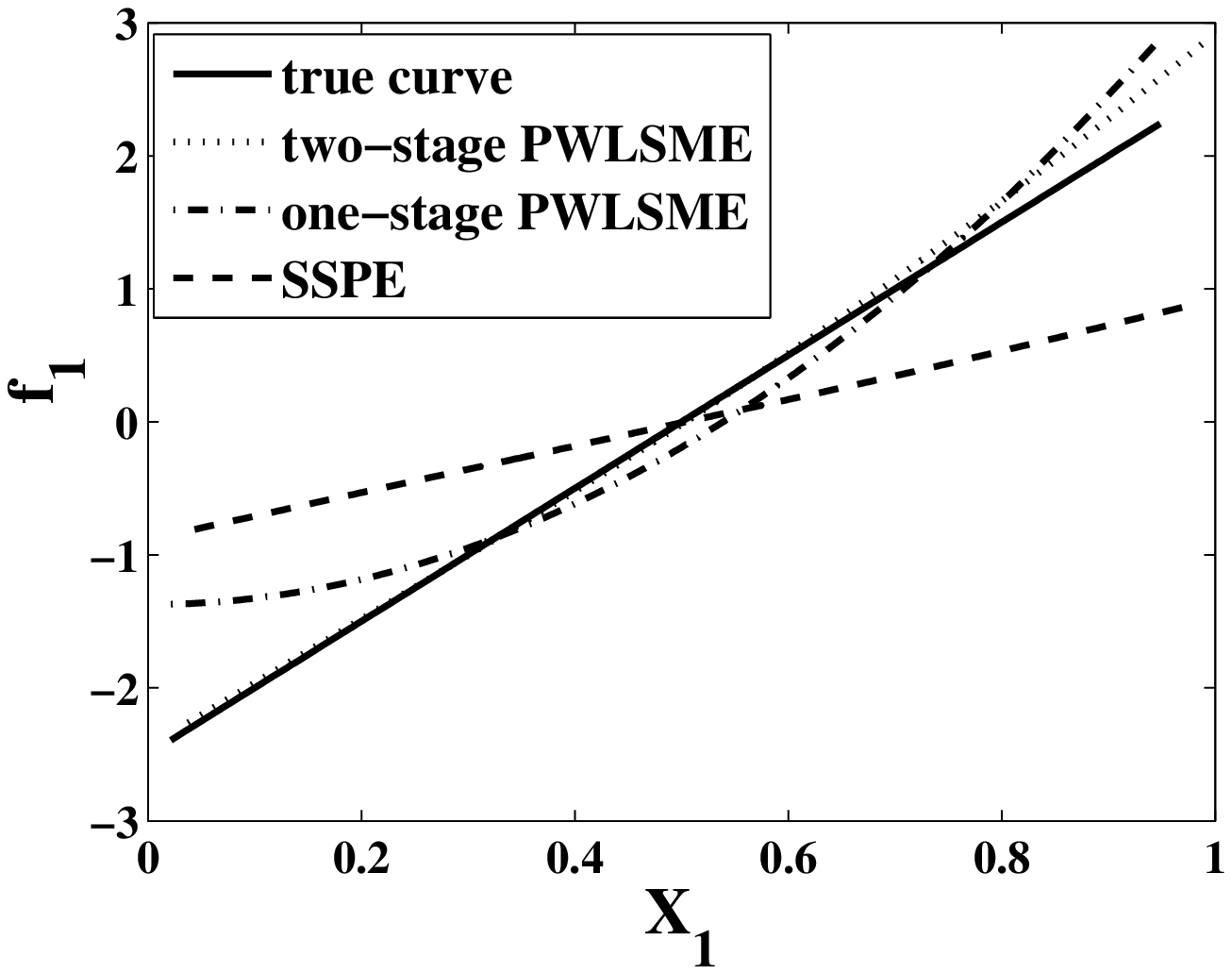,width=1.65in,height=1.85in}\hspace{-0.5cm}
\psfig{file=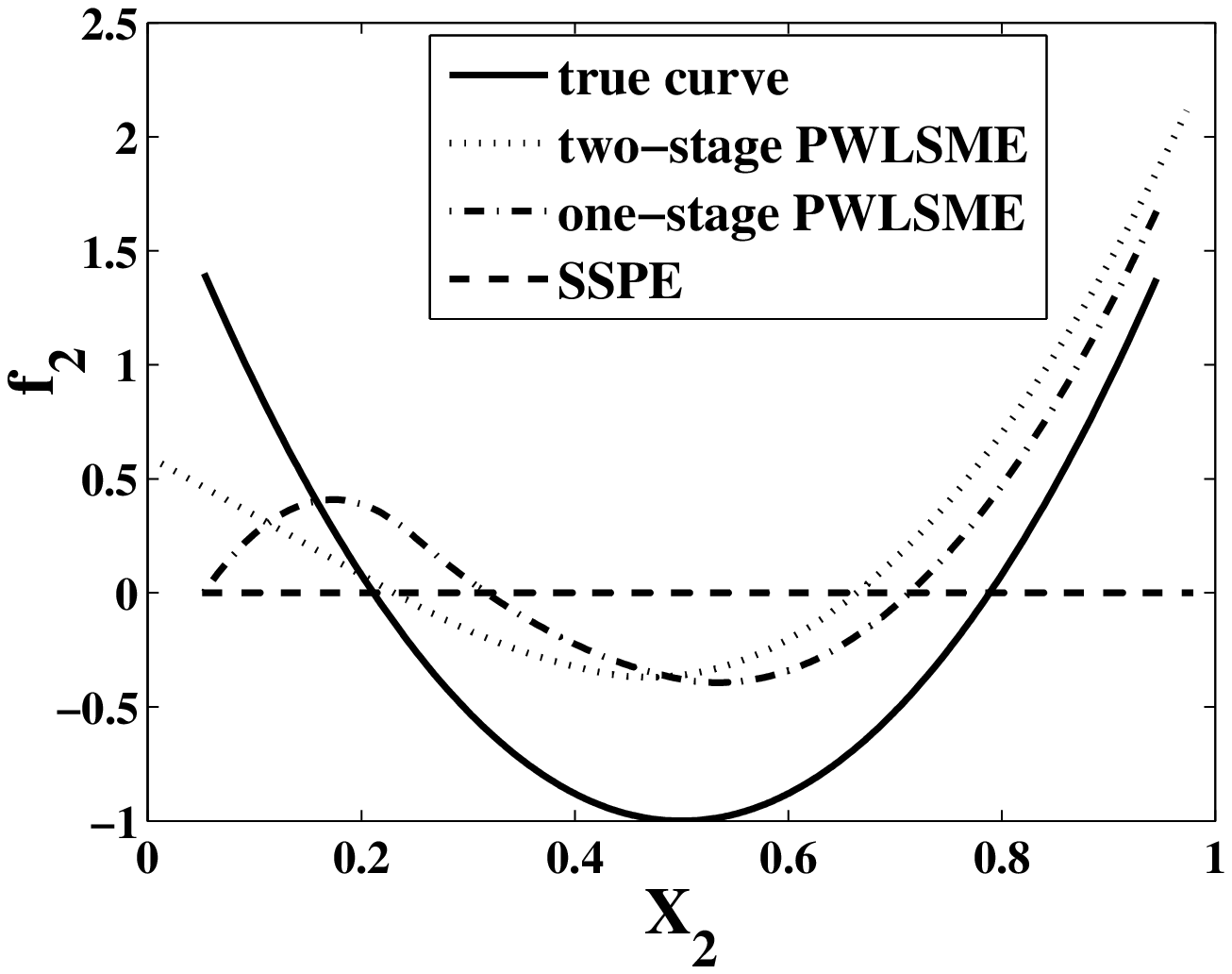,width=1.65in,height=1.85in}\hspace{-0.5cm}
\psfig{file=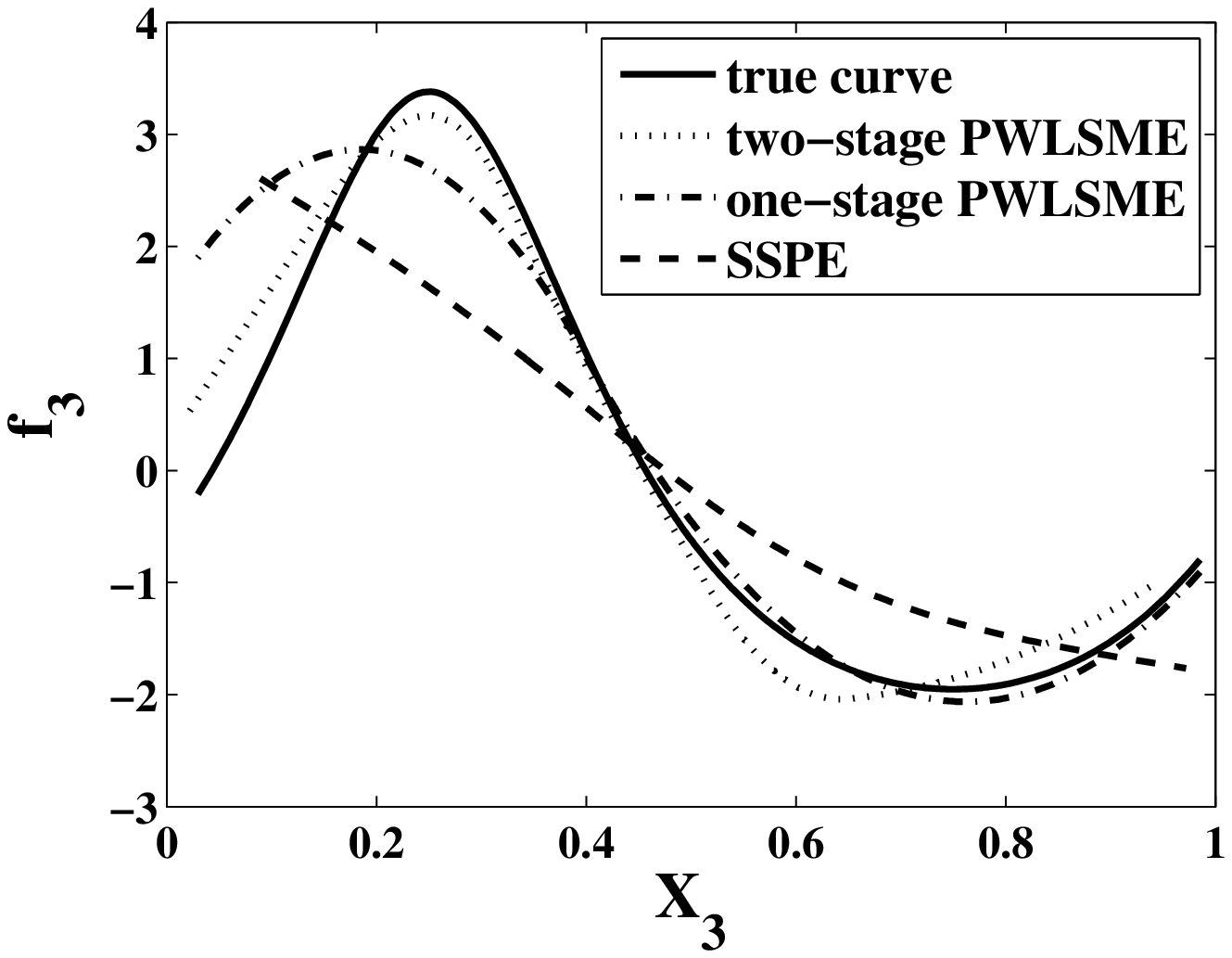,width=1.65in,height=1.85in}\hspace{-0.5cm}
\psfig{file=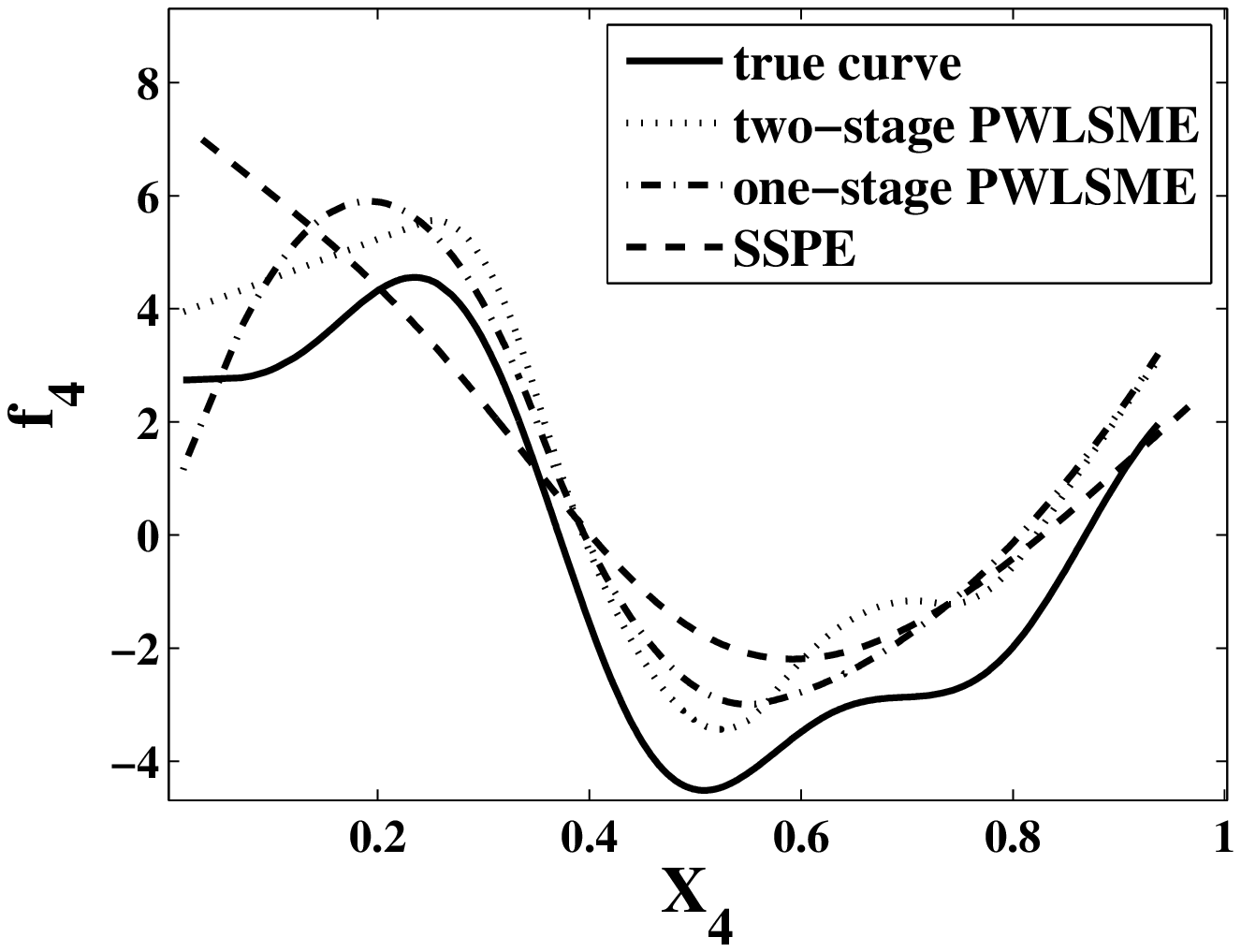,width=1.65in,height=1.85in}\hspace{-0.5cm}
} \centerline{
\psfig{file=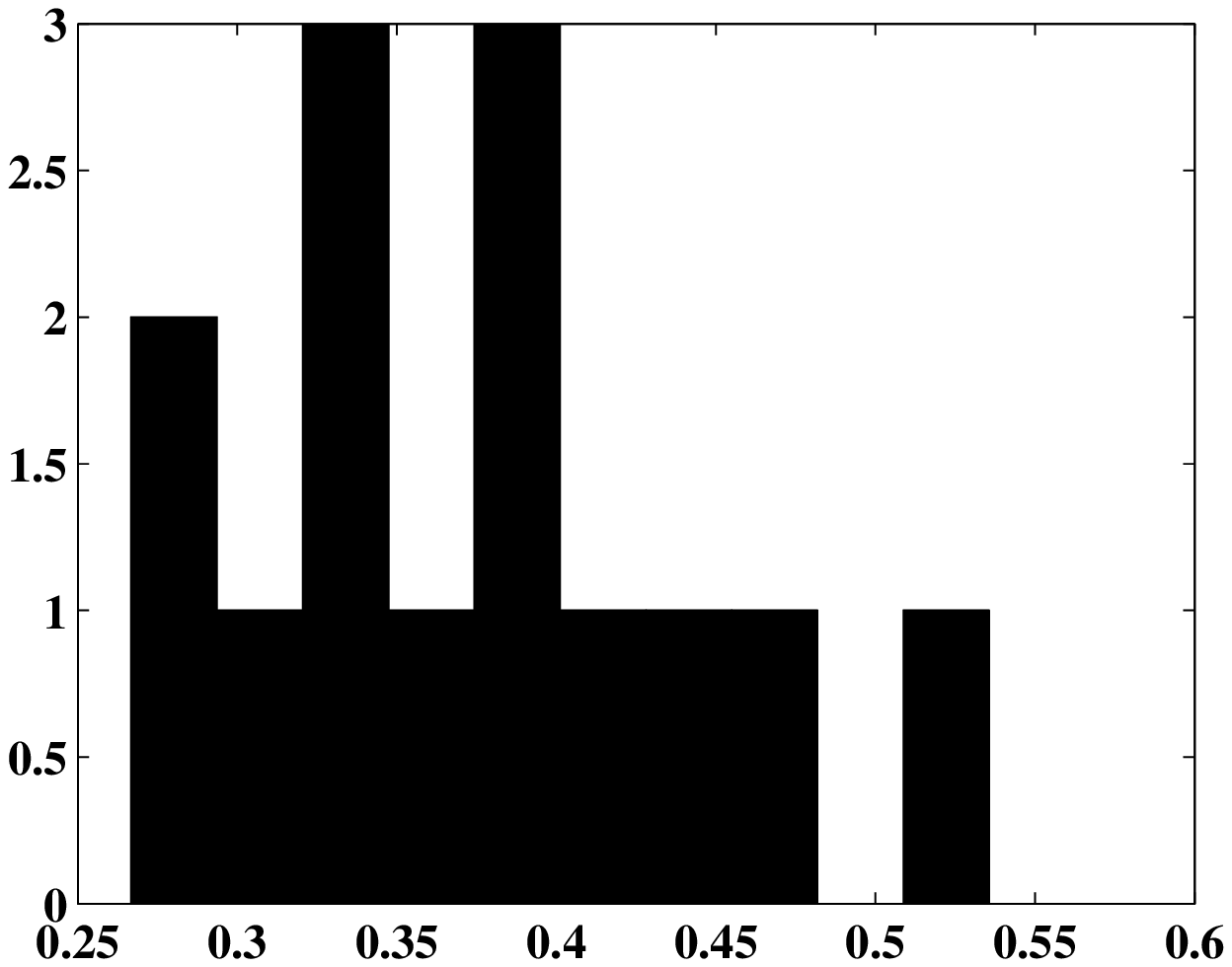,width=1.65in,height=1.35in}\hspace{-0.5cm}
\psfig{file=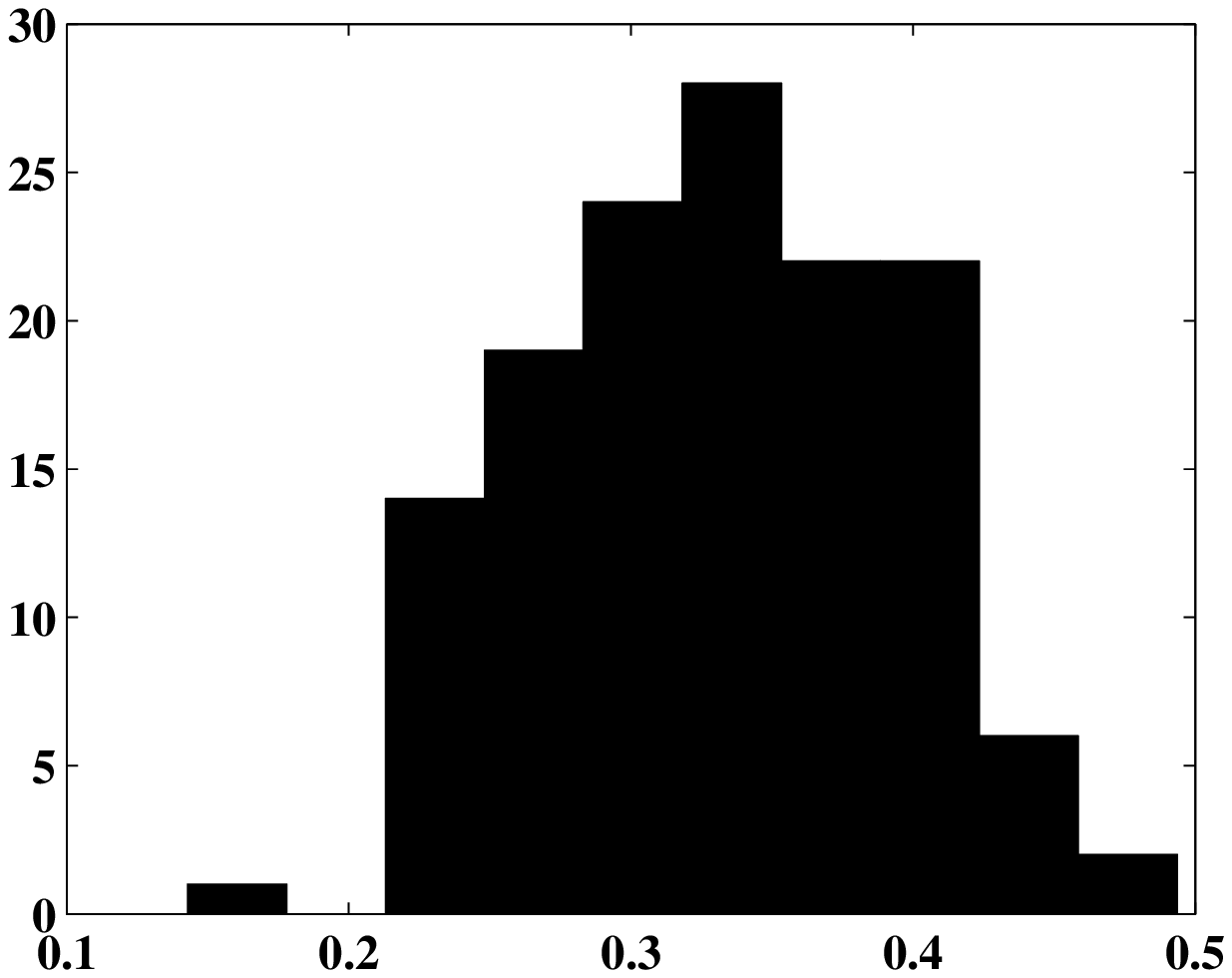,width=1.65in,height=1.35in}\hspace{-0.5cm}
\psfig{file=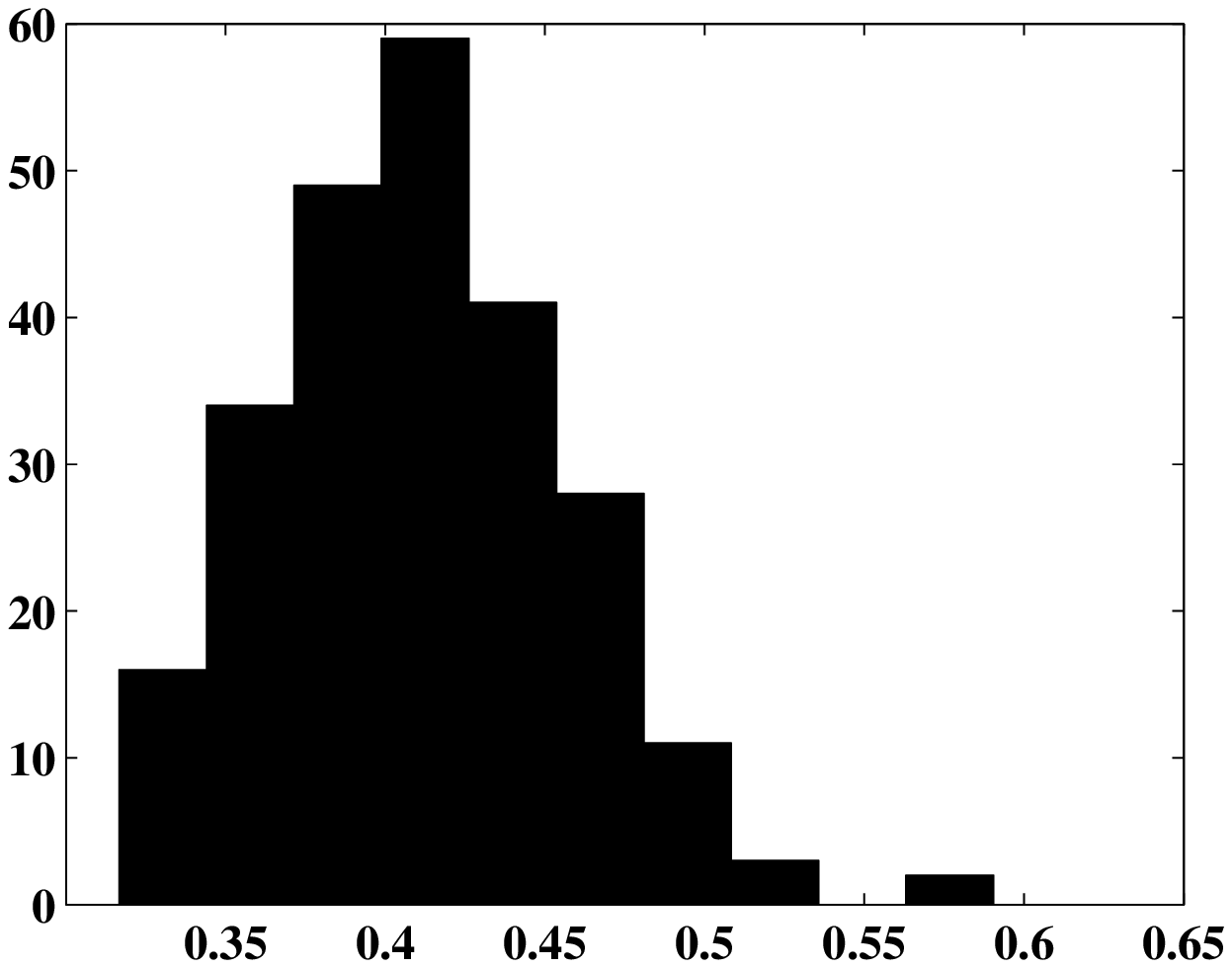,width=1.65in,height=1.35in}\hspace{-0.5cm}
\psfig{file=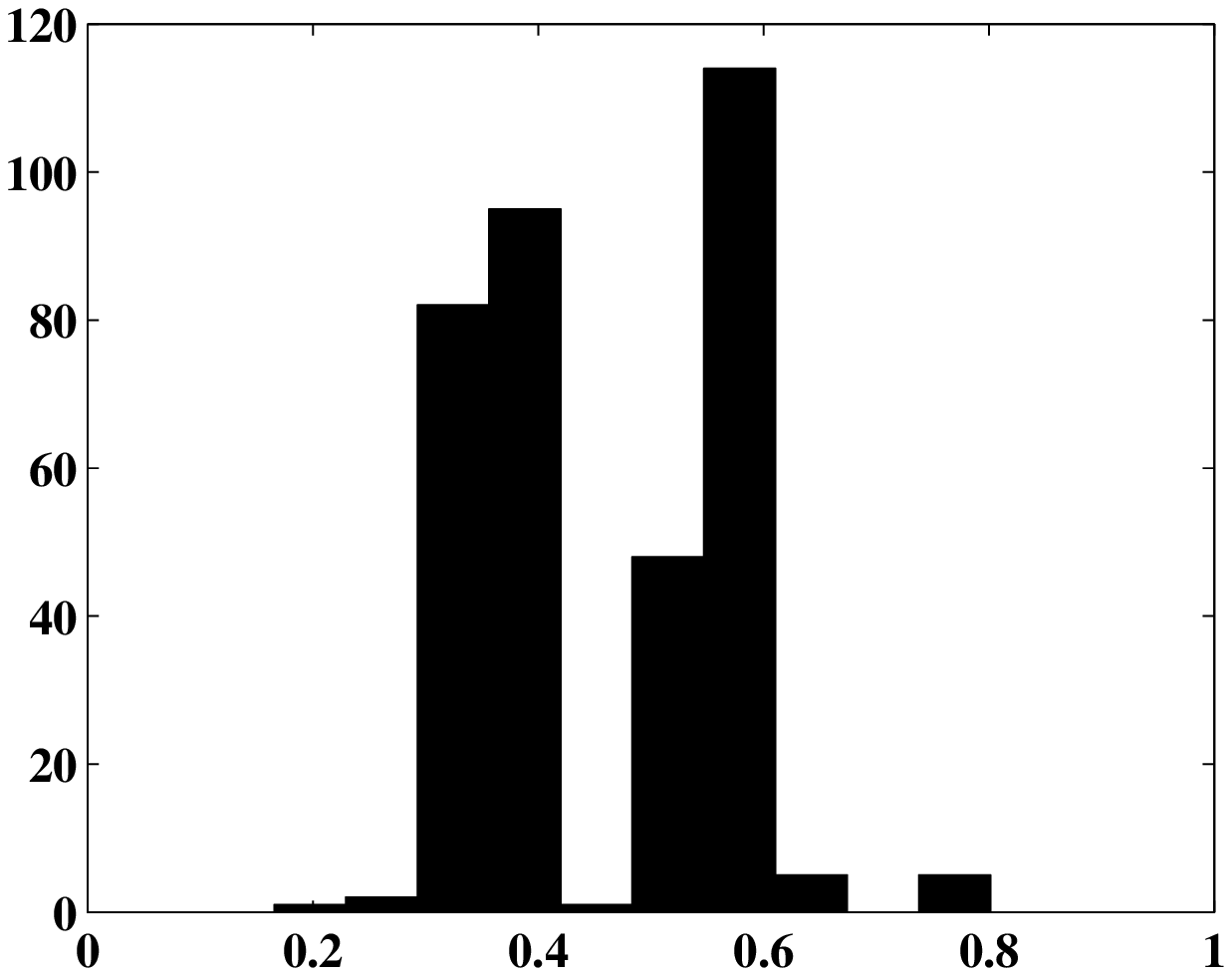,width=1.65in,height=1.35in}\hspace{-0.5cm}
} \vspace{-0.2in} \caption{\footnotesize Estimation curves from the
projected weighted Lasso-type spline method (PWLSM) for Example
\ref{example-cor}. True functions $f_j$ and estimation curves
$\hat{f}_j$ for the first four components of the simulation run that
achieved the median of the MSE are presented.  The pictures in the
second row are histograms of the knots used in the one-stage PWLSM
estimation for the first four components, respectively.}
\label{PWLSM2}
\end{figure}

The  simulation results are reported in Tables \ref{table-ind} and
\ref{table-cor}. The median of the MSE and the robust standard
deviation of the MSE (the ratio of the interquartile and the
standard normal interquartile
$(\Phi^{-1}(0.75)-\Phi^{-1}(0.25))$)  are reported.
``$\mbox{MSE}_{f_k}$'' means the MSE value of the estimates for
$f_k(\cdot)$, and ``$\mbox{MSE}$'' means the MSE for the full model.
The row ``SSP'' shows the results of the SSP method developed by
Meier, van de Geer, and B\"{u}hlmann (2009). The rows ``OLSM'',
``WLSM'', and ``PWLSM'' respectively summarize the results that are
based on ``oracle'' (assuming that all of the functions are known
except that to be estimated), ``one-stage'', and ``two-stage''
estimates for the original lasso-type spline method, the weighted
lasso-type spline method, and the projected weighted lasso-type
method, respectively. The TPR and FPR results for each method are
reported in Table \ref{table-TP}. The curve estimations for the
component functions are respectively summarized in Figures
\ref{OLSM1}---\ref{PWLSM2}.

\ctable[caption = Mean squared error (MSE) for Example
\ref{example-ind} (the numbers in  parentheses are the robust
standard deviations estimations)., label= table-ind,
pos=h]{lccccccccc}{}
{
\FL
\hline
       &            & $\mbox{MSE}_{f_1}$ & $\mbox{MSE}_{f_2}$ & $\mbox{MSE}_{f_3}$ & $\mbox{MSE}_{f_4}$ & MSE        \NN 
       \hline
SSP    &            & 0.122(0.189)     & 0.105(0.210)     & 0.105(0.189)     & 0.124(0.188)     & 1.358(0.277)   \NN
       \hline
       & Oracle     & 0.024(0.110)     & 0.013(0.108)     & 0.008(0.080)     & 0.028(0.126)     & 0.082(0.217) \NN
OLSM   & One-stage  & 0.530(0.142)     & 0.119(0.217)     &
0.190(0.217)     & 0.511(0.253)     & 1.357(0.295)    \NN
       & Two-stage  & 0.526(0.145)     & 0.022(0.143)     & 0.035(0.188)     & 0.366(0.165)     & 0.949(0.267)\NN
       \hline
       & Oracle     & 0.020(0.105)     & 0.013(0.111)     & 0.008(0.083)     & 0.027(0.128)     & 0.081(0.219)  \NN
WLSM   & One-stage  & 0.479(0.292)     & 0.025(0.133)     &
0.052(0.204)     & 0.071(0.175)     & 0.615(0.338)   \NN
       & Two-stage  & 0.040(0.592)     & 0.014(0.118)     & 0.013(0.093)     & 0.038(0.142)     & 0.141(0.575)   \NN
       \hline
       & Oracle     & 0.011(0.101)     & 0.015(0.112)     & 0.006(0.084)     & 0.024(0.130)     & 0.078(0.210)   \NN
PWLSM  & One-stage  & 0.115(0.141)     & 0.016(0.115)     &
0.008(0.094)     & 0.042(0.158)     & 0.194(0.265)   \NN
       & Two-stage  & 0.022(0.111)     & 0.016(0.125)     & 0.011(0.108)     & 0.027(0.146)     & 0.090(0.251)  \LL }

\ctable[caption = Mean square error (MSE) for Example
\ref{example-cor} (the numbers in  parentheses are the robust
standard deviations estimations)., label= table-cor,
pos=h]{lccccccccc} {}
{
\FL\hline
       &            & $\mbox{MSE}_{f_1}$ & $\mbox{MSE}_{f_2}$ & $\mbox{MSE}_{f_3}$ & $\mbox{MSE}_{f_4}$ & MSE              \NN 
       \hline
SSP    &            & 0.225(0.373)     & 0.430(0.624)     & 0.755(0.328)     & 3.020(0.580)     & 7.224(0.764)  \NN
       \hline
       & Oracle     & 0.025(0.116)     & 0.282(0.198)     & 0.047(0.148)     & 1.566(0.635)     & 3.127(0.795) \NN
OLSM   & One-stage  & 0.133(0.277)     & 0.594(0.139)     &
0.394(0.305)     & 2.315(0.611)     & 4.475(0.789)    \NN
       & Two-stage  & 0.045(0.218)     & 0.589(0.181)     & 0.163(0.256)     & 1.696(0.621)     & 3.585(0.787) \NN
       \hline
       & Oracle     & 0.018(0.110)     & 0.281(0.187)     & 0.050(0.154)     & 1.558(0.623)     & 3.112(0.780)  \NN
WLSM   & One-stage  & 0.072(0.140)     & 0.470(0.412)     &
0.309(0.284)     & 1.880(0.623)     & 3.842(0.831)    \NN
       & Two-stage  & 0.027(0.149)     & 0.312(0.464)     & 0.046(0.161)     & 1.574(0.631)     & 3.216(0.818)   \NN
       \hline
       & Oracle     & 0.025(0.112)     & 0.295(0.210)     & 0.055(0.156)     & 1.573(0.630)     & 3.178(0.778)   \NN
PWLSM  & One-stage  & 0.066(0.124)     & 0.403(0.364)     &
0.195(0.212)     & 1.759(0.606)     & 3.590(0.784)   \NN
       & Two-stage  & 0.022(0.134)     & 0.295(0.236)     & 0.052(0.159)     & 1.555(0.628)     & 3.176(0.760)   \LL} 

\ctable[caption = Median of the number of true positives (TP) and
false positives (FP) (the numbers in  parentheses are the standard
robust deviations estimations).,
label=table-TP,  pos=h]{lccccccccc} {} { 
\FL\hline
                             &       & TP              & FP                \NN
\hline
                             & SSP   & 4.000(0.000)   & 0.000(0.000)     \NN
                             & OLSM  & 3.000(0.000)   & 0.000(0.000)    \NN
Example \ref{example-ind}    & WLSM  & 4.000(0.741)   & 0.000(0.000)    \NN
                             & PWLSM & 4.000(0.000)   & 0.000(0.000)    \NN
\hline
                             & SSP   & 4.000(0.741)   & 0.000(0.741)    \NN
                             & OLSM  & 3.000(0.000)   & 1.000(1.483)    \NN
Example \ref{example-cor}    & WLSM  & 4.000(0.741)   & 0.000(0.741)     \NN
                             & PWLSM & 4.000(0.000)   & 1.000(0.741)     \LL }

The results tabulated in Tables~\ref{table-ind} and \ref{table-TP}
and plotted in Figures~\ref{WLSM1} and \ref{PWLSM1} with independent
covariates in Example \ref{example-ind} 
show the differences among the ``oracle'' cases and the  OLSM, WLSM,
and PWLSM cases to be nearly negligible. The MSE of the two-stage
estimates is significantly smaller  than  that of the corresponding
one-stage estimates and approximate that in the ``oracle'' case. Of
all of these methods, the  two-stage estimates of the PWLSM are
superior to the others. The numbers of true positives and false
positives for the PWLSM are the same as those of the SSP approach.
However, the OLSM has a smaller number of true positives and the
WLSM has a larger variation in true positives than the SSP, although
the true positives for the WLSM is the same as that for the SSP.
Figures \ref{OLSM1}, \ref{WLSM1}, and \ref{PWLSM1} show that the
OLSM and WLSM may fail to select the first component function. This
is because the first function is small in magnitude,  and is easily
selected out from the model due to the penalty. In contrast, the
PWLSM always selects all of the non-zero component functions into
the model, and the knots used in the spline basis are very sparse.
Furthermore, the PWLSM selects the linear function (for example, the
third panel on the top) as linear, whereas the SSP fails to do so.
The PWLSM thus outperforms the other methods in this setting.

For the correlated covariate case in Example \ref{example-cor}, the
results presented in Tables \ref{table-cor} and \ref{table-TP} and
Figures \ref{OLSM2}, \ref{WLSM2} and \ref{PWLSM2}) suggest that  all
of  the ``oracle'' estimations perform similarly. Our proposed three
LSMs all apparently improve on the method the SSP in terms of the
MSE. The numbers of true positives and false positives for the WLSM
are the same as those for the SSP. However, the PWLSM does not
perform better in every respect, as it keeps selecting all of the
true components at the expense of including a component that is
slightly more noisy than the other insignificant components.

\subsection{Application}

In this section, we give more details of the analysis of the dataset
described in Section~\ref{sec-data}. As described, the dataset has
been analyzed by Fan and Peng (2004) with the linear model and the
additive model, respectively.
We now use the method proposed in this paper to analyze the dataset
and to make a comparison with the results of  Fan and Peng (2004).

\begin{figure}
\vspace{0in}
\begin{center}
\begin{tabular}{cc}
\psfig{file=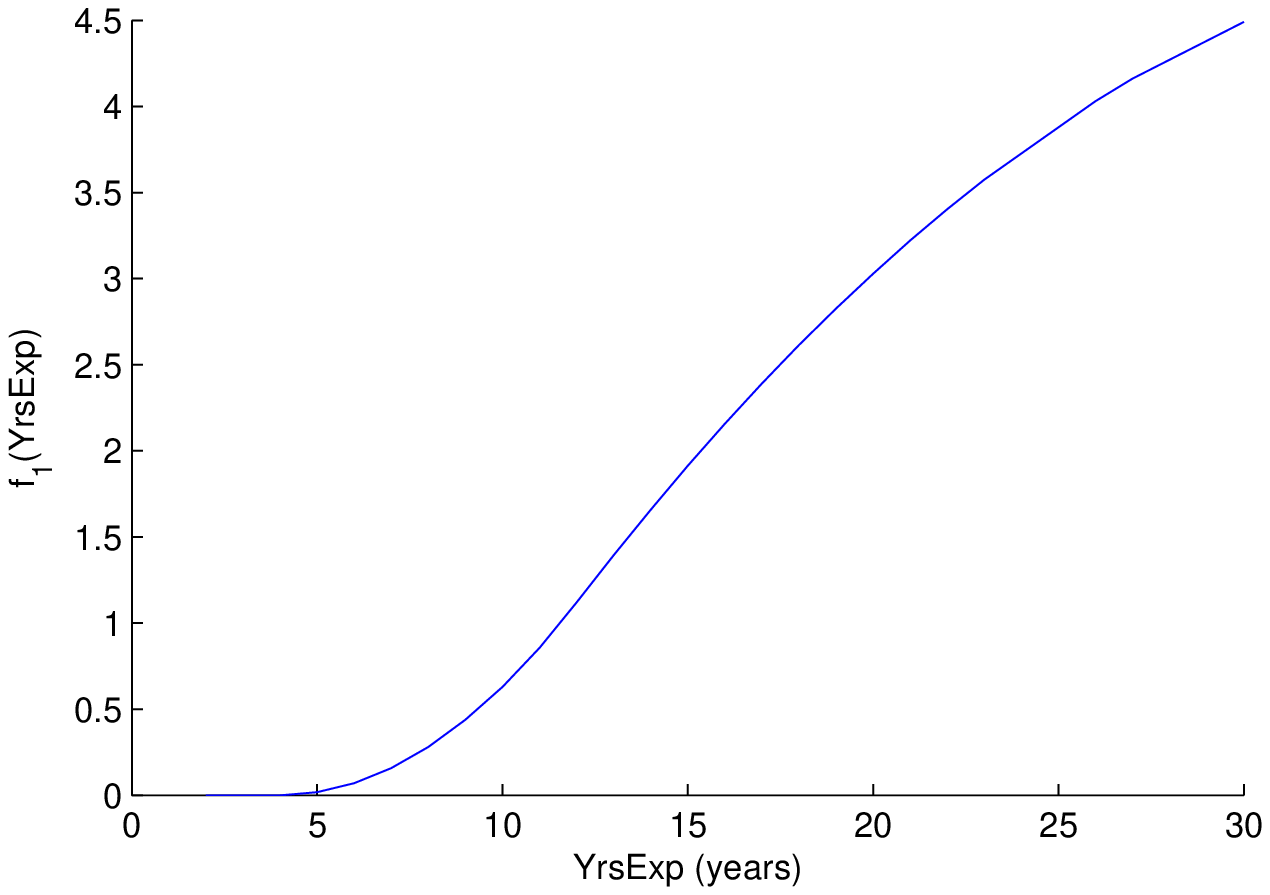,width=2.8in,height=1.8in}  &
\psfig{file=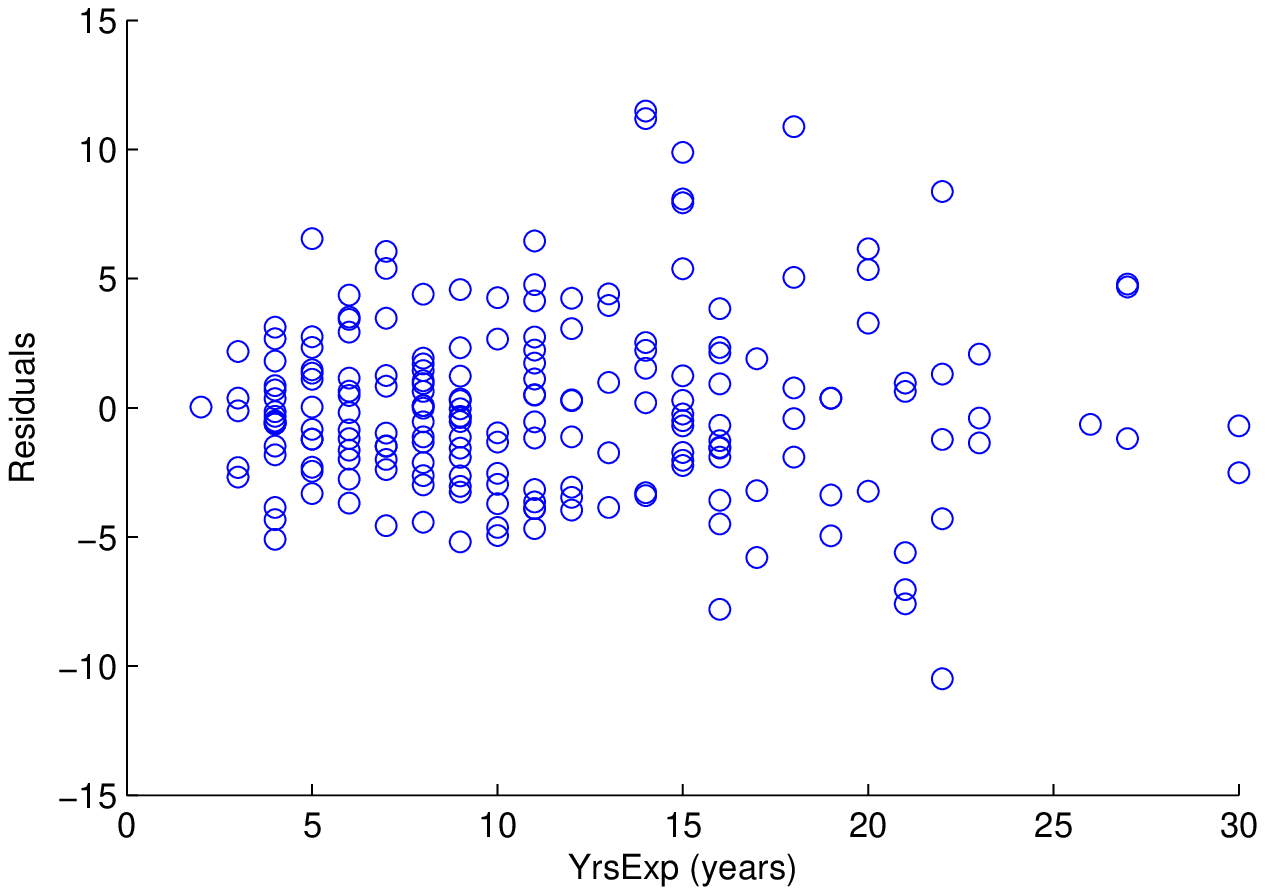,width=2.8in,height=1.8in} \\
(a) & (b)
\end{tabular}
\end{center}
 \vspace{-0.2in} \caption{\footnotesize(a) Regression components $f_1$,  and (b) Residuals versus Years Experience for
the partial additive linear
    model by SCAD penalized least-squares (\ref{d2}).}
\label{BK3}
\end{figure}

Similar to the approach of Fan and Peng (2004), we also move out the
outliers and use the 199 remaining observations for our analysis.
Consider the additive model
\begin{eqnarray}
\label{d2} \nonumber \mathrm{Salary} &=& \beta_0 + \beta_1
\mathrm{Female}
    +\beta_2 \mbox{PCJob} +
\sum\limits_{i=1}^4 \beta_{2+i} \mathrm{Edu}_i + \sum\limits_{i=1}^5
\beta_{6+i} \mathrm{JobGrd}_i \\ & \ &  + f_1(\mathrm{YrsExp})+ f_2(
\mathrm{Age}) +\varepsilon.
\end{eqnarray}
We use LASSO for the linear part and our method for the component
function selection. The $2/17, 3/17, \ldots, 16/17$ sample quantiles
of the variables ``YrsExp'' and ``Age'' are selected as knots, which
gives 15 initial knots to estimate the component functions. Fan and
Peng (2004) used only 5 knots to estimate each component function,
whereas our method gives 20 more parameters to model data. Despite
this, the computational complexity is not increased because the
quadratic approximation algorithm can be easily implemented,  and
most of the knots will be removed in an iterative fashion by our
component selection procedure. In line with Fan and Peng (2004) and
the foregoing discussion, we first weight the ``design matrix" such
that the original least-squares estimate has the standard deviation
of every estimate of the coefficients of the prediction variables
and a truncated power basis function close to the order of
$1/\sqrt{n}$. Two tuning parameters are then used to select the
variables in the linear part and the component functions. MGCV with
an inflation parameter of $1.5$ is used to select the tuning
parameters. The results are reported in the fourth column (WLSM, see
Section \ref{addsimu}) of Table \ref{table-data} and in Figure
\ref{BK3}.

Table \ref{table-data} shows that our method does not select the
component function of ``Age''. This is consistent with the result of
SCAD-PLS for the linear model, which is reported in the second
column of Table \ref{table-data}. The other estimates of the
coefficients are similar to those obtained with the first three
methods. The function of ``YrsExp''  is now estimated as an
increasing function (see Figure \ref{BK3}). Only two spline bases,
from among the 17 are selected to estimate the component function of
``YrsExp''. Hence, the selected model is much simpler than the
selected model derived with SCAD-PLS under the additive structure.
The $R^2$ value in the fourth column for the WLSM is larger than
that in the second or third column for SCAD-PLS. This means that,
compared with the SCAD-PLS method for either the linear model or the
additive model, our method provides a more reasonable estimation and
selects a simpler model in this real data example.

\ctable[caption = Estimates and standard errors for the Fifth
National Bank data., label=table-data, pos=h]
{rrcrcrcrcrcccccc} {} { 
\FL\hline
 &           & & {Least-Squares} & & {SCAD PLS}     & &   { SCAD PLS } & &{WLSM} \NN
\hline
 & Intercept & & 54.238(2.067)  & & 55.835(1.527)  & & 52.470(2.890)  & &55.820(1.437) \NN
 & Female    & & -0.556(0.637)  & & -0.624(0.639)  & & -0.933(0.708)  & &-0.693(0.656) \NN
 & PcJob     & & 3.982 (0.908)  & & 4.151(0.909)   & & 2.851(0.640)   & &3.935(0.908)\NN
 & Ed1       & & -1.739(1.049)  & &  0(---)        & & 0(---)         & & 0(---)     \NN
 & Ed2       & & -2.866(0.999)  & & -1.074(0.522)  & & -0.542(0.265)  & &-1.385(0.764)\NN
 & Ed3       & & -2.145(0.753)  & & -0.914(0.421)  & & 0(---)         & &-1.180(0.601) \NN
 & Ed4       & & -1.484(1.369)  & &  0(---)        & & 0(---)         & &  0(---)         \NN
 & Job1      & & -22.954(1.734) & & -24.643(1.535) & & -22.841(1.332) & & -23.325( 1.561) \NN
 & Job2      & & -21.388(1.686) & & -22.818(1.546) & & -20.591(1.370) & & -21.494(1.580) \NN
 & Job3      & & -17.642(1.634) & & -18.803(1.562) & & -16.719(1.391) & & -17.440(1.602)\NN
 & Job4      & & -13.046(1.578) & & -13.859(1.529) & & -11.807(1.359) & & -12.536(1.542)\NN
 & Job5      & & -7.462(1.551)  & & -7.770(1.539)  & & -5.235(1.150)  & & -6.477 (1.537)\NN
 & YrsExp    & & 0.215(0.065)   & & 0.193(0.046)   & & ---(---)       & & ---(---)\NN
 & Age       & & 0.030(0.039)   & &  0(---)        & & ---(---)       & & 0(---)    \NN
 & $R^2$     & & 0.8221         & & 0.8176         & &  0.8123        & & 0.8182     \LL }

\section{Conclusion}
In this paper, we propose a LASSO-type method for selecting
nonparametric components in the additive regression model. We can
use this method to simultaneously select and estimate components.
Simulations show that for a high-dimensional additive model, the
proposed methods can shrink the function components that correspond
to the nonsignificant predictors exactly to zero and produce a
parsimonious model. For an ultra-high dimensional additive model, we
follow the idea of Fan \etal \ (2009) and use then SIS method to
first reduce the ultra-dimension of the additive model to a high
dimension, and then use our proposed method to select and estimate
the significant components.  Intuitively, it is possible to extend
this idea to generalized additive models with binary response data
or poisson data, or with a given link function. Research in this
area is ongoing.

\section{Proof of Theorems}\label{addproof}

\noindent{\sc Proof of Theorem 1:} 
As $\mu=Ef_0(\X)$,  a natural consistent estimate of $\mu$ is
$\hat\mu=\frac{1}{n}\sum_{i=1}^ny_i$. Without loss of generality,
assume that $\mu=0$. By  the condition (2.6), there exist
$f^\ast_{k,n}$ such that
$$\|f^\ast_{k,n}-f_k\|_n=O(n^{-1/(2+w)} c_n^{w/(2+w)}) \quad \mbox{and}
\quad \mathcal{P}(f^\ast_{k,n}) \le c_n, \quad \mbox{for} \
k=1,\ldots,K.$$ Define $f_n^\ast=\sum\limits_{k=1}^K f_{k,n}^\ast
\in \F_n$ that satisfies
$$\|f-f_{n}^\ast\|_n=\|\sum\limits_{k=1}^K
f_k-\sum\limits_{k=1}^K f_{k,n}^\ast\|_n \le \sum\limits_{k=1}^K
\|f_{k}-f_{k,n}^\ast\|_n=O(n^{-1/(2+w)}c_n^{w/(2+w)})
$$
and
$$
\mathcal{P}(f_n^\ast)=\sum\limits_{k=1}^K
\mathcal{P}^{\frac12}(f_{n,k}^\ast) \le K \cdot c^\frac12_n.
$$
By the definition of $\hat{f}_n$, we have
$$
\frac{1}{n}\sum\limits_{i=1}^n (Y_i-\hat{f}_n(\bX_i))^2+\lambda_n
\mathcal{P}(\hat{f}_n) \le \frac{1}{n}\sum\limits_{i=1}^n
(Y_i-f_n^\ast(\bX_i))^2+\lambda_n \mathcal{P}(f^\ast_n),
$$
and
\begin{eqnarray*}
\|\hat{f}_n-f_n^\ast\|^2_n & \le &  \lambda_n
(\mathcal{P}(f_n^\ast)-\mathcal{P}(\hat{f}_n)
+\frac{2}{n}\sum\limits_{i=1}^n
\varepsilon_i(\hat{f}_n(\bX_i)-f_n^\ast(\bX_i)) \\ &+&
\frac{2}{n}\sum\limits_{i=1}^n
(f(\bX_i)-f_n^\ast(\bX_i))(\hat{f}_n(\bX_i)-f_n^\ast(\bX_i))
\\
& \le  & \lambda_n (\mathcal{P}(f_n^\ast)-\mathcal{P}(\hat{f}_n)
+\frac{2}{n}\sum\limits_{i=1}^n
\varepsilon_i(\hat{f}_n(\bX_i)-f_n^\ast(\bX_i)) + 2 \|f_n^\ast-f\|_n
\cdot \|\hat{f}_n-{f}_n^\ast\|_n,
\end{eqnarray*}
where the second inequality is derived from the Cauchy-Schwarz
inequality. Note  that the condition (2.7) on the entropy bound
implies
$$
\sup\limits_{g \in \F_n(1)} \frac{|n^{-1/2}\sum_{i=1}^n
\varepsilon_n g(\bX_i)|}{\|g\|_n^{1-\frac{1}{2w}}}=O_p(1).
$$
Define
$g=(\hat{f}_n-f_n^\ast)/(\mathcal{P}(\hat{f}_n)+Kc^\frac12_n)^2.$ As
$\F_n$ is a linear space, it is easy to see that $g\in \F_n$ and
$\mathcal{P}(g) \le 1$. Then, by the entropy bound, we have
$$
|\frac{1}{n} \sum\limits_{i=1}^n \varepsilon_i
(\hat{f}_n(\bX_i)-f_n^\ast(\bX_i))|\le
\|\hat{f}_n-f_n^\ast\|_n^{1-\frac{w}{2}}(\mathcal{P}(\hat{f}_n)+Kc^\frac12_n)^{w}|O_p(n^{-1/2})|,
$$
which implies that
\begin{eqnarray*}
\|\hat{f}_n-f_n^\ast\|_n^2  &\le &  \lambda_n
(\mathcal{P}(f^\ast_n)-\mathcal{P}(\hat{f}_n))+
\|\hat{f}_n-f_n^\ast\|_n^{1-\frac{w}{2}}(\mathcal{P}(\hat{f}_n)+Kc^\frac12_n)^{w}|O_p(n^{-1/2})|
\\
&+& 2\|f-f_n^\ast\|_n\|\hat{f}_n-f_n^\ast\|_n \\
&\triangleq & R_n +2\|f-f_n^\ast\|_n\|\hat{f}_n-f_n^\ast\|_n.
\end{eqnarray*}
This inequality  holds only when either one of the following
inequalities is fulfilled.
  $$\|\hat{f}_n-f_n^\ast\|^2_n \le 2 R_n, \quad \mbox{or}$$
   \begin{eqnarray*}\|\hat{f}_n-f_n^\ast \|_n \le   4 \|f-f_n^\ast\|_n,
  \quad \mbox{and} \quad
  R_n \le  2\|f-f_n^\ast\|_n\|\hat{f}_n-f_n^\ast\|_n.
  \end{eqnarray*}
To prove this, we consider each case separately.

(i)  If $\|\hat{f}_n-f_n^\ast \|_n \le 4 \|f-f_n^\ast\|_n$,
  and $R_n \le 2\|f-f_n^\ast\|_n\|\hat{f}_n-f_n^\ast\|_n$,
then we have
  $$\| \hat{f}_n- f_n^\ast\|_n =O_p(n^{-1/(2+w)}c_n^{w/(2+w)}), $$
and then
  $$ \| \hat{f}_n-f\|_n \le \|\hat{f}_n-f_n^\ast\|_n+\|
  f_n^\ast-f\|_n=
  O_p(n^{-1/(2+w)}c_n^{w/(2+w)}).
  $$
By $R_n \le 2\|f-f_n^\ast\|_n\|\hat{f}_n-f_n^\ast\|_n$ and $ \|
\hat{f}_n-f\|_n =
  O_p(n^{-1/(2+w)}c_n^{w/(2+w)}),$  it is clear that
  $$
  \mathcal{P}(\hat{f}_n) \le n^\frac12 \cdot
  ( n^{-1/(2+w)}c_n^{w/(2+w)} )^{1+\frac{w}{2}}=c_n^{\frac12}.
  $$

 (ii) When  $\|\hat{f}_n-f_n^\ast\|_n^2 \le 2 R_n$,
  we consider the two cases separately.

 Case (ii)-1. $\mathcal{P}(\hat{f}_n)> 2 \mathcal{P}(f^\ast_n).$
    The foregoing inequality  implies that
    \begin{eqnarray*}
    0 \le  \|\hat{f}_n-f\|_n^2  \le - \lambda_n
    \mathcal{P}(\hat{f}_n)+\|\hat{f}_n-f_n^\ast\|_n^{1-\frac{w}{2}}
    \left( \frac32
    \mathcal{P}(\hat{f})\right)^w |O_p(n^{-1/2})| .
    \end{eqnarray*} Then,
    $$
    \mathcal{P}(\hat{f}_n) \le \lambda_n^{-\frac{1}{1-w}}
    \|\hat{f}_n-f_n^\ast\|_n^{\frac{2-w}{2(1-w)}}|O_p(n^{-\frac{1}{2(1-w)}})|.
    $$
Substituting this into the preceding equation, and noting that
    $\mathcal{P}(f_n^\ast)-\mathcal{P}(\hat{f}_n) \le 0$, we obtain
    $$
    \| \hat{f}_n-f_n^\ast\|_n^2 \le
    \|\hat{f}_n-f_n^\ast\|_n^{1-\frac{w}{2}} \lambda_n^{-\frac{w}{1-w}}
    \|\hat{f}_n-f_n^\ast\|_n^{\frac{w(2-w)}{2(1-w)}}|O_p(n^{-\frac{w}{2(1-w)}-\frac12})|.
    $$
Invoking the condition for $\lambda_n$, we have
    $$
    \|\hat{f}_n-f_n^\ast\|_n \le
    \lambda_n^{-\frac{2w}{2-3w}}|O_p(n^{-\frac{1}{2-3w}})|=O_p(n^{-1/(2+w)}c_n^{w/(2+w)}).
    $$
    Again,  using this in the foregoing equation, we have
    $$
    \mathcal{P}(\hat{f}_n)=O(c_n^\frac12).
    $$

 Case (ii)-2. $\mathcal{P}(\hat{f}_n)< 2 \mathcal{P}(f_n^\ast).$
    The above equation yields
      $$\|\hat{f}_n-f_n^\ast\|^2_n \le 2 \lambda_n
      (\mathcal{P}(f_n^\ast)-\mathcal{P}(\hat{f}_n))+\|\hat{f}_n-f_n^\ast\|_n^{1-\frac{w}{2}}
      (\mathcal{P}(\hat{f}_n)+\mathcal{P}(f_n^\ast))^w|O_p(n^{-1/2})|,
      $$
    which implies that either
    $$
    \|\hat{f}_n-f_n^\ast\|_n^2 \le 4 \lambda_n
    |\mathcal{P}(f_n^\ast)-\mathcal{P}(f_n^\ast)|  \le 12 K c^\frac12_n
    \lambda_n
    $$
    or
    $$
     \|\hat{f}_n-f_n^\ast\|_n^2 \le 2 \|
     \hat{f}_n-f_n^\ast\|_n^{1-\frac{w}{2}}(\mathcal{P}(\hat{f}_n)+\mathcal{P}(f_n^\ast))^w |O_p(n^{-1/2})|.
    $$
    As $\mathcal{P}(f_n^\ast)=O(c_n^{\frac12})$,  both inequalities give
    $$\|\hat{f}_n-f\|_n=O_p(n^{-1/(2+w)}c_n^{w/(2+w)}).$$

Following this equation and Proposition 1 in Stone (1985) and the
definition of $\mathcal{P}(\hat{f}_n),$ we have
$$\|\hat{f}_{nk}-f_k\|_n=O_p(n^{-1/(2+w)}c_n^{w/(2+w)}) \quad \mbox{and} \quad \mathcal{P}(\hat{f}_k)=O(c_n) \quad \mbox{for} \quad 1 \le k \le K $$
This complete the proof. $\Box$

\noindent {\sc Proof of Proposition~\ref{lemm1}:} It is easy to
verify that the functions in $\mathscr F_n(1)$ are uniformly
bounded. By applying the results for entropy bounds in Birman and
Solmjak (1967), we can easily obtain the solution. $\Box$

\noindent {\sc Proof of Theorem 2:} By Conditions~1 and 2, and the
results for the spline approximation (see de Boor, 1978), when the
number of initial knots is sufficiently large, $\min_{1 \le k \le K}
p_k>n^{\frac{1}{2p+1}}$, it is obvious that the condition (2.6) of
Theorem 1, $\|f_{k,n}-f_k\|_n=O(n^{-{1/(2+w)}}c_n^{w/{(2+w)}})$ and
$\mathcal{P}(f_{k,n}) \le c_n, k=1,\ldots,K$ are satisfied when
$c_n$ is a constant and $w=1/p$.  By Proposition~\ref{lemm1}, the
entropy condition (2.7) in Theorem 1 is also satisfied by the spline
approximation function $\sum_{k=1}^K f_{k,n}$ when $w=1/p$. Then,
letting $\lambda_n=Cn^{-2p/(1+2p)}$, it is easy to see that Theorem
2 is a corollary of Theorem 1. $\Box$

\section*{Acknowledgement}

Heng Peng's research is supported by CERG grants of Hong Kong
Research Grant Council (HKBU 201809 and HKBU 201610), FRG grant from
Hong Kong Baptist University FRG/08-09/II-33, and a grant from
National Nature Science Foundation of China (NNSF 10871054).

\end{document}